\documentclass[12pt]{article} 

\usepackage{latexsym} % Gets \Box etc
\usepackage{amssymb}  % \gtrsim, \geqslant, etc etc: see amsguide.ps
\usepackage{amsbsy}   % \pmb and \boldsymbol
\usepackage{graphicx}       % For PostScript figures
\usepackage{wasysym}
\usepackage{psfrag}
\usepackage{subfigure}

\newcommand{\txt}{\textstyle}

\newcommand{\beq}{\begin{equation}}
\newcommand{\eeq}{\end{equation}}
\newcommand{\ba}{\begin{array}}
\newcommand{\bea}{\begin{eqnarray}}
\newcommand{\ea}{\end{array}}
\newcommand{\eea}{\end{eqnarray}}

\newcommand\comment[1]{ \hbox{[{\it Comment suppressed here.}\/]} }
\newcommand\hide[1]{}

\newcommand{\tr}{\hbox{tr}}

\renewcommand{\Re}{{\rm Re}\,}

\newcommand{\skipover}[1]{}

\newcommand{\quarter}{{\txt\frac{1}{4}}}

\makeatletter %\catcode`\@=11

\def\appendix{\par                              % Have \appendix say
    \setcounter{section}{0}                     % `Appendix A', not just `A'
    \setcounter{subsection}{0}
    \renewcommand{\theequation}{\Alph{section}.\arabic{equation}}
    \renewcommand{\thesection}{Appendix \Alph{section}
                \setcounter{equation}{0}  } %Have eqns numbered (A.1) etc
}

\def\applabel#1{\@bsphack
  \protected@write\@auxout{}%
         {\string\newlabel{#1}{{\Alph{section}}{\thepage}}}%
  \@esphack}
\def\section{
\setcounter{equation}{0}        % Reset eqn numbers at start of section
\@startsection {section}{1}{\z@}{-3.5ex plus -1ex minus 
 -.2ex}{2.3ex plus .2ex}{\large\bf}}
\renewcommand{\theequation}{\arabic{section}.\arabic{equation}}

\def\subsection{\@startsection{subsection}{2}{\z@}{-3.25ex plus -1ex minus 
 -.2ex}{1.5ex plus .2ex}{\normalsize\bf}}

\def\subsubsection{\@startsection{subsubsection}{3}{\z@}{-3.25ex plus
 -1ex minus -.2ex}{1.5ex plus .2ex}{\normalsize}}

\makeatother   %\catcode`\@=12

\newsavebox{\eqlabel}
\makeatletter  %\catcode`\@=11
\newlength{\numblen}
\newsavebox{\eqnumb}
\def\@eqnnum{\savebox{\eqnumb}{\rm (\theequation)}%
\settowidth{\numblen}{\usebox{\eqnumb}}%
\makebox[\numblen][l]{\usebox{\eqnumb}~~~\usebox{\eqlabel}}}
\makeatother   %\catcode`\@=12

\newenvironment{equationwithlabel}[1]{ %
  \begin{equation}\label{#1} }{\end{equation}} %\savebox{\eqlabel}{~}}
\newcommand{\beql}[1]{\begin{equationwithlabel}{#1}}
\newcommand{\eeql}{\end{equationwithlabel}}
\newcommand{\vp}{{\mathbf p}}    % vector p
\newcommand{\vk}{{\mathbf k}}    % vector k
\newcommand{\vq}{{\mathbf q}}    % vector q
\newcommand{\vx}{{\mathbf x}}    % vector x
\newcommand{\vl}{{\boldsymbol{\ell}}}    % vector \ell
\newcommand{\dm}{{\delta\mu}} % delta mu
\newcommand{\mubar}{{\bar\mu}}% mu bar

\newcommand{\pslash}{{\not\! p}}
\newcommand{\qslash}{{\not\! q}}
\newcommand{\muslash}{{\not\! \mu}}
\newcommand{\delslash}{{\not\! \partial}}

\begin{document}

\title{{\bf The Crystallography of Color Superconductivity}}

\author{
	Jeffrey A. Bowers and Krishna Rajagopal \\[0.5ex]
{\normalsize Center for Theoretical Physics}\\
{\normalsize Massachusetts Institute of Technology}\\
%{\normalsize 77 Mass Ave }\\ 
{\normalsize Cambridge, MA 02139 }
}

\newcommand{\preprintno}{
  \normalsize MIT-CTP-3259,\ \ NSF-ITP-02-25 % \\ DRAFT VERSION
}

\date{April 5, 2002 \\[1ex] \preprintno}

\begin{titlepage}
\maketitle
\def\thepage{}          % No page number on title page

\begin{abstract}
We develop the Ginzburg-Landau approach to comparing
different possible crystal structures for the crystalline
color superconducting phase of QCD, the QCD incarnation of 
the Larkin-Ovchinnikov-Fulde-Ferrell phase.  In this phase, quarks 
of different flavor with differing Fermi momenta form Cooper
pairs with nonzero total momentum, yielding a condensate that
varies in space like a sum of plane waves.  We work at zero
temperature, as is relevant for compact star physics.
The Ginzburg-Landau
approach predicts a strong first-order phase transition 
(as a function of the chemical potential difference
between quarks) and
for this reason is not under quantitative control.  Nevertheless, 
by organizing
the comparison between different possible arrangements of
plane waves ({\it i.e.}~different crystal structures) it provides
considerable qualitative insight into what makes a crystal
structure favorable.  Together, the qualitative insights and
the quantitative, but not controlled, calculations  
make a compelling case that the favored pairing pattern 
yields a condensate which 
is a  sum of eight plane waves forming a face-centered cubic
structure. They also predict that the phase is quite robust, with 
gaps comparable in magnitude to the BCS gap that would 
form if the Fermi momenta were degenerate.
These predictions may be tested in ultracold gases made of fermionic 
atoms.  In a QCD context, our results lay the foundation for
a calculation of vortex pinning
in a crystalline color superconductor, and thus for the 
analysis of pulsar glitches that may originate
within the core of a compact star.
\end{abstract}

\end{titlepage}

\renewcommand{\thepage}{\arabic{page}}
%\setcounter{page}{1}

%------------------------------------------------------------------------
\section{Introduction}
\label{sec:intro}

Since quarks that are antisymmetric in color attract,
cold dense quark matter is unstable to the formation of
a condensate of Cooper pairs, making it a color 
superconductor~\cite{Barrois,BailinLove,ARW1,RappETC,Reviews}.  
At asymptotic densities, the ground state of QCD with 
quarks of three flavors ($u$, $d$ and $s$) 
is expected to be the color-flavor locked (CFL) 
phase~\cite{CFL,Reviews}.
This phase features a condensate of Cooper pairs of
quarks that includes $ud$, $us$, and $ds$ pairs. Quarks
of all colors and all flavors participate in the
pairing, and all excitations with quark quantum numbers are
gapped. 
As in any BCS state, the Cooper pairing in the CFL
state pairs quarks whose momenta are equal in 
magnitude and opposite in direction, and pairing is strongest
between pairs of quarks whose momenta are both near their respective
Fermi surfaces.  

Pairing persists even in the face of a stress (such
as a chemical potential difference or a mass difference)
that seeks to push the quark Fermi surfaces apart, although
a stress that is too 
strong will 
ultimately disrupt BCS pairing.
The CFL phase is the ground
state for real QCD, assumed to be in equilibrium
with respect to the weak interactions, as long
as the density is high enough. 
Now, imagine decreasing the quark number chemical potential $\mu$ 
from asymptotically large values. 
The quark matter 
at first remains color-flavor locked, 
although the CFL condensate may rotate
in flavor space as terms of order $m_s^4$
in the free energy become important~\cite{BedaqueSchaefer}. 
Color-flavor locking is maintained
until an ``unlocking transition'', which  
must be first order~\cite{ABR2+1,SW2}, occurs 
when~\cite{ABR2+1,SW2,neutrality,ARRW,AlfordRajagopal}
\begin{equation}\label{ApproxUnlocking}
\mu \approx m_s^2 / 4\Delta_0\ .
\end{equation}
In this expression, $\Delta_0$ is the BCS pairing gap,
estimated in both models and asymptotic analyses to be of order 
tens to 100 MeV~\cite{Reviews}, and  
$m_s$ is the strange quark mass parameter.  
Note that $m_s$ includes the contribution
from any $\langle \bar s s \rangle$ condensate 
induced by the nonzero current strange quark mass, making it a 
density-dependent effective mass.
At densities that may occur at the center of compact stars, 
corresponding to $\mu\sim 400-500$ MeV, $m_s$ is certainly significantly
larger than the current quark mass, and its value is not well known.
In fact, $m_s$ decreases discontinuously at the unlocking 
transition~\cite{BuballaOertel}.
Thus, the criterion (\ref{ApproxUnlocking}) can only be used
as a rough guide to the location of the unlocking transition
in nature~\cite{BuballaOertel,AlfordRajagopal}.
Given this quantitative uncertainty, there remain two logical possibilities
for what happens as a function of decreasing $\mu$.
One possibility is a first-order phase transition directly from
color-flavor locked quark matter to hadronic matter,
as explored in Ref.~\cite{ARRW}.  The second possibility
is an unlocking transition, 
followed only at a lower $\mu$ by a transition to hadronic matter.
We assume the second possibility here, and explore its consequences.

One may think that $ud$ BCS pairing could persist below the 
unlocking transition. However, this does not occur in
electrically neutral bulk matter~\cite{AlfordRajagopal}.
The unlocking transition is a transition from the CFL
phase to ``unpaired'' quark matter in which electrical neutrality
is enforced by an electrostatic potential $\mu_e\sim m_s^2/4\mu$,
to lowest order in $m_s/\mu$, meaning that the Fermi momenta
are related (to this order) by
\begin{eqnarray}
p_F^d &=& p_F^u + \frac{m_s^2}{4\mu}\nonumber\\
p_F^s &=& p_F^u - \frac{m_s^2}{4\mu}\ .
\label{UnpairedFermiMomenta}
\end{eqnarray}

If this ``unpaired'' quark matter exists in some window
of $\mu$ between hadronic matter and CFL quark matter, there {\it will}
be pairing 
in this window: all that Ref.~\cite{AlfordRajagopal}
shows is that there will be no BCS pairing between quarks of
different flavors.  One possible pattern of pairing is the 
formation of $\langle uu \rangle$, 
$\langle dd \rangle$ and $\langle ss \rangle$ condensates.
These must be either $J=1$ or symmetric in color, whereas
the QCD interaction favors the formation of color
antisymmetric $J=0$ pairs. The gaps in these phases may be
as large as of order 1 MeV~\cite{Schaefer1Flavor}, or
could be much smaller~\cite{ARW1}.  Another possibility
for pairing in the ``unpaired'' quark matter is crystalline
color 
superconductivity~\cite{BowersLOFF,ngloff,LOFFphonon,pertloff,massloff,Giannakis},
which involves pairing between quarks whose momenta do not
add to zero, as first considered in a terrestrial condensed
matter physics context by Larkin, Ovchinnikov,
Fulde and Ferrell (LOFF)~\cite{LO,FF}.
Unpaired quark matter with Fermi momenta (\ref{UnpairedFermiMomenta})
is susceptible to the formation of a crystalline color superconducting
condensate constructed from pairs of quarks that are antisymmetric
in color and flavor, where both quarks have momenta near their respective
unpaired Fermi surfaces.
We shall argue in this paper
that the crystalline color superconducting
phase is more robust than previously thought, with 
gaps comparable to $\Delta_0$. In fact, crystalline color
superconducting pairing lowers the free energy so much that
its pairing energy should be taken into account in the
comparison (\ref{ApproxUnlocking}). This comparison should
be redone as a competition between the crystalline color superconducting
and CFL phases: crystalline color superconductivity in
the ``unpaired'' quark matter is sufficiently robust that it
will delay the transition to the CFL phase to a significantly
higher density than that in (\ref{ApproxUnlocking}).

The idea behind crystalline color superconductivity is 
that 
once CFL pairing is disrupted, leaving quarks with 
differing Fermi momenta that are unable to participate
in BCS pairing,  
it is natural to ask whether there is some generalization
of the pairing ansatz in which pairing 
between two species of quarks persists even once their
Fermi momenta differ.   
It may 
be favorable for quarks with differing Fermi momenta 
to form pairs whose momenta are {\it not} equal
in magnitude and opposite in sign~\cite{LO,FF,BowersLOFF}.  
This generalization of the pairing ansatz (beyond BCS ans\"atze
in which only quarks with momenta which add to zero pair) is favored because
it gives rise to a region of phase space where {\it both}
of the quarks
in a pair are close to their respective Fermi surfaces,
and such pairs can be created at low cost in free energy.
Condensates of this sort spontaneously
break translational and rotational invariance, leading to
gaps that vary in a crystalline pattern.

Crystalline color superconductivity may occur within
compact stars.  With this context in mind, it is
appropriate for us to work at zero temperature because
compact stars that are more than a few minutes old
are several orders of magnitude colder than the 
tens-of-MeV-scale critical temperatures for BCS or crystalline
color superconductivity.
As a function of increasing depth in a compact star,
$\mu$ increases, $m_s$ decreases, and $\Delta_0$ changes also.
This means that in some shell within
the quark matter core of a neutron star (or within a strange
quark star),  $m_s^2/\mu\Delta_0$ may lie within the
appropriate window where
crystalline color superconductivity is favored.
Because this phase is a (crystalline) superfluid,
it will be threaded with vortices in
a rotating compact star.  Because these rotational
vortices may pinned in place by 
features of the crystal structure, such a shell may be
a locus for glitch phenomena~\cite{BowersLOFF}.

To date, crystalline color superconductivity has only
been studied in simplified models with
pairing between two quark species whose
Fermi momenta are pushed apart
by a chemical
potential difference~\cite{BowersLOFF,ngloff,LOFFphonon,pertloff,Giannakis}
or a mass difference~\cite{massloff}.
We suspect that in reality, in three-flavor quark matter
whose unpaired Fermi momenta are split as in (\ref{UnpairedFermiMomenta}),
the color-flavor pattern of pairing
in the crystalline phase will involve $ud$,
$us$ and $ds$ pairs, just as in the CFL phase.  
Our focus here, though, is on the 
form of the crystal structure. That is, we wish to focus on
the dependence of the crystalline condensate on position
and momenta. As in previous work,  
we therefore simplify the color-flavor pattern to one involving
massless $u$ and $d$ 
quarks only, with Fermi momenta split by introducing
chemical potentials
\begin{eqnarray}
\mu_d &=& \bar \mu + \delta\mu\nonumber\\
\mu_u &=& \bar \mu - \delta\mu \ .
\end{eqnarray}
In this toy model, we vary $\delta\mu$ by hand. 
In three-flavor quark matter, the analogue of
$\delta\mu$ is controlled by the nonzero strange
quark mass and the requirement of electrical neutrality and
would be of order $m_s^2/4\mu$ as in (\ref{UnpairedFermiMomenta}).

In our toy model, we shall take the interaction between
quarks to be pointlike, with the quantum numbers of
single-gluon exchange.  
This $s$-wave interaction is a reasonable starting point
at accessible densities but 
is certainly {\it in}appropriate at asymptotically
high density, where the interaction between quarks (by gluon
exchange) is dominated by forward scattering.  
The crystalline color superconducting state
has been analyzed at asymptotically high densities in
Refs.~\cite{pertloff,Giannakis}.  As we shall explain,
the analysis of Ref.~\cite{pertloff} indicates that
the crystal structure in this circumstance would be different
from that we obtain.

Our toy model may
turn out to be a better model for the analysis of LOFF
pairing in atomic systems. (There,
the phenomenon could be called ``crystalline superfluidity''.)
Recently, ultracold gases of fermionic atoms have been cooled
down to the degenerate regime, with temperatures less
than the Fermi energy~\cite{ColdFermions}, and reaching
the pairing transition (perhaps by increasing the atom-atom
interaction rather than by further reducing the temperature) seems
a reasonable possibility~\cite{AtomicBCS}.
In such systems, there really are only two 
species of atoms (two spin states)
that pair with each other,
whereas in QCD our model is a toy model for a system
with nine quarks.  In the atomic systems, the interaction
will be $s$-wave dominated whereas in QCD, it remains to be
seen how good  this approximation is at accessible densities.
Furthermore, in the atomic physics context experimentalists
can control the densities of the two different atoms that pair,
and in particular can  tune their density difference. This means
that experimentalists wishing to search for crystalline
superfluidity have the ability to dial the 
most relevant control parameter~\cite{AtomicLOFF,CombescotMora}.
In QCD, in contrast, $\delta\mu$ is
controlled by $m_s^2/\mu$, meaning that it is up to nature
whether, and if so at what depth in a compact star, crystalline
color superconductivity occurs.

\begin{figure}
\centering
\psfrag{muu}[r][r]{$\mu_u$}
\psfrag{mud}[l][l]{$\mu_d$}
\psfrag{p}[l][l]{$\vp$}
\psfrag{mp}[r][r]{$-\vp$}
\psfrag{p2q}[r][r]{$-\vp+2\vq$}
\psfrag{pp}[r][r]{$\vp'$}
\psfrag{mpp}[l][l]{$-\vp'$}
\psfrag{pp2q}[l][l]{$-\vp'+2\vq$}
\psfrag{2q}[c][c]{$2\vq$}
\psfrag{psiu}[c][l]{$\psi_u$}
\psfrag{psid}[c][l]{$\psi_d$}
\includegraphics[width=3.5in]{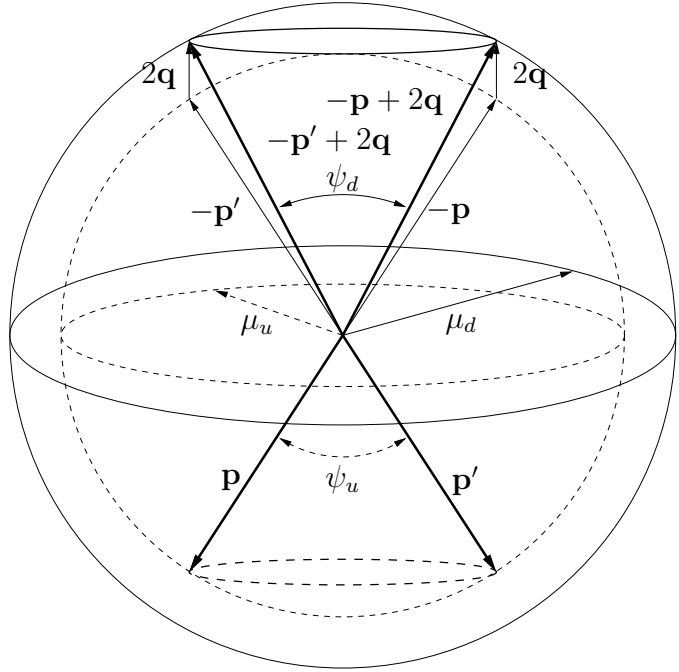}
\caption{
\label{ringsfig}
In the simplest LOFF state, in which every Cooper
pair carries momentum $2{\bf q}$, an up quark
with momentum $\vp$ (or $\vp'$) pairs with
a down quark with momentum $-\vp+2\vq$ (or $-\vp'+2\vq$).
In the figure $\vq$ has been taken to point upwards.
The pairing is dominated by pairing between up quarks
in a band centered
on the dashed ring shown on the up Fermi surface,
and down quarks in a band centered on 
the solid ring shown on the down Fermi surface.
The vector sum of the momenta in any pair
is $2\vq$, and the pairing bands are defined
as those pairs such that both quarks in
a pair are within $\Delta$ of their respective Fermi surface.
The thickness of the pairing bands is thus of order $\Delta$.  
The width of the pairing bands is of order $\Delta\bar\mu/\dm$,
meaning that their angular width is of order $\Delta/\dm$.
We take a weak coupling limit in which $\Delta_0\rightarrow 0$
while $\dm/\Delta_0$ and $|\vq|/\Delta_0$ are held fixed
and nonzero.  In this limit, $\Delta/\Delta_0$
also stays fixed, and therefore so does the angular
width of the pairing bands. In this limit, 
the two Fermi surfaces in the figure get closer and closer
together and 
the angles $\psi_u$ and $\psi_d$ become degenerate, and
take on the value $\psi_0=2\cos^{-1}(\dm/|\vq|)\simeq 67.1^\circ$.
If, in addition, we  take the further
limit in which $\dm/\Delta_0 \rightarrow \dm_2/\Delta_0$,
approaching the second-order phase transition, then
in this Ginzburg-Landau limit the angular extent of the pairing
bands shrinks to zero, reducing them to only the rings shown.
}
\end{figure}

In the simplest LOFF state, each Cooper pair carries 
momentum $2{\bf q}$.
Although the magnitude $|{\bf q}|$ is determined
energetically, the direction $\hat{\bf q}$
is chosen spontaneously.  
The condensate is dominated by
those regions in momentum space in which a quark pair
with total momentum $2{\bf q}$ has both members of
the pair within of order $\Delta$ of their respective 
Fermi surfaces. These regions form circular bands
on the two Fermi surfaces, as shown in Fig.~\ref{ringsfig}.  
Making the ansatz that all Cooper pairs make the same choice of
direction $\hat{\bf q}$ corresponds
to choosing a single circular band on each Fermi surface, as
in Fig.~\ref{ringsfig}.
In position space, it corresponds to a condensate that
varies in space like 
\begin{equation}\label{PlaneWaveCondensate}
\langle \psi({\bf x}) \psi({\bf x})\rangle \propto \Delta e^{2i{\bf q}
\cdot {\bf x}}\ .  
\end{equation} 
This ansatz is certainly {\it not} the best choice, because
it only allows a small fraction of all the quarks near their
Fermi surfaces to pair.
If a single plane wave is favored, why not two? That is,
if one choice of $\hat{\bf q}$ is favored, why not add 
a second ${\bf q}$, with the same $|{\bf q}|$ but
a different $\hat{\bf q}$, to allow
more quarks near their Fermi surfaces to pair?  
If two are favored, why not three?
Why not eight?  If eight plane waves are favored, how should
their $\hat{\bf q}$'s be oriented?  These are the sort of questions
we seek to answer in this paper.

Upon making the plane-wave ansatz (\ref{PlaneWaveCondensate}),
we know from previous work~\cite{LO,FF,Takada,BowersLOFF} that this simplest
LOFF phase is favored over the BCS state and over no pairing
at all in a range $\delta\mu_1< \delta\mu<\delta\mu_2$,
where $\dm_1=\Delta_0/\sqrt{2}\simeq 0.707\Delta_0$ and
$\dm_2\simeq 0.754\Delta_0$.
At $\delta\mu_2$, there is a second-order phase transition
at which $\Delta/\Delta_0\rightarrow 0$ and
$|{\bf q}|/\Delta_0$ tends to a nonzero limit,
which we shall denote $q_0/\Delta_0$, where 
$q_0\simeq 0.90\Delta_0\simeq 1.2 \delta\mu_2$.\footnote{In 
high density QCD
in the limit of a large number of colors
$N_c$, the ground state is not a color superconductor. Instead,
it features a
chiral condensate that varies in some crystalline
pattern~\cite{DGR},
although at least at weak coupling it seems
that this phase only occurs for $N_c$'s of order
thousands~\cite{ShusterSon,PRWZ}.  For this chiral crystal,
which is qualitatively different from a crystalline
color superconductor because $|\vq|=\mu$ rather than
$|\vq|\sim \dm$, several possible crystal
structures have been analyzed in Ref.~\cite{RappCrystal}.}
At the second-order phase transition,
the quarks that participate in the crystalline pairing lie
on circular rings on their Fermi surfaces that are characterized
by an opening angle $\psi_0=2\cos^{-1}(\dm/|\vq|)\simeq 67.1^\circ$ 
and an angular width that is of
order $\Delta/\dm$, which therefore tends to zero.  
At $\delta\mu_1$
there is a first-order phase transition at which the LOFF 
solution with gap $\Delta$ is superseded by the BCS solution
with gap $\Delta_0$. 
(The analogue in QCD would be a LOFF window in $m_s^2/\mu$, with
CFL below and unpaired quark matter above.)

Solving the full gap equation for more general crystal structures
is challenging.  In order to proceed, we take advantage of 
the second-order phase transition at $\delta\mu=\delta\mu_2$.
Because $\Delta$ tends to zero there, we can write the gap
equation as a Ginzburg-Landau expansion, working order by 
order in $\Delta$.  In Section~2, we develop this expansion
to order $\Delta^6$ for a crystal made up of the sum of
arbitrarily many plane waves, all with the same $|{\bf q}|=q_0$
but with different patterns of $\hat \vq$'s.  In Section~3
we present results for a large number of crystal structures.

The Ginzburg-Landau
calculation finds many crystal structures that are much
more favorable than the single plane wave (\ref{PlaneWaveCondensate}).
For many crystal structures, it 
predicts 
a strong first-order phase transition, at
some $\delta\mu_* \gg \delta\mu_2$, between unpaired quark
matter and a crystalline phase with a $\Delta$ that
is comparable in magnitude to $\Delta_0$. 
Once 
$\delta\mu$ is reduced to $\delta\mu_2$, where the single
plane wave would just be beginning to develop, these 
more favorable solutions already have very robust 
condensation energies, perhaps even larger
than that of the BCS phase.  These results are exciting,
because they suggest that the crystalline phase is much
more robust than previously thought.  However, they cannot
be trusted quantitatively because the Ginzburg-Landau 
analysis is controlled in the limit $\Delta\rightarrow 0$,
and we find a first-order phase transition to a state
with $\Delta\neq 0$.  

Even though it is quite a different problem, we can look
for inspiration to the Ginzburg-Landau analysis of the crystallization
of a solid from a liquid~\cite{Chaikin}. There too, a
Ginzburg-Landau analysis predicts a first-order phase
transition, and thus predicts its own quantitative
downfall.  But, qualitatively it is correct: it predicts
the formation of a body-centered-cubic crystal and experiment
shows that most elementary solids are body-centered cubic
near their first-order crystallization transition.

Thus inspired, let us ask what crystal structure our
Ginzburg-Landau analysis predicts for the crystalline
color superconducting phase.  
To order $\Delta^2$, we learn that $|\vq|\simeq 1.2\dm$
even when $\dm\neq \dm_2$. We also 
learn 
that, apparently, the more  plane waves  the better.  We learn nothing
about the preferred arrangement of the set of ${\hat\vq}$'s.
By extending the calculation 
to order $\Delta^4$ and $\Delta^6$, we find:
\begin{itemize} 
\item
Crystal structures with intersecting pairing rings are strongly
disfavored.
Recall that each $\hat\vq$ is associated with pairing among 
quarks that lie on
one ring of opening angle $\psi_0 \simeq 67.1^\circ$ 
on each Fermi surface. We find
that any crystal structure in which such rings 
intersect pays a large free energy price.
The favored crystal structures should therefore 
be those whose
set of $\hat\vq$'s are characterized by the maximal number of
nonintersecting rings.  The maximal number of nonintersecting
rings with opening angle $67.1^\circ$ that fit on a sphere
is nine~\cite{extremal,tammes}.
\item
We also find that those crystal structures characterized by
a set of ${\bf q}$'s within which there are a large number
of combinations satisfying ${\bf q}_1 - {\bf q}_2 +{\bf q}_3 -
{\bf q}_4=0$ and a large number of combinations 
satisfying ${\bf q}_1 - {\bf q}_2 +{\bf q}_3 -
{\bf q}_4+{\bf q}_5 - {\bf q}_6 =0$  are favored.
Speaking loosely, ``regular'' structures are favored over
``irregular'' structures.
The only configurations of nine nonintersecting rings
are rather irregular, whereas if we limit ourselves to 
eight rings, there is a regular choice which is favored
by this criterion: choose eight ${\bf q}$'s pointing towards
the corners of a cube. In fact, a deformed cube which
is slightly taller or shorter than it is wide (a cuboid) 
is just as good.
\end{itemize}

These qualitative arguments are supported by the 
quantitative results of our Ginzburg-Landau analysis, which 
does indeed indicate that the most favored
crystal structure is a cuboid that is very close
to a cube. This crystal
structure is so favorable that the coefficient of 
$\Delta^6$ in the Ginzburg-Landau expression for the 
free energy, which
we call $\gamma$, is large and negative.  (In fact, we find several 
crystal structures with negative $\gamma$, but the cube has
by far the most negative $\gamma$.)  We could go on, to 
$\Delta^8$ or higher, until we found a Ginzburg-Landau
free energy for the cube which is bounded from below.
However, we know that this free energy would give a strongly
first-order phase transition, meaning that the Ginzburg-Landau
analysis would anyway not be under quantitative control.
A better strategy, then, is to use the Ginzburg-Landau
analysis to understand the physics at a qualitative level,
as we have done.  With an understanding of why
a crystal structure with eight plane waves whose wave vectors
point to the corners of a cube is strongly favored in hand,
the next step would be 
to make this ansatz and solve the gap equation without
making a Ginzburg-Landau approximation. We leave this to future work. 
The eight ${\bf q}$'s that describe the crystal structure
we have found to be most favorable are eight of the vectors
in the reciprocal lattice of a face-centered-cubic crystal.
Thus, as we describe and depict in Section~4,
the crystal structure we propose is one in which the
condensate forms a face-centered-cubic
lattice in position space.

\section{Methods}
\label{sec:method}

\subsection{The gap equation}

We study the crystalline superconducting phase in a toy model
for QCD that has two massless flavors of quarks and a pointlike
interaction.  
%Keep in mind that although our calculations proceed in
%this particular context, our results are quite generic and apply to
%any binary system where two species of fermion have a weak,
%attractive, pointlike interaction.  
The Lagrange function 
%for our toy model 
is
\begin{equation}
\mathcal{L} = \bar \psi (i \delslash + \muslash) \psi - \frac{3}{8} \lambda 
(\bar\psi \Gamma^A \psi)( \bar\psi \Gamma_A \psi)
\end{equation}
where $\muslash = \gamma^0 (\mubar - \tau_3 \dm)$.  The $\tau$'s are
Pauli matrices in flavor space, so the up and down quarks have
chemical potentials as in (\ref{UnpairedFermiMomenta}).            
The vertex is $\Gamma^A =
\gamma^\mu T^a$ so that our pointlike interaction mimics the spin,
color, and flavor structure of one-gluon exchange. (The $T^a$ are color
$SU(3)$ generators normalized so that $\tr (T^a T^b) = 2
\delta^{ab}$.) We denote the coupling
constant in the model by $\lambda$.

\begin{figure}
\centering
\begin{psfrags}
\psfrag{deltalabel}[cc][lc]{$\mathbf{\Delta}$}
\parbox{1.5in}{\includegraphics[width=1.25in]{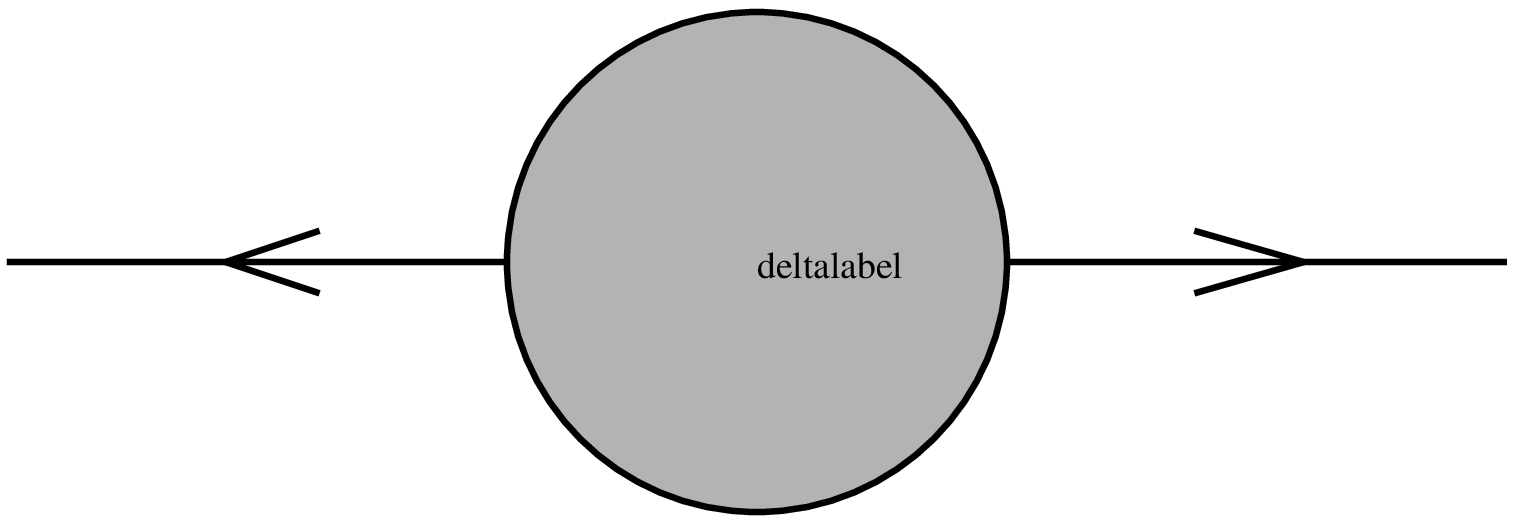}}
\end{psfrags}
$=$ \hspace{0.25in}
\parbox{1.25in}{\includegraphics[width=1in]{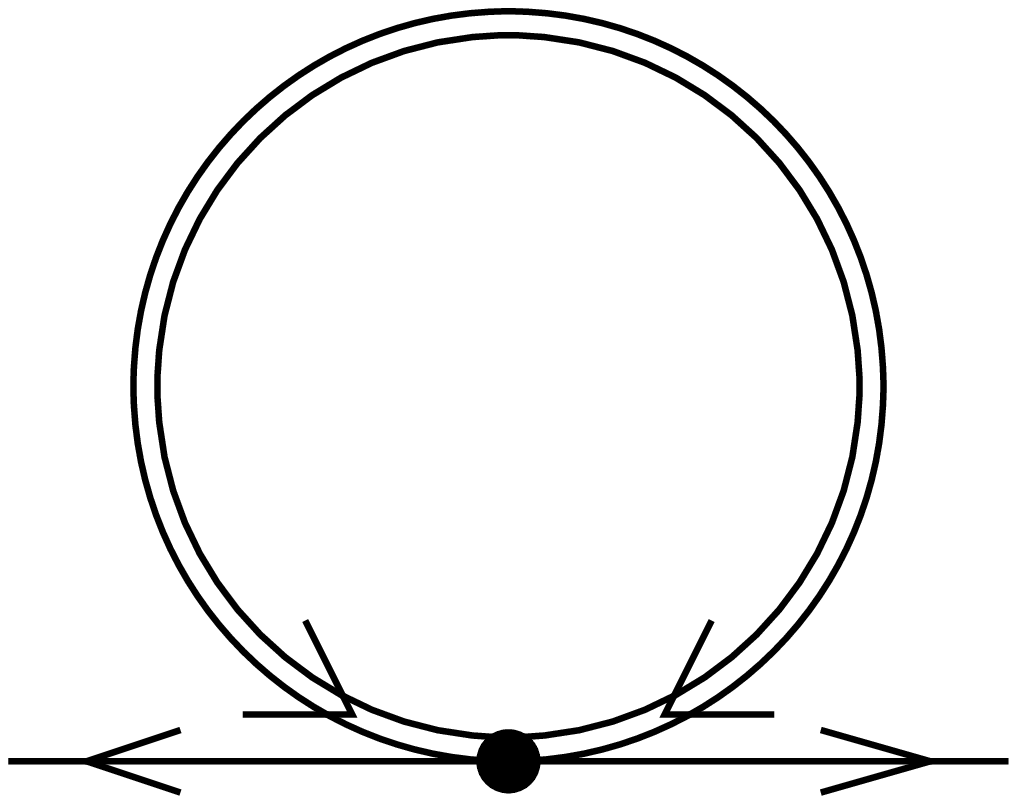}}
\vspace{0.2in}
\caption{
\label{sdeqnfig}
The Schwinger-Dyson graph for the LOFF gap parameter
$\mathbf{\Delta}$.  The black dot is the pointlike interaction vertex and the
double line represents the full anomalous propagator $F$, which
is given in terms of $\mathbf{\Delta}$ in Fig.~\ref{expansionfig}.  }
\end{figure}

It is convenient to use a Nambu-Gorkov diagrammatic method to obtain
the gap equation for the crystalline phase.  Since we are
investigating a phase with spatial inhomogeneity, we begin in position
space.  We introduce the two-component spinor $\Psi(x) = (\psi(x),
\bar\psi^T(x))$ and the quark propagator $i S(x,x') = \langle \Psi(x)
\bar\Psi(x') \rangle$, which has ``normal'' and ``anomalous''
components $G$ and $F$, respectively:
\begin{equation}
i S(x,x') = \left( \begin{array}{cc} iG(x,x') & i F(x,x') \\ i \bar F(x,x') & i\bar G(x,x') \end{array} \right) = \left( \begin{array}{cc} \langle \psi(x) \bar\psi(x') \rangle & \langle \psi(x) \psi^T(x') \rangle \\ \langle \bar\psi^T(x) \bar\psi(x') \rangle & \langle \bar\psi^T(x) \psi^T(x') \rangle \end{array} \right).
\end{equation}
The conjugate propagators $\bar F$ and $\bar G$ satisfy 
\bea
\label{barprops}
i \bar G(x,x') & = & \gamma^0 (i G(x',x))^\dag \gamma^0 \\  
i \bar F(x,x') & = & \gamma^0 (i F(x',x))^\dag \gamma^0.
\eea
The gap parameter $\mathbf\Delta(x)$ that describes the
diquark condensate
is related to the 
anomalous propagator $F$ 
%is related to the gap parameter
%$\mathbf\Delta(x)$ of the diquark condensate 
by a Schwinger-Dyson equation
\begin{equation}
\mathbf\Delta(x) = i \frac{3}{4} \lambda \Gamma^A F(x,x) \Gamma_A^T
\label{sdeqn}
\end{equation}
illustrated diagrammatically in Fig.~\ref{sdeqnfig}.
In our toy model, we are neglecting quark masses and
thus the normal part of the one-particle-irreducible
self-energy is zero; the anomalous part of the 1PI self energy
is just $\mathbf\Delta(x)$.
The crystal order parameter $\mathbf\Delta(x)$ defined
by (\ref{sdeqn}) is a 
matrix in spin, flavor and color space.  In the mean-field
approximation, we can use the equations of motion for $\Psi(x)$ to
obtain a set of coupled equations that determine the propagator
functions in the presence of the diquark condensate 
characterized by $\mathbf\Delta(x)$: 
\beql{ngeqns}
\left(
\begin{array}{cc} 
i \delslash + \muslash & \mathbf\Delta(x) \\
 \bar\mathbf\Delta(x) & (i \delslash - \muslash)^T 
\end{array}
\right) 
\left(
\begin{array}{cc}
G(x,x') & F(x,x') \\  \bar F(x,x') & \bar G(x,x') 
\end{array}
\right) = 
%S(x,x') = 
\left( 
\begin{array}{cc}
1 & 0 \\
0 & 1 
\end{array}
\right)
\delta^{(4)}(x-x')
\eeql
%\beql{ngeqns}
%\ba{rcl}
%(i \delslash + \muslash) G(x,x') - \mathbf\Delta(x) \bar F(x,x') & = & \delta^{(4)}(x-x') \\
%(-i \delslash^T + \muslash) \bar F(x,x') + \bar\mathbf\Delta(x) G(x,x') & = & 0
%\ea
%\eeql
where $\bar\mathbf\Delta(x) = \gamma^0 \mathbf\Delta(x)^\dag
\gamma^0$.  Any function $\mathbf\Delta(x)$ that solves equations
(\ref{sdeqn}) and (\ref{ngeqns}) is a stationary point of the free energy
functional $\Omega [ \mathbf\Delta(x) ]$; of these stationary points,
the one with the lowest $\Omega$ describes the 
ground state of the system.  
Our task, then, is to invert (\ref{ngeqns}), obtaining
$F$ in terms of $\mathbf\Delta(x)$, substitute in (\ref{sdeqn}),
find 
solutions for $\mathbf\Delta(x)$, and then evaluate $\Omega$
for all solutions we find.

There are some instances where analytic solutions to equations
(\ref{sdeqn}) and (\ref{ngeqns}) can be found.  The simplest case is that
of a spatially uniform condensate.  Translational invariance then
implies that the propagators are diagonal in momentum space: $S(p,p')
= S(p) (2 \pi)^4 \delta^{(4)}(p-p')$.  In this case, 
Eqs.~(\ref{ngeqns}) immediately yield
\begin{equation}
S(p)^{-1} = 
\left( 
\begin{array}{cc}
\pslash + \muslash & \mathbf\Delta \\
\bar\mathbf\Delta & (\pslash - \muslash)^T 
\end{array}
\right),
\end{equation}
which is easily inverted to obtain $S$, which can then
be substituted on the right-hand side of Eq.
(\ref{sdeqn}) to obtain a self-consistency equation ({\it i.e.}~a gap
equation) for $\mathbf\Delta$.  The solution of this gap equation 
describes the
familiar ``2SC'' phase~\cite{Barrois,BailinLove,ARW1,RappETC,Reviews}, 
a two-flavor, two-color BCS condensate,
\begin{equation}
\mathbf\Delta = T^2 \tau_2 C \gamma_5 \Delta_0,
\label{2SCchannel}
\end{equation}
where $T^2$, $\tau_2$, and $C \gamma_5$ indicate that the condensate
is a color antitriplet, flavor singlet, and Lorentz scalar,
respectively.\footnote{The QCD interaction, and thus the
interaction in our toy model,
is attractive in the color
and flavor antisymmetric channel and this dictates the
color-flavor pattern of (\ref{2SCchannel}).  Our toy model
interaction  does not distinguish
between the Lorentz scalar (\ref{2SCchannel}) 
and the pseudoscalar possibility.
But, the instanton interaction in QCD favors the scalar condensate.}
The remaining factor $\Delta_0$, which without loss of generality
can be taken to be real,  gives the magnitude of
the condensate.  In order to solve the resulting gap equation
for $\Delta_0$, we must complete the specification of our
toy model by introducing a cutoff.  In previous
work~\cite{ARW1,RappETC,CFL,Reviews},  it has been shown that
if, for a given cutoff, the coupling
$\lambda$ is chosen so that the model describes a reasonable
vacuum chiral condensate, then at $\mu\sim 400-500$ MeV
the model describes a diquark
condensate which has $\Delta_0$ of order 
tens to 100 MeV.
Ratios between physical observables
depend only weakly on the cutoff, meaning that when $\lambda$ 
is taken to vary with the cutoff such that one observable is
held fixed, others depend only weakly on the cutoff.
For this reason, we are free to 
make a convenient choice of cutoff so long as we then choose
the value of $\lambda$ that yields the ``correct'' $\Delta_0$.
Since we do not really know the correct value of $\Delta_0$
and since this is after all only a toy model, we simply think of
$\Delta_0$ as the single free parameter in the model,
specifying the strength of the interaction and thus the 
size of the BCS condensate.
Because the quarks near the Fermi surface 
contribute most to pairing, it is convenient 
to introduce a cutoff $\omega$ defined so as to 
restrict the gap integral to 
momentum modes near the Fermi surface ($||\vp| - \bar\mu | \leq \omega$).
In the weak coupling (small $\lambda$) 
limit, the explicit solution to the gap equation is then
\begin{equation}
\label{bcsgapeqn}
\Delta_0 = 2 \omega e^{-\pi^2/2 \lambda \bar\mu^2}\ .
\end{equation}
This is just the familiar BCS result for the gap. (Observe
that the density of states at the Fermi surface is
$N_0 = 2\bar\mu^2/\pi^2$.)  We denote the gap 
for this BCS solution by $\Delta_0$,
reserving the symbol $\Delta$ for the gap parameter
in the crystalline phase.
We shall see explicitly below that when
we express our results for $\Delta$
relative to $\Delta_0$, they are completely 
independent the cutoff $\omega$ 
as long as $\Delta_0/\mu$ is small.

The BCS phase, with $\Delta_0$ given by (\ref{bcsgapeqn}),
has a lower free energy than unpaired quark matter as
long as $\dm<\dm_1= \Delta_0/\sqrt{2}$~\cite{Clogston}.
The first-order unpairing transition at $\dm=\dm_1$
is the analogue in our two-flavor toy model of the unlocking
transition in QCD.
For $\dm>\dm_1$, the free energy of any crystalline
solution we find below must be compared to that of unpaired
quark matter; for $\dm<\dm_1$, crystalline solutions should be
compared to the BCS phase.  We shall work at $\dm>\dm_1$.

The simplest example of a LOFF condensate is one that
varies like a
plane wave: $\mathbf\Delta(x) = \mathbf\Delta \exp(- i 2 q \cdot x)$.
The condensate is static, meaning that $q = (0, \vq)$.  
We shall denote $|\vq|$ by $q_0$. In this condensate, the
momenta of two quarks in a Cooper pair is $(\vp+\vq,-\vp+\vq)$
for some $\vp$, meaning that the total momentum of each and every
pair is $2\vq$.
See Refs.~\cite{ngloff,massloff} for an analysis of this condensate
using the Nambu-Gorkov formalism. Here, we sketch the results.
If we shift the definition of $\Psi$ in momentum space to
$\Psi_q(p)\equiv(\psi(p+q),\bar\psi(-p+q))$, then in
this shifted basis the propagator is diagonal:
\begin{equation}
i S_q(p,p')= \langle\Psi_q(p)\bar\Psi_q(p')\rangle = i S_q(p)
\delta^4(p-p')
\label{momentumshift}
\end{equation}
and the inverse propagator is simply
\begin{equation}
\label{planewave2}
S_q(p)^{-1} = 
\left( 
\begin{array}{cc}
\pslash + \qslash + \muslash & \mathbf\Delta \\
\bar\mathbf\Delta & (\pslash - \qslash - \muslash)^T 
\end{array}
\right) \ .
\end{equation}
See Refs.~\cite{ngloff,massloff}
for details and to see
how this equation can be inverted and
substituted into Eq.~(\ref{sdeqn}) to obtain a gap equation for $\Delta$.   
This gap equation has nonzero solutions for 
$\dm<\dm_2\simeq 0.7544 \Delta_0$, and has a second-order
phase transition at $\dm=\dm_2$ with $\Delta \sim (\dm_2-\dm)^{1/2}$.
We rederive these results below.

\begin{figure}[t]
\centering
\begin{psfrags}
\psfrag{deltalabel}[cc][lc]{$\mathbf{\Delta}_{\vq_2}$}
\psfrag{deltabarlabel}[cc][lc]{$\bar\mathbf{\Delta}_{\vq_1}$}
\psfrag{p1}{$\vp$}
\psfrag{p2}{$-\vp + 2\vq_1$}
\psfrag{p3}{$\vp - 2\vq_1 + 2\vq_2$}
\parbox{3.25in}{\includegraphics[width=3.25in]{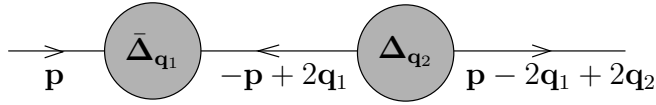}}
\end{psfrags}
\vspace{0.2in}
\caption{
\label{scatterfig}
A process whereby a quark with momentum $\vp$ scatters by interactions
with two plane-wave condensates and acquires a momentum $\vp - 2\vq_1 + 2\vq_2$.
}
\end{figure}

If the system is unstable to the formation of a single
plane-wave condensate, we might expect that a condensate of multiple
plane waves is still more favorable.  
Again our goal is to find gap
parameters $\mathbf\Delta(x)$ that are self-consistent solutions of
Eqs.~(\ref{sdeqn}) and (\ref{ngeqns}).  We use an ansatz that
retains the Lorentz, flavor, and color structure of the 2SC phase:
\begin{equation}
\mathbf\Delta(x) = T^2 \tau_2 C \gamma_5 \Delta(x)
\end{equation}
but now $\Delta(x)$ is a scalar function that characterizes the
spatial structure of the crystal.  We write this function as a
superposition of plane waves:
\begin{equation}
\Delta(x) = \sum_{\vq} \Delta_\vq e^{-i 2 q \cdot x} 
\label{spatialansatz}
\end{equation}
where, as before,  $q = (0,\vq)$.  
The $\{ \Delta_\vq \}$ constitute a set
of order parameters for the crystalline phase.  Our task is to
determine for which set of $\vq$'s the $\Delta_\vq$'s are nonzero.
Physically, for each
$\Delta_\vq\neq 0$ the condensate includes some Cooper pairs for which the
total momentum of a pair is $2 \vq$.  This is indicated by the
structure of the anomalous propagator $F$ in momentum space: 
Eqs.~(\ref{sdeqn}) and (\ref{spatialansatz}) together imply that
\begin{equation}
%F(p,p') = -i \langle \bar\psi^T(-p) \bar\psi(p') \rangle  = \sum_\vq  F_\vq(p) (2\pi)^4 \delta^{(4)}(p-p'+2q)
F(p,p') = -i \langle \psi(p) \psi^T(-p') \rangle  = \sum_\vq  F_\vq(p) (2\pi)^4 \delta^{(4)}(p-p'-2q)
\end{equation}
and 
\begin{equation}
\mathbf{\Delta}_\vq = i \frac{3}{4} \lambda \int \frac{d^4 p}{(2\pi)^4} \Gamma^A F_\vq(p) \Gamma_A^T
\label{psdeqn}
\end{equation}
where $\mathbf{\Delta}_\vq = T^2 \tau_2 C \gamma_5 \Delta_\vq$.
Eq.~(\ref{psdeqn}) yields an infinite set of coupled gap equations, one
for each $\vq$.  (Note that each $F_\vq$ depends on all 
the $\Delta_\vq$'s.)
It is not consistent to choose only a 
finite set of $\Delta_\vq$ to be nonzero
because when multiple plane-wave condensates are present, these
condensates induce an infinite ``tower'' (or lattice) of 
higher momentum condensates.  This is easily understood by noting that
a quark with momentum $\vp$ can acquire an additional
momentum $2 \vq_2 - 2 \vq_1$ by 
interacting with two different plane-wave condensates as
it propagates through the medium, as shown in Fig.~\ref{scatterfig}.  
Note that this
process cannot occur when there is only a single plane-wave
condensate.  The analysis of the single plane-wave condensate
closes with only a single nonzero $\Delta_\vq$, and is therefore
much easier than the analysis of a generic crystal structure.
Another way that this difficulty manifests itself is that 
once we move beyond the single plane-wave
solution to a more generically nonuniform condensate, 
it is no longer possible to diagonalize the propagator
in momentum space by a shift, as was possible 
in Eqs.~(\ref{momentumshift}, \ref{planewave2}).
 
\subsection{The Ginzburg-Landau approximation}

The infinite system of equations 
(\ref{psdeqn}) has been solved analytically only in one
dimension, 
where it turns out that the gap parameter can be
expressed as a Jacobi elliptic function that, as promised, is composed
of an infinite number of plane waves~\cite{LOFF1D}.  
In three dimensions, the crystal structure of the LOFF
state remains unresolved~\cite{BuzdinKachkachi,Houzet,CombescotMora}.
In the vicinity of the second-order transition at $\dm_2$,
however, we can simplify the calculation considerably by
utilizing the smallness of $\Delta$ to make a controlled
Ginzburg-Landau approximation.  
This has the advantage of providing a controlled 
truncation of the
infinite series of plane waves, because near $\dm_2$ the system is
unstable to the formation of plane-wave condensates only for $\vq$'s
that fall on a sphere of a certain radius $q_0$, as we shall see
below.  This was in fact the
technique employed by Larkin and Ovchinnikov in their
original paper~\cite{LO}, and it has been further developed
in Refs.~\cite{BuzdinKachkachi,Houzet,CombescotMora}.
As far as we know, though, no previous authors
have done as complete a study of possible crystal structures
in three dimensions as we attempt. Most have limited
their attention to, at most, structures 1, 2, 5 and 9 from
the 23 structures we describe in Fig.~\ref{stereographicfig}
and Table~1 below. As
far as we know, no previous authors have investigated
the crystal structure that we find to be most favorable.

The authors of Refs.~\cite{BuzdinKachkachi,Houzet,CombescotMora}
have focused on using the Ginzburg-Landau
approximation at nonzero temperature, near the critical
temperature at which the LOFF condensate vanishes.
Motivated by our interest in compact stars, 
we follow Larkin and Ovchinnikov in staying at $T=0$
while using the fact that, for
a single plane-wave condensate, $\Delta\rightarrow 0$ for
$\dm\rightarrow \dm_2$ to motivate the Ginzburg-Landau
approximation.  
The down side of this is that,
in agreement with previous authors,
we find that the $T=0$ phase transition becomes first order
when we generalize beyond a single
plane wave.
In the end, therefore, the lessons of our Ginzburg-Landau
approximation must be taken qualitatively.  We nevertheless
learn much that is of value.
\begin{figure}[t]
\psfrag{deltalabel}[cc][lc]{$\mathbf{\Delta}$}
\psfrag{deltabarlabel}[cc][lc]{$\bar\mathbf{\Delta}$}
\psfrag{delta}[cc][lc]{$\mathbf{\Delta}$}
\psfrag{deltabar}[cc][lc]{$\bar\mathbf{\Delta}$}
\begin{eqnarray}
\parbox{.8in}{\includegraphics[width=.8in]{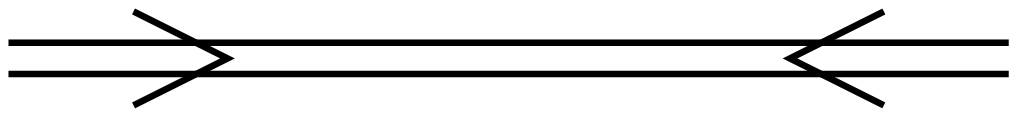}}
& = & 
-
\hspace{0.1in}
\parbox{1.2in}{\includegraphics[width=1.2in]{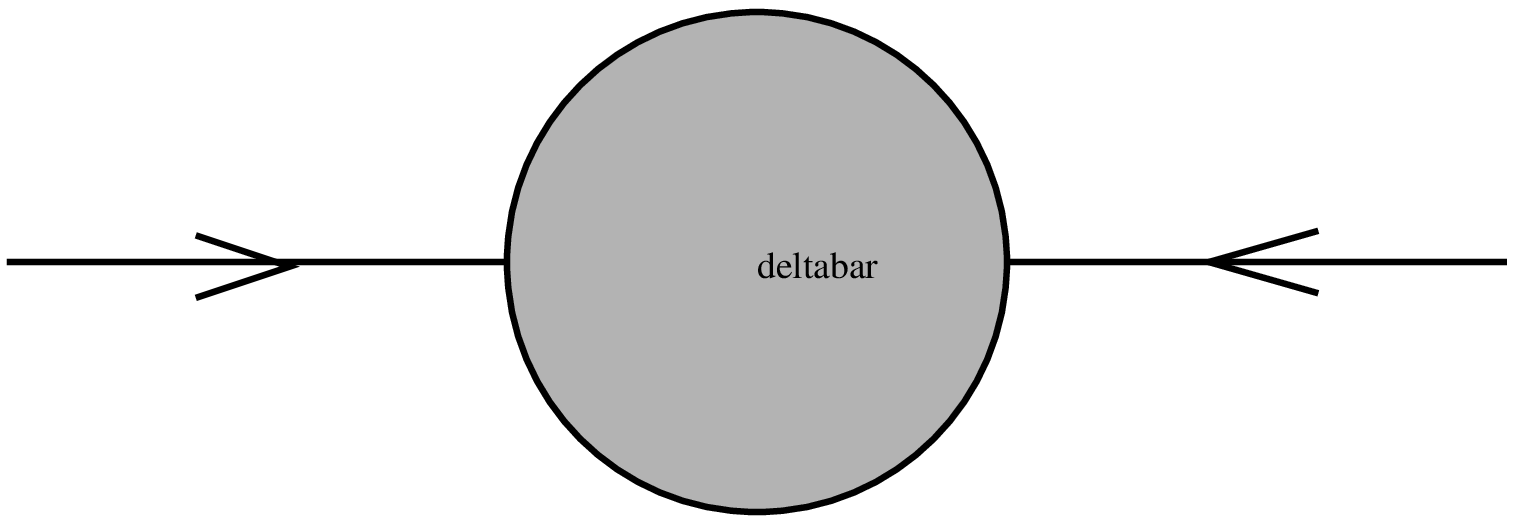}} 
\hspace{0.1in}
+
\hspace{0.1in}
\parbox{2.4in}{\includegraphics[width=2.4in]{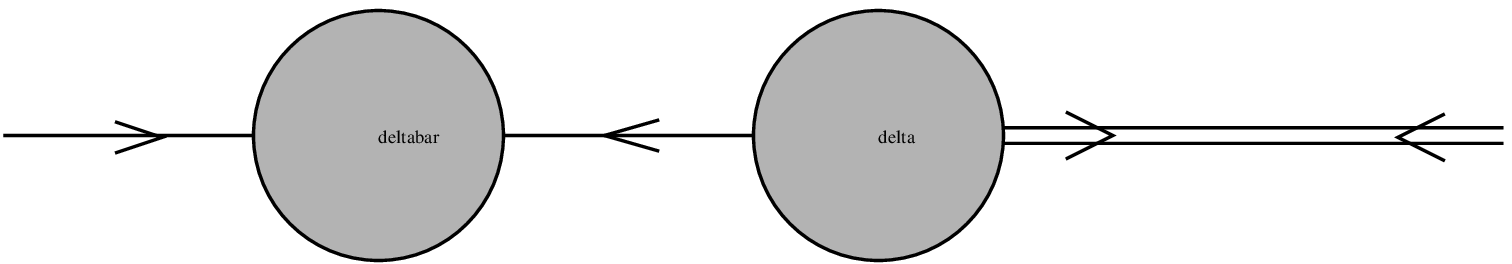}} \nonumber \\
 & & \parbox{3in}{} \nonumber \\
& = & 
-
\hspace{0.1in}
\parbox{1.2in}{\includegraphics[width=1.2in]{deltabar.eps}} 
\hspace{0.1in}
-
\hspace{0.1in}
\parbox{2.8in}{\includegraphics[width=2.8in]{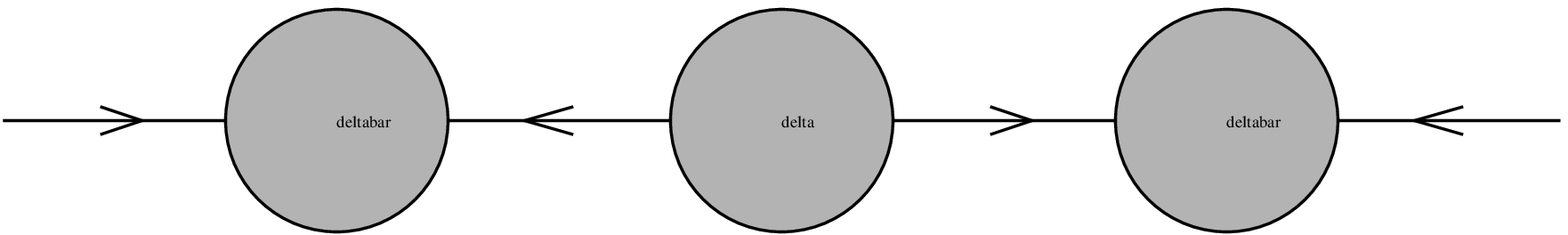}} \nonumber \\
& &
- 
\hspace{0.1in}
\parbox{4.4in}{\includegraphics[width=4.4in]{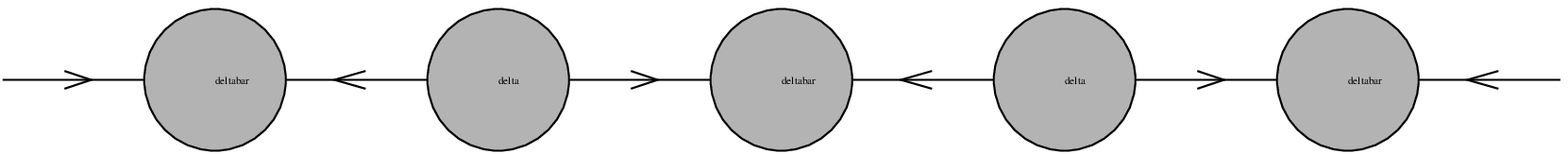}} \nonumber \\
& & 
- 
\hspace{0.1in}
\cdots \nonumber
\end{eqnarray}
\caption{
\label{expansionfig}
The diagrammatic expression for the full anomalous propagator $\bar F$, and the
first three terms in the series expansion in powers of $\Delta$.  }
\end{figure}

To proceed with the Ginzburg-Landau expansion, we first integrate
equations (\ref{ngeqns}) to obtain
\bea
\label{intngeqns}
G(x,x') & = & G^{(0)}(x,x') - \int d^4 z \ G^{(0)}(x,z) \mathbf\Delta(z) \bar F(z,x') \\
\bar F(x,x') & = & -\int d^4 z \ \bar G^{(0)}(x,z) \bar\mathbf\Delta(z) G(z,x')
\eea
where $G^{(0)} = (i \delslash + \muslash)^{-1}$, $\bar G^{(0)} = ((i
\delslash - \muslash)^T)^{-1}$.  Then we expand these equations in
powers of the gap function $\mathbf\Delta(x)$.  For $\bar F(x,x')$ we
find (suppressing the various spatial
coordinates and integrals for notational simplicity)
\beql{} 
\ba{rcl} 
\bar F & = & -\bar G^{(0)} \bar\mathbf\Delta G^{(0)} - \bar G^{(0)}
\bar\mathbf\Delta G^{(0)} \mathbf\Delta \bar G^{(0)} \bar\mathbf\Delta
G^{(0)} \\ & & - \bar G^{(0)} \bar\mathbf\Delta G^{(0)} \mathbf\Delta
\bar G^{(0)} \bar\mathbf\Delta G^{(0)} \mathbf\Delta \bar G^{(0)}
\bar\mathbf\Delta G^{(0)} + \mathcal{O}(\Delta^7) 
\ea 
\eeql 
as expressed diagrammatically in Fig.~\ref{expansionfig}.  We then
substitute this expression for $\bar F$
into the right-hand side of the Schwinger-Dyson 
equation (actually the conjugate of equation (\ref{psdeqn})).  After some spin,
color, and flavor matrix manipulation, the result in momentum space is
\vfill\eject

\begin{eqnarray}
\label{fullgapeqn}
\Delta_\vq^* & = & - \frac{2 \lambda \bar\mu^2}{\pi^2}\Pi(\vq) \Delta_\vq^* - \frac{2 \lambda \bar\mu^2}{\pi^2} \sum_{\vq_1,\vq_2,\vq_3} J(\vq_1\vq_2\vq_3\vq) \Delta_{\vq_1}^* \Delta_{\vq_2} \Delta_{\vq_3}^* \delta_{\vq_1-\vq_2+\vq_3-\vq} \nonumber \\
 & & - \frac{2 \lambda \bar\mu^2}{\pi^2} \sum_{\vq_1,\vq_2,\vq_3,\vq_4,\vq_5}K(\vq_1\vq_2\vq_3\vq_4\vq_5\vq) \Delta_{\vq_1}^* \Delta_{\vq_2} \Delta_{\vq_3}^* \Delta_{\vq_4} \Delta_{\vq_5}^* \delta_{\vq_1-\vq_2+\vq_3-\vq_4+\vq_5-\vq} \nonumber \\
 & & + \mathcal{O}(\Delta^7) 
\end{eqnarray}

\begin{figure}[t]
\Large
\psfrag{deltabar}[cc][lc]{\small $\Delta_\vq^*$}
\psfrag{delta}[cc][lc]{\small $\Delta_\vq^*$}
\psfrag{delta1}[cc][lc]{\small $\Delta_{\vq_1}^*$}
\psfrag{delta2}[cc][lc]{\small $\Delta_{\vq_2}$}
\psfrag{delta3}[cc][lc]{\small $\Delta_{\vq_3}^*$}
\psfrag{delta4}[cc][lc]{\small $\Delta_{\vq_4}$}
\psfrag{delta5}[cc][lc]{\small $\Delta_{\vq_5}^*$}
\begin{eqnarray}
\parbox{1in}{\includegraphics[width=1in]{deltabar.eps}} & = & - \parbox{0.9in}{\includegraphics[width=0.9in]{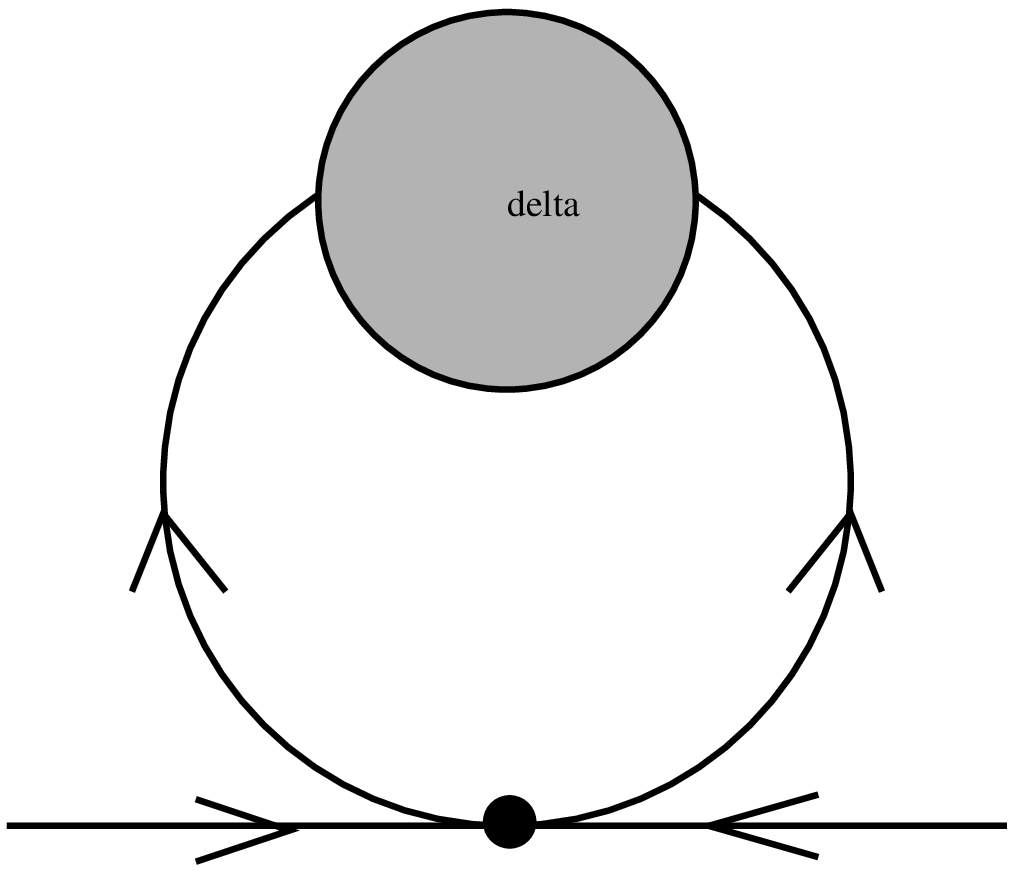}} -  \sum_{\mbox{\small $\vq_1 - \vq_2 + \vq_3 = \vq$}} \parbox{1.25in}{\includegraphics[width=1.25in]{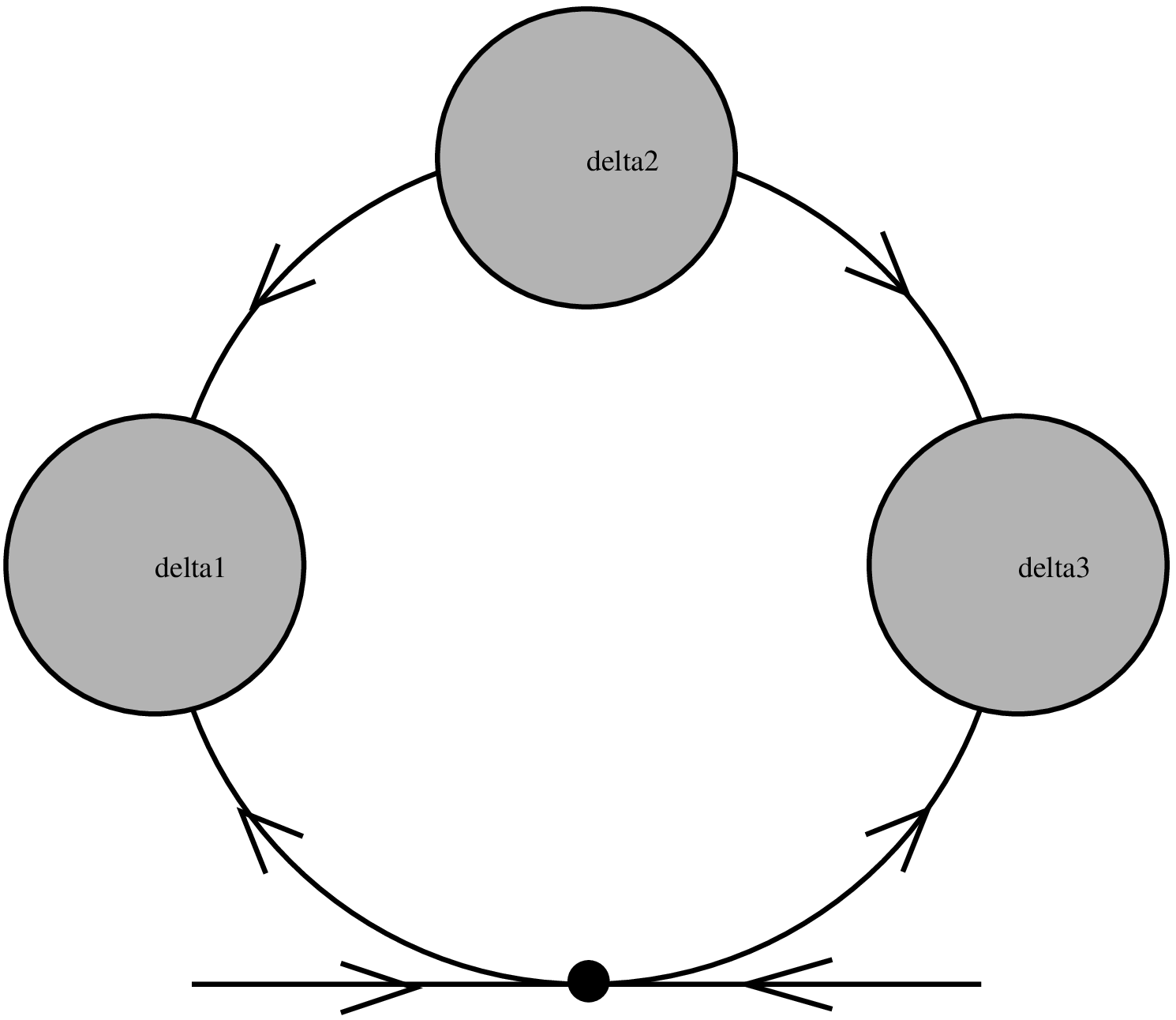}} \nonumber \\
 & & - \sum_{\mbox{\small $\vq_1 - \vq_2 + \vq_3 - \vq_4 + \vq_5 = \vq$}} \parbox{1.4in}{\includegraphics[width=1.4in]{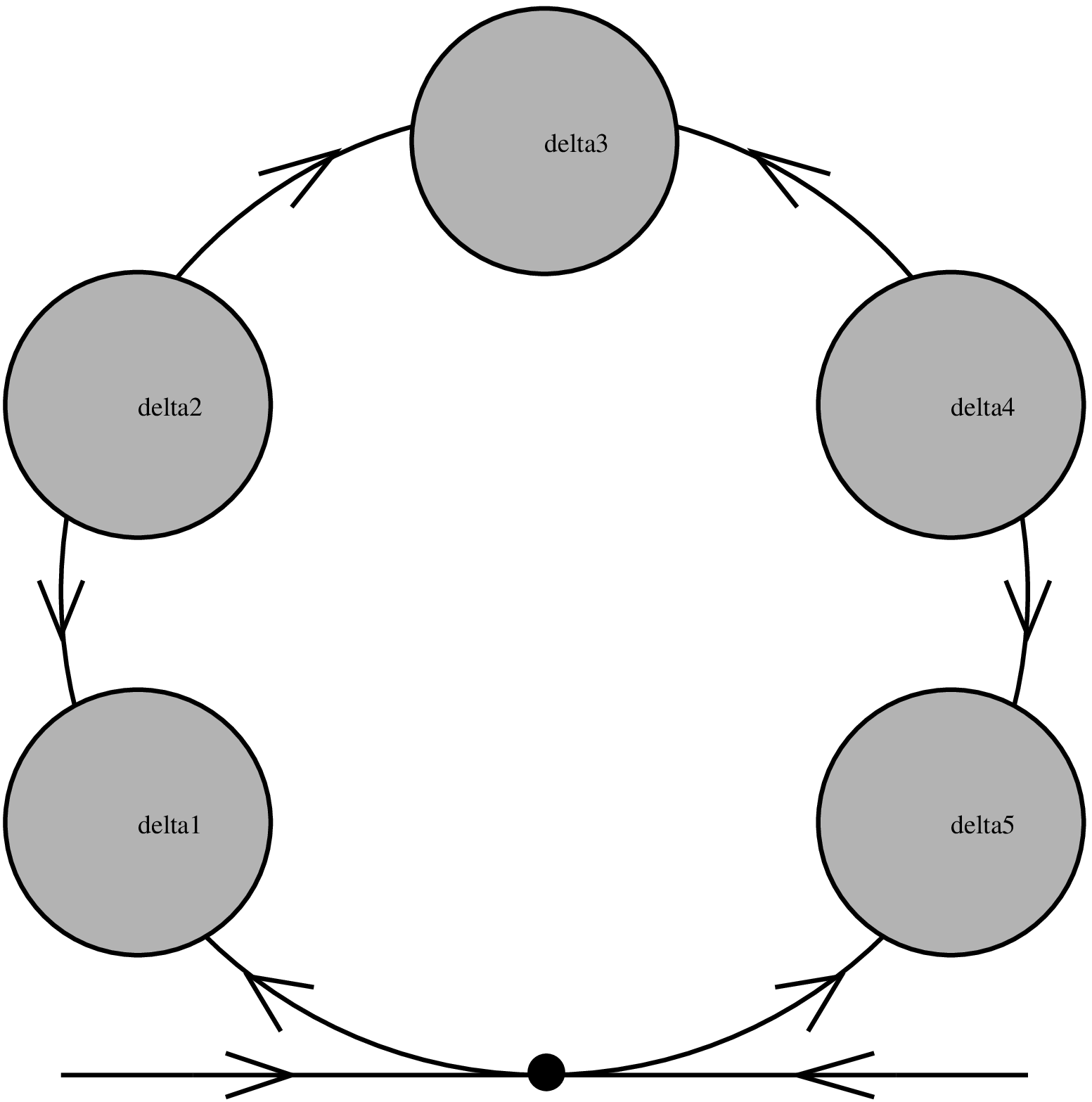}} + \cdots \nonumber
\end{eqnarray}
\caption{
\label{gapeqnfig}
The series expansion corresponding to Eqs.~(\ref{fullgapeqn})
and (\ref{integrals}).  
This diagrammatic equation is obtained 
by substituting the series expansion of Fig.~\ref{expansionfig} 
into the Schwinger-Dyson equation of Fig.~\ref{sdeqnfig}.  
}
\end{figure}

\noindent
as shown in Fig.~\ref{gapeqnfig}.  The prefactors have been chosen
for later convenience.  The functions $\Pi$, $J$, and $K$
corresponding to the three graphs in Fig.~\ref{gapeqnfig}
are given by:
\begin{eqnarray}
\label{integrals}
\Pi(\vq) & = & -i \frac{\pi^2}{\bar\mu^2} \int \frac{d^4 p}{(2\pi)^4} \gamma_\mu (\pslash - \muslash_d)^{-1} (\pslash + 2 \qslash + \muslash_u)^{-1} \gamma^\mu \nonumber \\
J(\vq_1\vq_2\vq_3\vq_4) & = & -i\frac{\pi^2}{\bar\mu^2} \int \frac{d^4 p}{(2\pi)^4} \gamma_\mu (\pslash - \muslash_d)^{-1} (\pslash + 2 \qslash_1 + \muslash_u)^{-1} \nonumber \\
 & & \times (\pslash + 2 \qslash_1 - 2\qslash_2 - \muslash_d)^{-1} (\pslash + 2 \qslash_1 - 2\qslash_2 + 2\qslash_3 + \muslash_u)^{-1} \gamma^\mu \nonumber \\ 
K(\vq_1\vq_2\vq_3\vq_4\vq_5\vq_6) & = & -i\frac{\pi^2}{\bar\mu^2} \int \frac{d^4 p}{(2\pi)^4} \gamma_\mu (\pslash - \muslash_d)^{-1} (\pslash + 2 \qslash_1 + \muslash_u)^{-1} \nonumber \\
 & & \times (\pslash + 2 \qslash_1 - 2\qslash_2 - \muslash_d)^{-1} (\pslash + 2 \qslash_1 - 2\qslash_2 + 2\qslash_3 + \muslash_u)^{-1}  \nonumber \\
 & & \times  (\pslash + 2 \qslash_1 - 2\qslash_2 +2\qslash_3 - 2\qslash_4 - \muslash_d)^{-1} \nonumber \\
 & & \times (\pslash + 2 \qslash_1 - 2\qslash_2 + 2\qslash_3 - 2\qslash_4 + 2\qslash_5 + \muslash_u)^{-1} \gamma^\mu. \nonumber \\
\end{eqnarray}

We shall see that $\dm$ and $|\vq|$ are both of order $\Delta$
which in turn is of order $\Delta_0$.  This means that all these
quantities are much less than
$\bar\mu$ in the weak coupling limit.
Thus, in the weak coupling limit we can choose the cutoff $\omega$
such that $\dm, |\vq| \ll \omega \ll \bar\mu$. In this
limit, $J$ and $K$ are independent of the cutoff $\omega$,
as we shall see in the appendix where we present their
explicit evaluation.  In this limit, 
\begin{eqnarray}
\label{Piandalpha}
\Pi(\vq) &=& 
\left[-1+ \frac{\dm}{2|\vq|} \log\left( \frac{|\vq| + \dm}{|\vq| - \dm}\right)
- \frac{1}{2}\log \left( \frac{\omega^2}{\vq^2 - \dm^2} \right) \right]
\nonumber\\
&=& -\frac{\pi^2}{2\lambda\bar\mu^2} +  
\left[-1+ \frac{\dm}{2|\vq|} \log\left( \frac{|\vq| + \dm}{|\vq| - \dm}\right)
- \frac{1}{2}\log \left( \frac{\Delta_0^2}{4(\vq^2 - \dm^2)} \right) \right]
\nonumber\\
&=& -\frac{\pi^2}{2\lambda\bar\mu^2} +
\alpha\left(\frac{|\vq|}{\Delta_0},\frac{\dm}{\Delta_0}\right)\ ,
\end{eqnarray}
where we have used the explicit solution to the
BCS gap equation (\ref{bcsgapeqn}) to eliminate the cutoff $\omega$
in favor of the BCS gap $\Delta_0$, and where the last equation
serves to define $\alpha$.  Note that $\alpha$
depends on the cutoff $\omega$ only through $\Delta_0$,
and depends only on the
ratios $|\vq|/\Delta_0$ and $\dm/\Delta_0$.

It will prove convenient to use the definition of $\alpha$ to
rewrite the Ginzburg-Landau
equation 
Eq.~(\ref{fullgapeqn}) as
\begin{eqnarray}
\label{newfullgapeqn}
0 & = & \alpha(|\vq|)\Delta_\vq^* \,+
\sum_{\vq_1,\vq_2,\vq_3} J(\vq_1\vq_2\vq_3\vq) 
\Delta_{\vq_1}^* \Delta_{\vq_2} \Delta_{\vq_3}^* 
\delta_{\vq_1-\vq_2+\vq_3-\vq} 
\nonumber \\   
&  & + \sum_{\vq_1,\vq_2,\vq_3,\vq_4,\vq_5}K(\vq_1\vq_2\vq_3\vq_4\vq_5\vq) 
\Delta_{\vq_1}^* \Delta_{\vq_2} \Delta_{\vq_3}^* 
\Delta_{\vq_4} \Delta_{\vq_5}^* \delta_{\vq_1-\vq_2+\vq_3-\vq_4+\vq_5-\vq} 
\nonumber \\
&  & + \mathcal{O}(\Delta^7).
\end{eqnarray}

\begin{figure}[t]
\centering
\psfrag{0.2}[tc][tc]{\small 0.2}
\psfrag{0.4}[tc][tc]{\small 0.4}
\psfrag{0.6}[tc][tc]{\small 0.6}
\psfrag{0.8}[tc][tc]{\small 0.8}
\psfrag{1}[tc][tc]{\small 1.0}
\psfrag{1.2}[tc][tc]{\small 1.2}
\psfrag{1.4}[tc][tc]{\small 1.4}
\psfrag{1.6}[tc][tc]{\small 1.6}
\psfrag{y1}[rc][rc]{\small 0.2}
\psfrag{y2}[rc][rc]{\small 0.4}
\psfrag{y3}[rc][rc]{\small 0.6}
\psfrag{y4}[rc][rc]{\small 0.8}
\psfrag{y5}[rc][rc]{\small 1.0}
\psfrag{xlabel}{$|\vq|/\Delta_0$}
\psfrag{ylabel}{$\dm/\Delta_0$}
\includegraphics[width=4in]{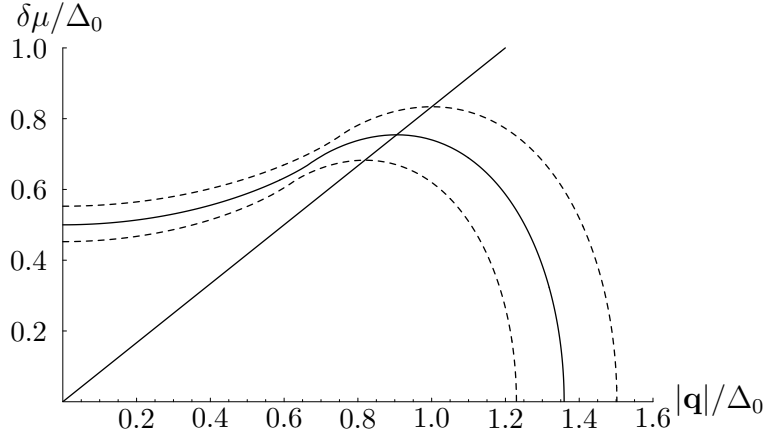}
\caption{
\label{zerogapfig}
Along the solid curve,
$\alpha(|\vq|,\dm)=0$.  The maximum
$\dm$ reached by this curve is $\dm=0.754\Delta_0\equiv\dm_2$,
which occurs at $|\vq|=0.9051\Delta_0=1.1997\dm_2$.  The diagonal line
is $|\vq|=1.1997\dm$.  Along the
upper and lower dashed curves, $\alpha=+0.1$  and $\alpha=-0.1$,
respectively.
}
\end{figure}

To learn how to interpret $\alpha$, consider 
the single plane-wave condensate in which $\Delta_\vq\neq 0$
only for a single $\vq$.
If we divide equation (\ref{fullgapeqn}) by $\Delta_\vq^*$, 
we see that the equation
$\Pi = -\pi^2/2\lambda\bar\mu^2$, 
which is to say $\alpha=0$, defines a curve
in the space of $(|\vq|,\dm)$ 
where we can find a solution to the gap equation with
$\Delta_\vq\rightarrow 0$, with $|\vq|$ on the curve and
for any $\hat{\bf q}$. 
This curve is shown in Fig.~\ref{zerogapfig}.  
We shall see below that when only one $\Delta_\vq$ is nonzero,
the ``$J$ sum'' and ``$K$ sum'' in (\ref{fullgapeqn}) are both
positive. This means that wherever $\alpha<0$, {\it i.e.} below
the solid curve in Fig.~\ref{zerogapfig}, there are solutions with 
$\Delta_\vq\neq 0$ for these values of $|\vq|$,
and wherever $\alpha>0$, {\it i.e.}~above the solid
curve, there are no single
plane-wave solutions to the gap equation.  
The solid curve in Fig.~\ref{zerogapfig} therefore 
marks the boundary of the instability towards the formation
of a single plane-wave condensate.
The highest point on this curve is special, as it denotes
the maximum value of $\dm$ for which a single plane-wave
LOFF condensate can arise. This second-order critical point
occurs at
$(|\vq|,\dm) = (q_0, \dm_2)$ with $\dm_2
\simeq 0.7544 \Delta_0$ and $q_0/\dm_2 \simeq 1.1997$, where
$\Delta_0$ is the BCS gap of Eq.~(\ref{bcsgapeqn}).  

As $\dm\rightarrow\dm_2$ from above, only those plane waves
lying on a sphere in momentum space with $|\vq|=q_0$ are becoming 
unstable to condensation.  If we analyze them one by one,
all these plane waves are equally unstable. That is, 
in the
vicinity of the critical point $\dm_2$, 
the LOFF gap equation admits plane-wave
condensates with $\Delta_\vq\neq 0$ for a single $\vq$ lying
somewhere on the sphere
$|\vq| = q_0$. For each such plane wave, the
paired quarks occupy a ring with opening angle   
$\psi_0 = 2 \cos^{-1}(\dm/|\vq|) \simeq 67.1^\circ$
on each Fermi surface, as shown
in Fig.~\ref{ringsfig}.

\subsection{The free energy}

In order
to compare different crystal structures,
with (\ref{newfullgapeqn}) in hand, 
we can now derive a 
Ginzburg-Landau free energy functional
$\Omega[\Delta(x)]$ which characterizes the system in the vicinity of
$\dm_2$, where $\Delta\rightarrow 0$. 
This is most readily obtained by noting that the 
gap equations (\ref{newfullgapeqn}) must be equivalent to 
\begin{equation}
\frac{\partial\Omega}{\partial\Delta_{\vq}} = 0
\end{equation}
because solutions to the gap equations are stationary points
of the free energy.  This determines the free energy
up to an overall multiplicative constant, which can be found by
comparison with the single plane-wave solution previously known.  
The
result is
\begin{eqnarray} 
\label{freeenergy}
\frac{\Omega}{N_0} & = &   \alpha(q_0) 
\sum_{\vq,\ |\vq|=q_0} \Delta_\vq^* \Delta_\vq + \frac{1}{2} \sum_{\vq_1\cdots\vq_4, \ |\vq_i|=q_0} J(\vq_1\vq_2\vq_3\vq_4) \Delta_{\vq_1}^* \Delta_{\vq_2} \Delta_{\vq_3}^* \Delta_{\vq_4} \delta_{\vq_1-\vq_2+\vq_3-\vq_4} \nonumber \\
 & & + \frac{1}{3} \sum_{\vq_1\cdots\vq_6, \ |\vq_i|=q_0} K(\vq_1\vq_2\vq_3\vq_4\vq_5\vq_6) \Delta_{\vq_1}^* \Delta_{\vq_2} \Delta_{\vq_3}^* \Delta_{\vq_4} \Delta_{\vq_5}^* \Delta_{\vq_6} \delta_{\vq_1-\vq_2+\vq_3-\vq_4+\vq_5-\vq_6} \nonumber \\
 & & + \mathcal{O}(\Delta^8) 
\end{eqnarray}
where $N_0 = 2\bar\mu^2/\pi^2$ and where we have restricted
our attention to modes with $|\vq|=q_0$, as we now explain.
Note that in the vicinity of $|\vq|=q_0$  and $\dm=\dm_2$,
\begin{equation}
\alpha \approx \left( \frac{\dm - \dm_2}{\dm_2} \right)\ .
\end{equation}
We see that for $\alpha>0$ (that is, for $\dm>\dm_2$)
$\Delta_\vq=0$ is stable whereas for $\alpha<0$ (that is, $\dm<\dm_2$),
the LOFF instability sets in.  In the limit $\alpha\rightarrow 0^-$,
only those plane waves on the sphere $|\vq| = q_0$ are unstable.
For this reason we only include these plane waves in the 
expression (\ref{freeenergy}) for the free energy.  

We shall do most of our analysis in the vicinity of
$\dm=\dm_2$, where we choose $|\vq|=q_0=1.1997\dm_2$ 
as just described.
However, we shall also
want to apply our results at $\dm>\dm_2$.
At these values of
$\dm$, we shall choose $|\vq|$ in such a way as to minimize 
$\alpha(|\vq|)$, because this minimizes the quadratic term in 
the free energy $\Omega$ and thus minimizes the free
energy in the vicinity of $\Delta\rightarrow 0$, which
is where the Ginzburg-Landau analysis is reliable.
As Fig.~\ref{zerogapfig} indicates, for any given $\dm$
the minimum value of $\alpha$ is to be found at $|\vq|=1.1997 \dm$.
Therefore, when we apply Eq.~(\ref{freeenergy}) 
away from $\dm_2$, we shall set $q_0=1.1997 \dm$, just as at 
$\dm=\dm_2$. As a consequence, the opening angle of the
pairing rings, $\psi_0=2\cos^{-1}(\dm/|\vq|)$, is unchanged
when we move away from $\dm=\dm_2$.

%------------------------------------------------------------------------

\section{Results}
\label{sec:results}

\subsection{Generalities}

All of the modes on the sphere
$|\vq|=q_0$ become unstable
at $\dm=\dm_2$.  The quadratic term in the free energy
includes no interaction between modes with different $\vq$'s,
and so predicts that $\Delta_\vq\neq 0$ for all modes
on the sphere. 
Each plane-wave mode corresponds to a ring of paired
quarks on each Fermi surface, so we would obtain
a cacophony of multiple overlapping rings,
favored by the quadratic term because this allows more 
and more of the quarks
near their respective Fermi surfaces to pair. 
Moving beyond lowest order,
our task is to evaluate the quartic
and sextic terms in the free energy.
These higher order terms 
characterize the effects of interactions
between $\Delta_\vq$'s with differing $\vq$'s
(between the different pairing rings) 
and thus determine how condensation in one mode
enhances or deters condensation in other modes.
The results we shall present rely on our ability
to evaluate $J$ and $K$, defined in Eqs.~(\ref{integrals}).
We describe the methods we use to
evaluate these expressions in an appendix
and focus here on 
describing and understanding the
results.
We shall see, for example, that although the quadratic
term favors adding
more rings, the higher order terms strongly disfavor
configurations in which $\Delta_\vq$'s corresponding
to rings that intersect          are nonzero. 
Evaluating the quartic and sextic terms in the free energy will
enable us to evaluate the free energy of 
condensates with various configurations of several plane waves and thereby
discriminate between candidate crystal structures.

A given crystal structure can be described by a set of vectors
$\mathcal{Q} = \{ \vq_a, \vq_b, \cdots \}$, specifying which plane
wave modes are present in the condensate, and a set of gap parameters
$\{ \Delta_{\vq_a}, \Delta_{\vq_b}, \cdots \}$, indicating the
amplitude of condensation in each of the modes.  Let us define
$\mathcal{G}$ as the group of proper and improper rotations that
preserve the set $\mathcal{Q}$.  We make the assumption that
$\mathcal{G}$ is also the point group of the crystal itself; this
implies that $\Delta_{\vq} = \Delta_{\vq'}$ if $\vq'$ is in the orbit
of $\vq$ under the group action.  For most (but not all) of the
structures we investigate, $\mathcal{Q}$ has only one orbit and
therefore all of the $\Delta_\vq$'s are equal.

\begin{figure}
\begin{center}
\psfrag{q1}{$\vq_1$}
\psfrag{q2}{$\vq_2$}
\psfrag{q3}{$\vq_3$}
\psfrag{q4}{$\vq_4$}
\psfrag{q5}{$\vq_5$}
\psfrag{q6}{$\vq_6$}
\psfrag{psi}{$\psi$}
\psfrag{chi}{$\chi$}
%\parbox{1.75in}{(a)} \hspace{0.75in} \parbox{2in}{(b)}
\parbox{1.75in}{\includegraphics[width=1.75in]{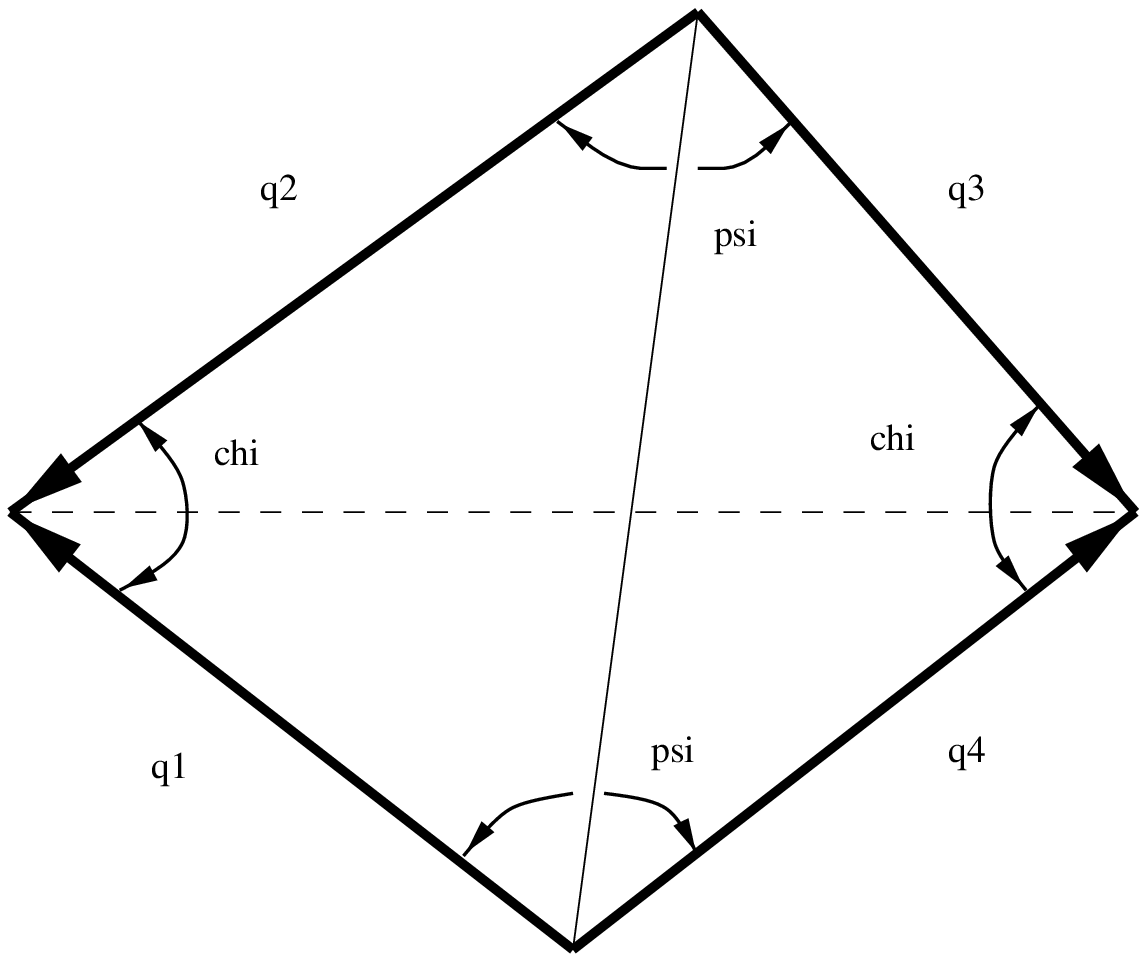}} 
\hspace{0.75in}
\parbox{2in}{\includegraphics[width=2in]{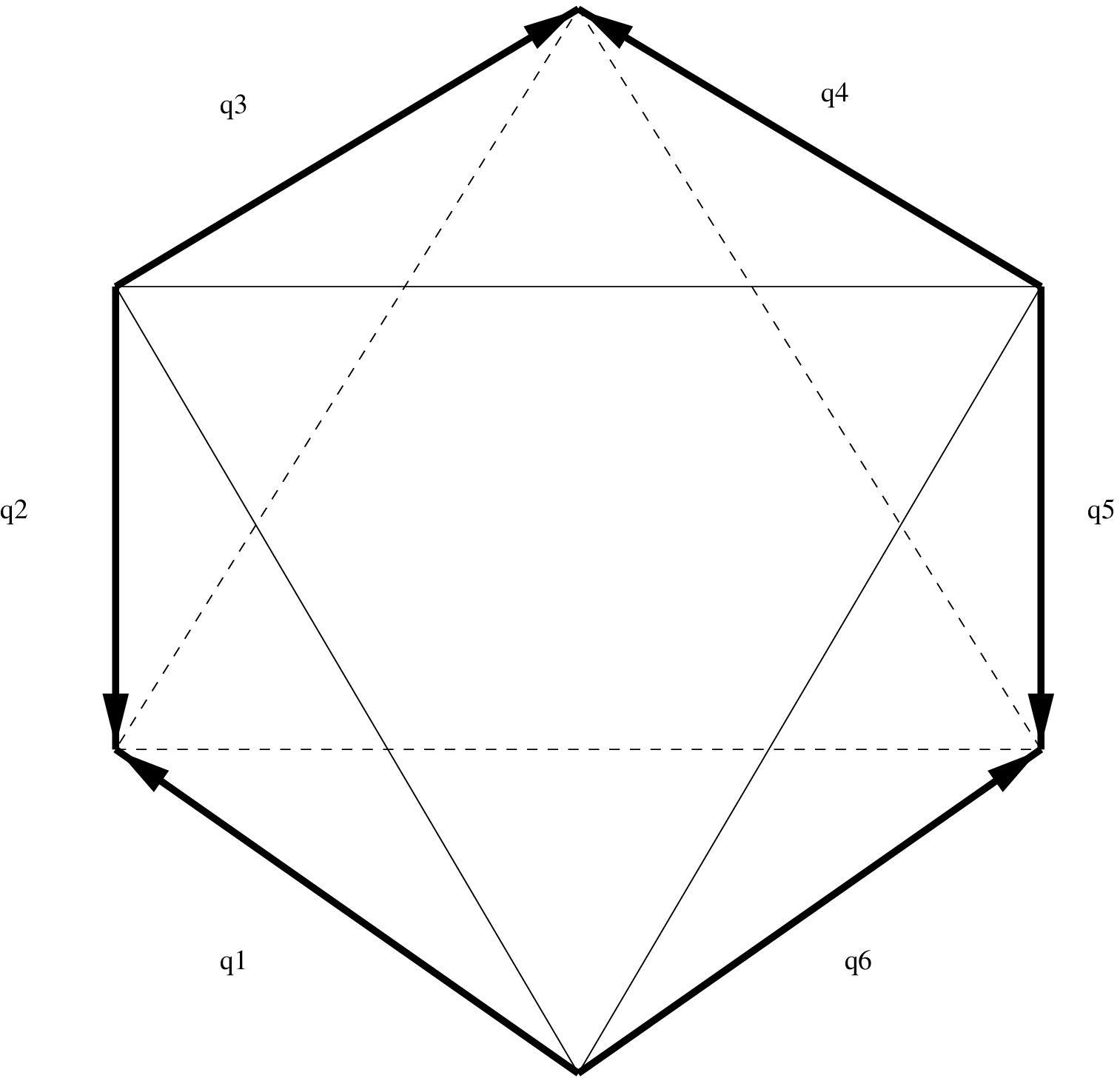}} 
%\parbox{5in}{(c)} \\
%\parbox{3.5in}{\includegraphics[width=4in]{twowaveshapes.eps}}
%\includegraphics[width=2.8in]{rhombus.eps}
%\parbox{4.5in}{\includegraphics[width=3in]{gammafig.eps}} 
%\includegraphics[width=4in]{gammafig.eps} 
\end{center}
\caption{
\label{rhombushexagonfig}
Rhombic and hexagonal combinations of $\vq$'s.  On the left is a rhombus
with $\vq_1-\vq_2+\vq_3-\vq_4 = 0$.  On the right is a hexagon
with $\vq_1-\vq_2+\vq_3-\vq_4+\vq_5-\vq_6 = 0$.  The edges have equal
lengths ($|\vq_i| = q_0$). The shapes are in general nonplanar.
}
\end{figure}

For a given set $\mathcal{Q}$, the quartic term in the free energy
(\ref{freeenergy}) is a sum over all combinations of four 
$\vq$'s that
form closed ``rhombuses'', as shown in Fig.~\ref{rhombushexagonfig}.  
The four $\vq$'s are chosen from the set $\mathcal{Q}$
and they need not be distinct.
By a rhombus we mean a closed figure composed of four equal
length vectors which will in general be nonplanar.
A rhombus  is therefore characterized by two
internal angles $(\psi,\chi)$ with the constraint $0 \leq \psi + \chi
\leq \pi$.  Each shape corresponds to a value of the $J$ function (as
defined in equations (\ref{integrals})); the rotational invariance of
the $J$ function implies that congruent shapes give the same value and
therefore $J(\vq_1\vq_2\vq_3\vq_4) = J(\psi,\chi)$.  So, each unique
rhombic combination of $\vq$'s in the set $\mathcal{Q}$
that characterizes a given crystal structure
yields a unique contribution to the quartic coefficient in
the Ginzburg-Landau 
free energy of that crystal
structure.  The continuation to next order is
straightforward: the sextic term in the free
energy (\ref{freeenergy}) is a sum
over all combinations of six $\vq$'s that form closed ``hexagons'', as
shown in Fig.~\ref{rhombushexagonfig}.  Again these shapes are
generally nonplanar and each unique hexagonal combination of $\vq$'s
yields a unique value of the $K$ function and a unique 
contribution to the sextic
coefficient in the Ginzburg-Landau free energy of the crystal.

When all of the $\Delta_\vq$'s are equal, we can evaluate aggregate
quartic and sextic coefficients $\beta$ and $\gamma$, respectively, as
sums over all rhombic and hexagonal combinations of the $\vq$'s
in the set $\mathcal{Q}$:
\begin{equation}
\label{squarehexagonsums}
\beta = \sum_\square J(\square), \ \gamma =  \sum_{\hexagon}  K(\hexagon).
\end{equation}
Then, for a crystal with $P$ plane waves, the free energy has the simple
form
\begin{equation}
\frac{\Omega(\Delta)}{N_0} =  P \alpha \Delta^2 + \frac{1}{2} \beta \Delta^4 + \frac{1}{3} \gamma \Delta^6 + \mathcal{O}(\Delta^8)
\end{equation}
and we can analyze a candidate crystal structure by calculating the
coefficients $\beta$ and $\gamma$ and studying the resultant form of
the free energy function.  

If $\beta$ and $\gamma$ are both positive,
a second-order phase transition occurs at $\alpha = 0$; near the
critical point the value of $\gamma$ is irrelevant and the
minimum energy solution is
\begin{equation}
\Delta = \left( \frac{P |\alpha|}{\beta} \right)^{\frac{1}{2}}, \ \ \ \ 
\frac{\Omega}{N_0} = -\frac{P^2 \alpha^2}{2 \beta}\ , 
\end{equation}
for $\dm\leq \dm_2$ ({\it i.e.}~$\alpha\leq 0$). 

If $\beta$ is negative and $\gamma$ is positive, the phase transition
is in fact
first order and occurs at a new critical point defined by
\begin{equation}
\alpha =\alpha_* = \frac{3\beta^2}{16 P \gamma}\ .
\label{alphastar}
\end{equation}
In order to find the $\dm_*$ corresponding to $\alpha_*$,
we need to solve 
\begin{equation}
\alpha\left(|\vq|,\dm_*\right)=\alpha\left(1.1997\dm_*,\dm_*\right)
=\alpha_*= \frac{3\beta^2}{16 P \gamma}\ .
\end{equation}
Since $\alpha_*$ is positive, the critical point
$\dm_*$ at which the first-order phase
transition occurs
is larger than $\dm_2$. 
If $\alpha_*$ is small, then $\dm_*\simeq (1+\alpha_*)\dm_2$.
Thus, a crystalline
color superconducting state whose crystal structure
yields a negative $\beta$ and positive $\gamma$
persists as a possible ground state even above $\dm_2$,
the maximum $\dm$ at which the plane-wave
state is possible.  At the first-order critical
point (\ref{alphastar}),
the free energy has degenerate minima at 
\begin{equation}
\Delta = 0\ ,\ \ \ \ \Delta = \left(\frac{3|\beta|}{4\gamma}\right)
^{1/2}\ .
\end{equation}
If we reduce $\dm$ below $\dm_*$, the minimum with $\Delta\neq 0$
deepens.  Once $\dm$ is reduced to the point at which
the single plane wave would just be starting to form with
a free energy infinitesimally below zero, the free energy
of the crystal structure with negative $\beta$ and positive
$\gamma$ has
\begin{equation}
\Delta = \left( \frac{|\beta|}{\gamma} \right)^{\frac{1}{2}}\ ,\ \ \ \ 
\frac{\Omega}{N_0} = - \frac{|\beta|^3 }{6 \gamma^2} 
\label{DeltaOmegaatdm2}
\end{equation}
at $\dm=\dm_2$.

Finally, if $\gamma$ is negative, the order $\Delta^6$ 
Ginzburg-Landau free energy is unbounded from below. 
In this circumstance, we know that we have found a first
order phase transition but we do not know at what $\dm_*$
it occurs, because the stabilization of the Ginzburg-Landau
free energy at large $\Delta$ must come about at order
$\Delta^8$ or higher.

\subsection{One wave}

With these general considerations in mind we now proceed to look at 
specific examples of crystal structures.  We begin with
the
single plane-wave condensate ($P=1$).  The quartic coefficient
of the free energy is 
\begin{equation}
\label{J0eqn}
\beta = J_0 = J(0,0) = \frac{1}{4} \frac{1}{\vq^2-\dm^2} \simeq 
+ \frac{0.569}{\dm^2},
\end{equation}
and the sextic coefficient is 
\begin{equation}
\label{K0eqn}
\gamma = K_0 = K(\vq\vq\vq\vq\vq\vq) = \frac{1}{32} \frac{\vq^2 + 3 \dm^2}{(\vq^2 - \dm^2)^3} \simeq + \frac{1.637}{\dm^4},
\end{equation}
yielding a second-order phase transition at $\alpha = 0$.
These coefficients agree with those 
obtained by expanding the all-orders-in-$\Delta$
solution for the single plane wave which can be obtained
by variational methods~\cite{FF,Takada,BowersLOFF}
or by starting from 
(\ref{planewave2}), as in Refs.~\cite{ngloff,massloff}.
The coefficient $\beta$ in (\ref{J0eqn}) was 
first found by Larkin and Ovchinnikov~\cite{LO}.

\subsection{Two waves}

\begin{figure}[t]
\centering
\psfrag{psi}{$\psi$}
\parbox{3.5in}{\includegraphics[width=4in]{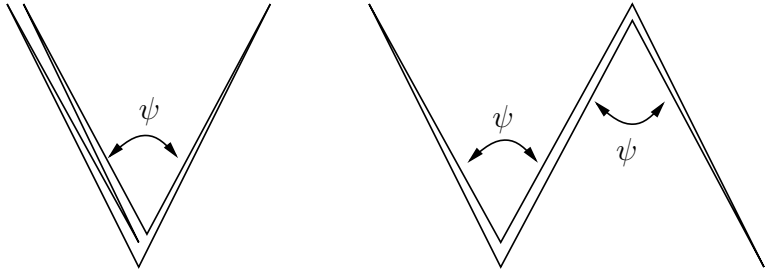}}
\caption{
\label{k1k2fig}
Two different ``hexagonal'' shapes (as in Fig.~\ref{rhombushexagonfig}) 
that can be constructed from
two vectors $\vq_a$ and $\vq_b$. These shapes correspond to the
functions $K_1(\psi)$ and $K_2(\psi)$ in Eq.~(\ref{gammaeqn}).  
}
\end{figure}

\begin{figure}[t]
\centering
\psfrag{xlabela}{$\psi$}
\psfrag{ylabela}{$\beta(\psi)\dm^2$}
\psfrag{xlabelb}{$\psi$}
\psfrag{ylabelb}{$\gamma(\psi)\dm^4$}
\psfrag{30}[tc][tc]{\small $30^{\circ}$}
\psfrag{60}[tc][tc]{\small $60^{\circ}$}
\psfrag{90}[tc][tc]{\small $90^{\circ}$}
\psfrag{120}[tc][tc]{\small $120^{\circ}$}
\psfrag{150}[tc][tc]{\small $150^{\circ}$}
\psfrag{180}[tc][tc]{\small $180^{\circ}$}
\psfrag{y1}[rc][rc]{\small $-2$}
\psfrag{y2}[rc][rc]{\small $2$}
\psfrag{y3}[rc][rc]{\small $4$}
\psfrag{y4}[rc][rc]{\small $6$}
\psfrag{y5}[rc][rc]{\small $8$}
\psfrag{y6}[rc][rc]{\small $10$}
\psfrag{y7}[rc][rc]{\small $10$}
\psfrag{y8}[rc][rc]{\small $20$}
\psfrag{y9}[rc][rc]{\small $30$}
\psfrag{y10}[rc][rc]{\small $40$}
\psfrag{y11}[rc][rc]{\small $50$}
\psfrag{psi0labela}{$\psi = \psi_0 \simeq 67.1^{\circ}$}
\psfrag{psi0labelb}{$\psi = \psi_0$}
\psfrag{arrowlabela}{$\beta(\psi_0) \simeq -1.138/\dm^2$}
\psfrag{arrowlabel}{$\gamma(\psi_0) \simeq +0.249/\dm^4$}
%\parbox{4.5in}{\includegraphics[width=3in]{betafig.eps}} 
\parbox{4in}{\includegraphics[width=4in]{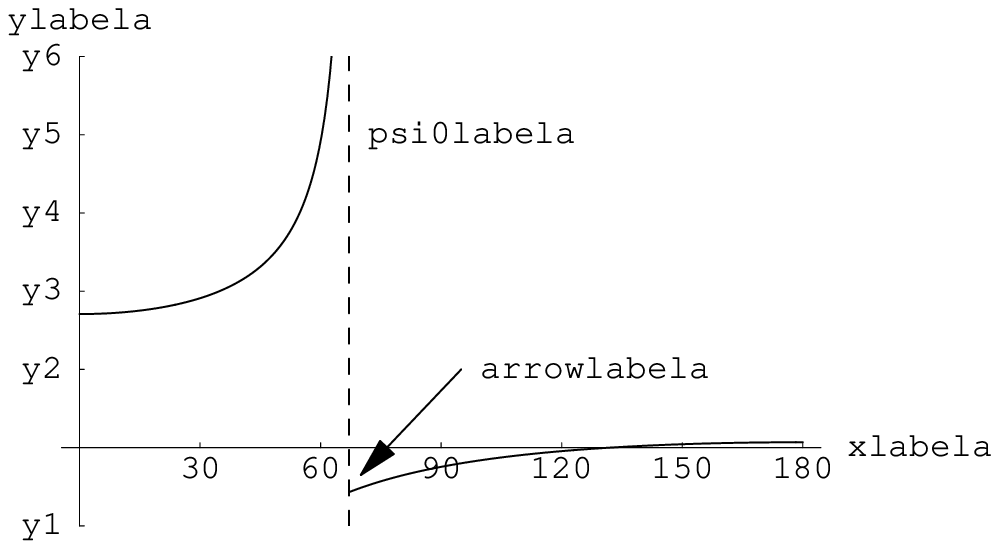}} \\
\parbox{4in}{\phantom{spacer}} \\
%\hspace{0.25in}
%\parbox{4.5in}{\includegraphics[width=3in]{gammafig.eps}} 
\parbox{4in}{\includegraphics[width=4in]{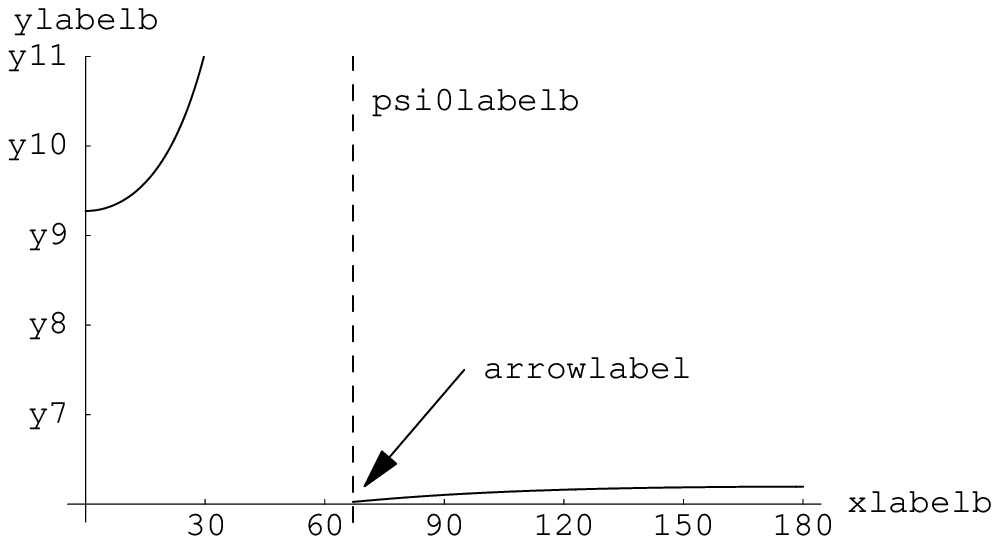} }
\caption{
\label{betagammafig}
$\beta(\psi)$
and $\gamma(\psi)$, the quartic and sextic coefficients in 
the Ginzburg-Landau free energy 
for a condensate consisting of two plane waves whose wave
vectors define an angle $\psi$.  
%The minimum value of $\gamma$ is $\gamma(\psi_0) \simeq 0.249/\dm_2^4$.
}
\end{figure}

Our next example is a condensate of two plane waves ($P=2$) with wave vectors
$\vq_a$ and $\vq_b$ and equal gaps $\Delta_{\vq_a} = \Delta_{\vq_b} =
\Delta$.  The most symmetrical arrangement is an antipodal pair
($\vq_b = -\vq_a$), which yields a cosine spatial variation
$\Delta(\vx) \sim \cos(2\vq_a\cdot\vx)$.  We will find it useful, however,
to study the generic case where $\vq_a$ and $\vq_b$ have the
same magnitude but define an
arbitrary angle $\psi$.  We find that the quartic coefficient
is
\begin{equation}
\beta(\psi) = 2 J_0 + 4 J(\psi,0) 
\end{equation}
and the sextic coefficient is 
\begin{equation}
\label{gammaeqn}
\gamma(\psi) = 2 K_0 + 12 K_1(\psi) + 6 K_2(\psi) 
\end{equation}
where $K_1(\psi) = K(\vq_a\vq_a\vq_a\vq_a\vq_b\vq_b)$ and $K_2(\psi) =
K(\vq_a\vq_a\vq_b\vq_a\vq_a\vq_b)$. ($K_1$
and $K_2$  arise from the ``hexagonal''
shapes shown in Fig.~\ref{k1k2fig}.)  The functions
$\beta(\psi)$ and $\gamma(\psi)$ are plotted in 
Fig.~\ref{betagammafig}.  
These functions manifest a number of interesting
features.  Notice that the functions are singular and discontinuous at
a critical angle $\psi = \psi_0 \simeq 67.1^\circ$, where $\psi_0$ is
the opening angle of a LOFF pairing ring on the Fermi surface.  For
the two-wave condensate we have two such rings, and the two rings are
mutually tangent when $\psi = \psi_0$.  For $\psi < \psi_0$, both
$\beta$ and $\gamma$ are large and positive, implying that an
intersecting ring configuration is energetically unfavorable.  For
$\psi > \psi_0$ the functions are relatively flat and small,
indicating some indifference towards any particular arrangement of the
nonintersecting rings.  There is a range of angles for which $\beta$
is negative and a first-order transition occurs (note that $\gamma$ is
always positive).  The favored arrangement is a pair of adjacent rings
that nearly intersect ($\psi = \psi_0 + \epsilon$).

It is unusual to find coefficients in a Ginzburg-Landau
free energy that behave discontinuously 
as a function of parameters describing
the state, as seen in Fig.~\ref{betagammafig}.
These discontinuities arise because, as we described in the caption
of Fig.~\ref{ringsfig}, we are taking two limits.  
We first take a weak coupling limit in which 
$\Delta_0$,~$\Delta$,~$\dm$,~$|\vq| \ll \omega\ll \bar\mu$ 
while $\dm/\Delta_0$, $|\vq|/\Delta_0$
and $\Delta/\dm$ (and thus the angular width of the pairing
bands) are held fixed.
Then, we take the Ginzburg-Landau limit in which 
$\dm/\Delta_0 \rightarrow \dm_2/\Delta_0$ and 
$\Delta/\Delta_0\rightarrow 0$ and the pairing bands shrink to 
rings of zero angular width.  In the Ginzburg-Landau limit,
there is a sharp distinction between $\psi< 67.1^\circ$ 
where the rings intersect and $\psi>67.1^\circ$ where they
do not.  Without taking the weak coupling limit, the
plots of $\beta(\psi)$ and $\gamma(\psi)$ would nevertheless
look like smoothed versions of those in Fig.~\ref{betagammafig},
smoothed on angular scales of order $\dm^2/\omega^2$ (for $\beta$)
and $\dm^4/\omega^4$ (for $\gamma$).
However, in the weak coupling limit these small angular scales
are taken to zero.
Thus,
the double limit sharpens what would otherwise be
distinctive but continuous features of the coefficients in the
Ginzburg-Landau free energy into discontinuities.

\subsection{Crystals}

From our analysis of the two-wave condensate, we can infer that for a
general multiple-wave condensate it is unfavorable to allow the
pairing rings to intersect on the Fermi surface.  For nonintersecting
rings, the free energy should be relatively insensitive to how the
rings are arranged on the Fermi surface.  However, 
Eqs.~(\ref{squarehexagonsums}) suggest that a combinatorial advantage is
obtained for exceptional structures that permit a large number of
rhombic and hexagonal combinations of wave vectors.  
That is, if there are many ways of picking four 
(not necessarily different) wave vectors
from the set of wave vectors that specify the crystal
structure for which $\vq_1-\vq_2+\vq_3-\vq_4 = 0$,
or if there are many ways of picking six wave vectors
for which $\vq_1 -\vq_2 +\vq_3 -
\vq_4 + \vq_5 - \vq_6 = 0$, such a crystal structure
enjoys a combinatorial advantage that will tend to make
the magnitudes of $\beta$ or $\gamma$ large. 
For a rhombic
combination $\vq_1-\vq_2+\vq_3-\vq_4 = 0$, the four $\vq$'s must be
the four vertices of a rectangle that is inscribed in a circle on the
sphere $|\vq| = q_0$. (The circle need not be a great circle, and the
rectangle can degenerate to a line or a point if the four $\vq$'s are
not distinct).  For a hexagonal combination $\vq_1 -\vq_2 +\vq_3 -
\vq_4 + \vq_5 - \vq_6 = 0$, the triplets $(\vq_1\vq_3\vq_5)$ and
$(\vq_2\vq_4\vq_6)$ are vertices of two inscribed triangles that have
a common centroid. In the degenerate case where only four of the six
$\vq$'s are distinct, the four distinct $\vq$'s must be the vertices of an
inscribed rectangle or an inscribed isoceles trapezoid for which one
parallel edge is twice the length of the other.  When five of the six
$\vq$'s are distinct, they can be arranged as a rectangle plus any fifth
point, or as five vertices of an inscribed cuboid
arranged as one antipodal pair plus the three corners adjacent to
one of the antipodes.

\begin{figure}[p]
\caption{
\label{stereographicfig}
Stereographic projections of the candidate crystal structures. 
The points ($\newmoon$) and circles ($\fullmoon$) are projections 
of $\vq$'s that are 
respectively above and below the equatorial plane of the 
sphere $|\vq| = q_0$.  
}
\vspace{0.1in}
\centering
\parbox{1.1in}{\centering point}
\parbox{1.1in}{\centering antipodal pair}
\parbox{1.1in}{\centering triangle}
\parbox{1.1in}{\centering tetrahedron}
\parbox{1.1in}{\centering square}
\parbox{1.1in}{\includegraphics[width=1.1in]{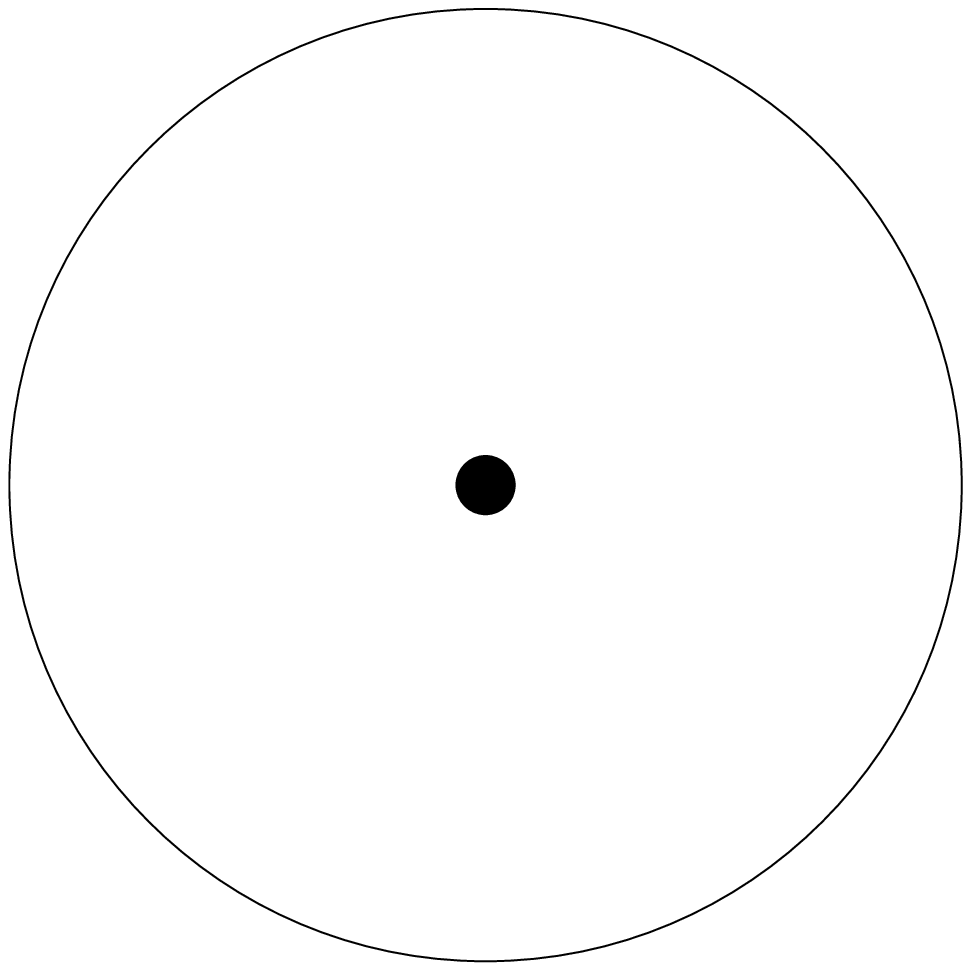}} 
\parbox{1.1in}{\includegraphics[width=1.1in]{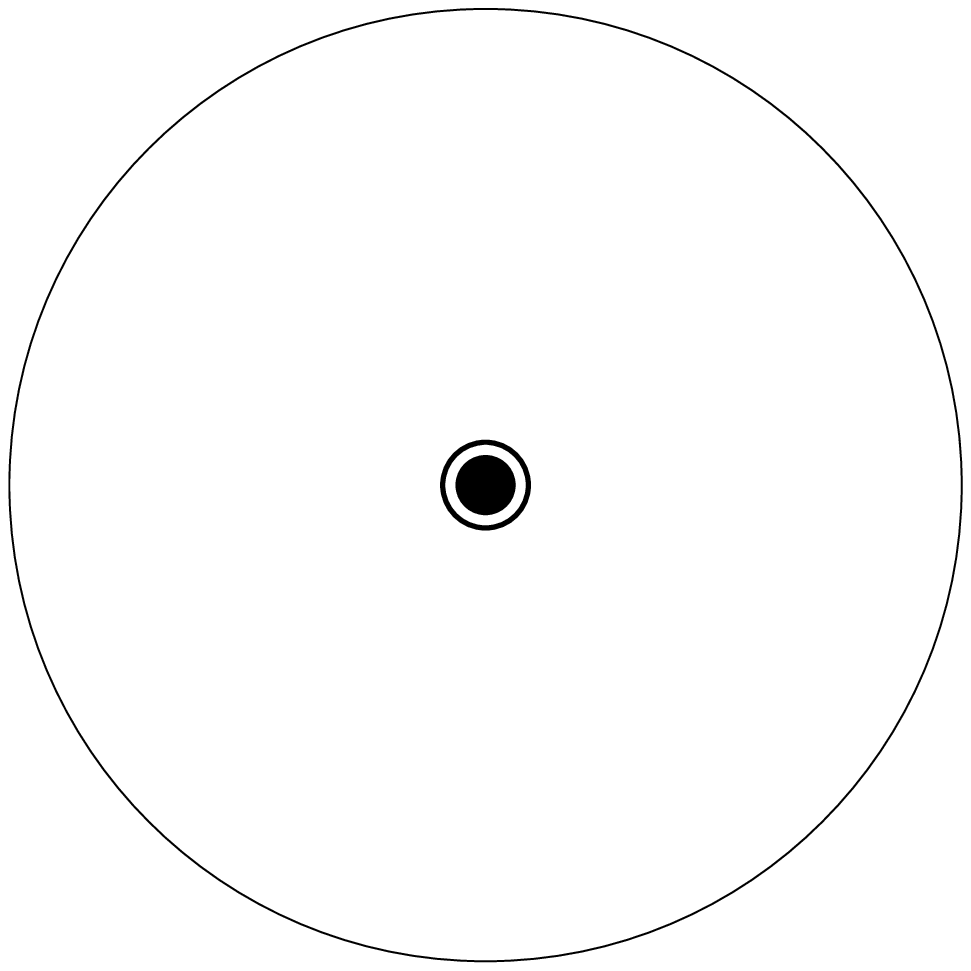}} 
\parbox{1.1in}{\includegraphics[width=1.1in]{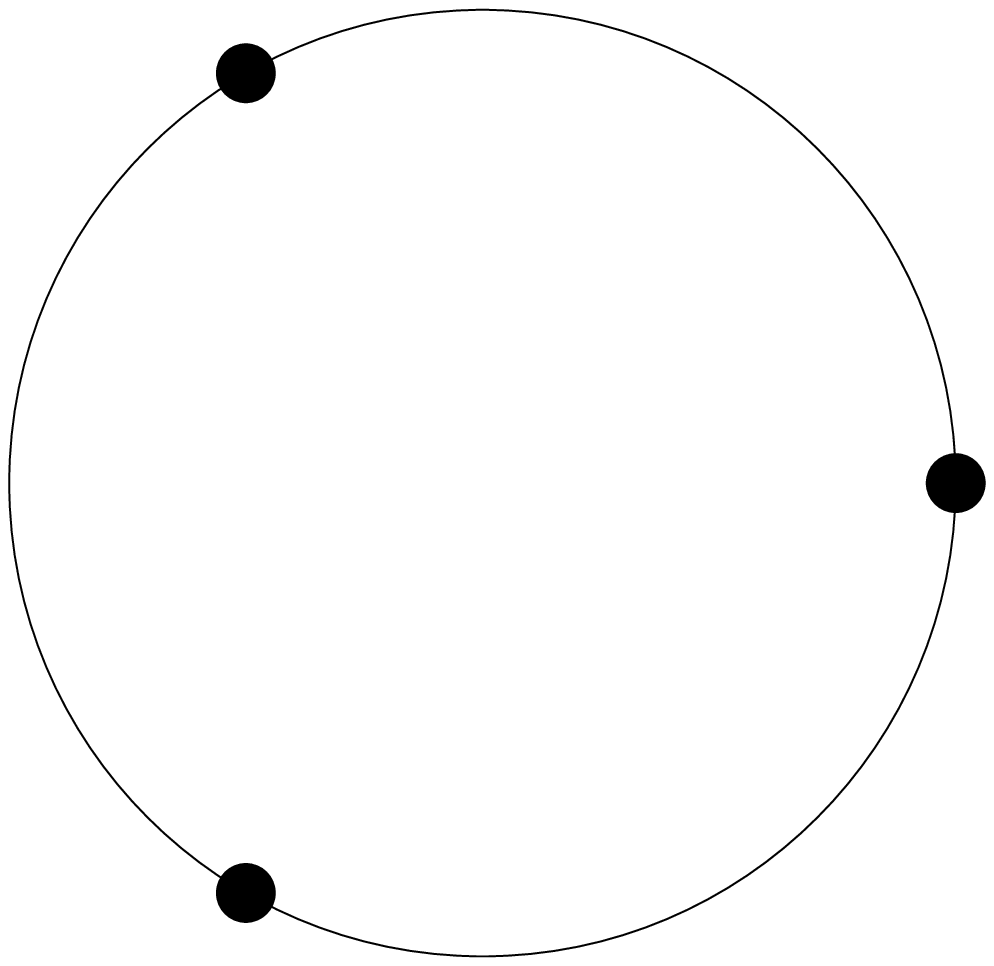}} 
\parbox{1.1in}{\includegraphics[width=1.1in]{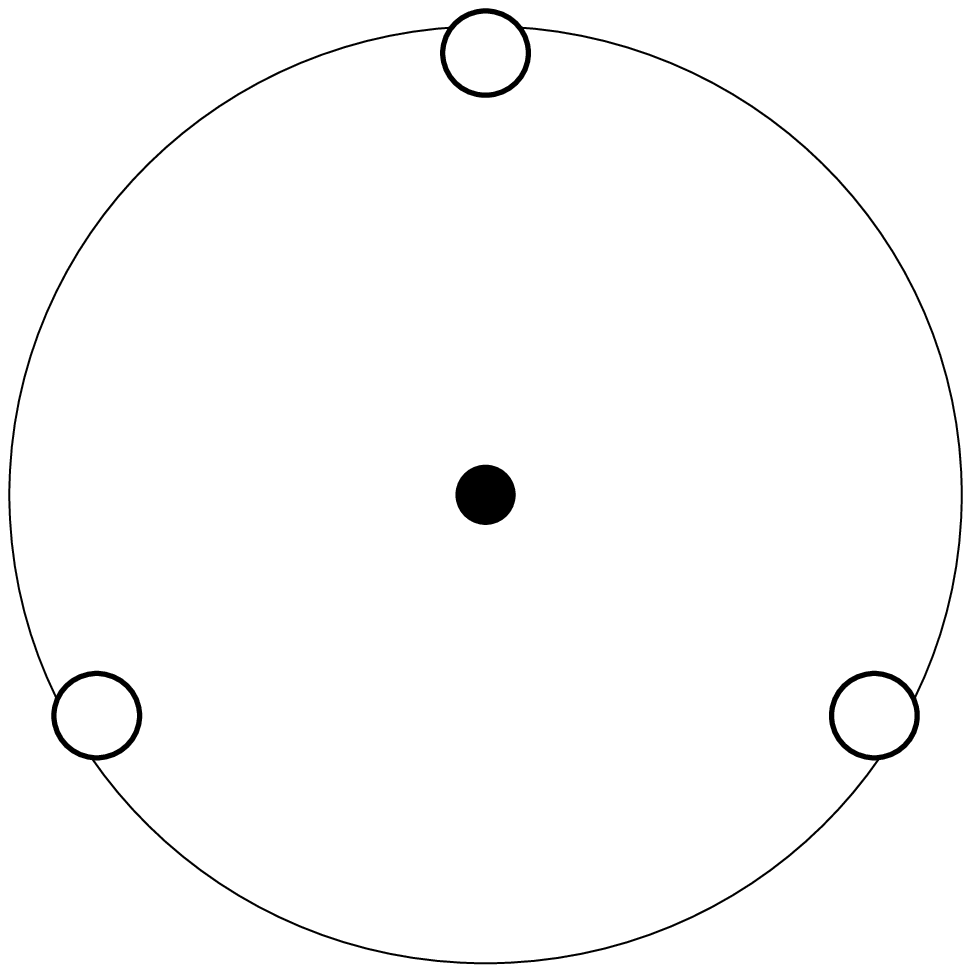}} 
\parbox{1.1in}{\includegraphics[width=1.1in]{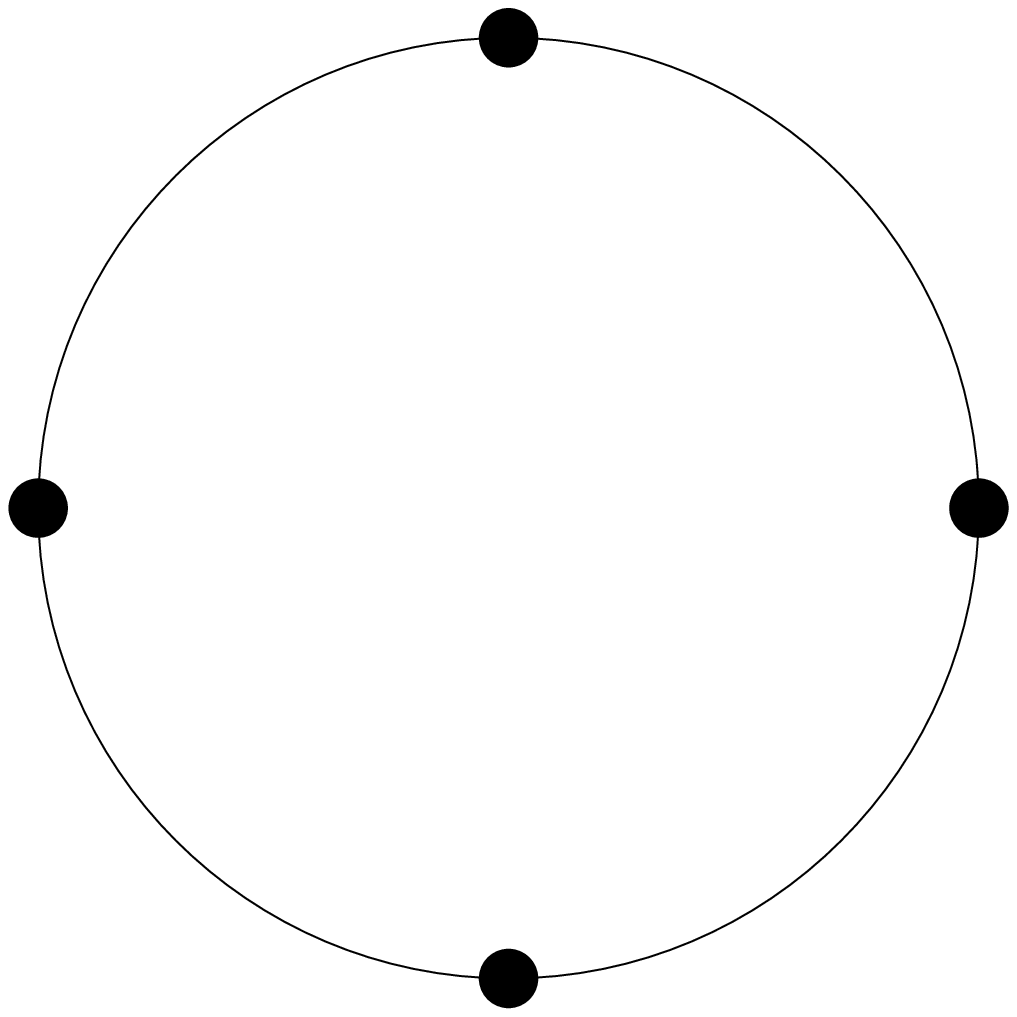}} \\
\parbox{5in}{\phantom{blah}} \\ 
\parbox{1.1in}{\centering \phantom{spacer}  pentagon}
\parbox{1.1in}{\centering trigonal bipyramid}
\parbox{1.1in}{\centering \phantom{spacer}  square pyramid}
\parbox{1.1in}{\centering \phantom{spacer}  octahedron}
\parbox{1.1in}{\centering \phantom{spacer}  trigonal prism}
\parbox{1.1in}{\includegraphics[width=1.1in]{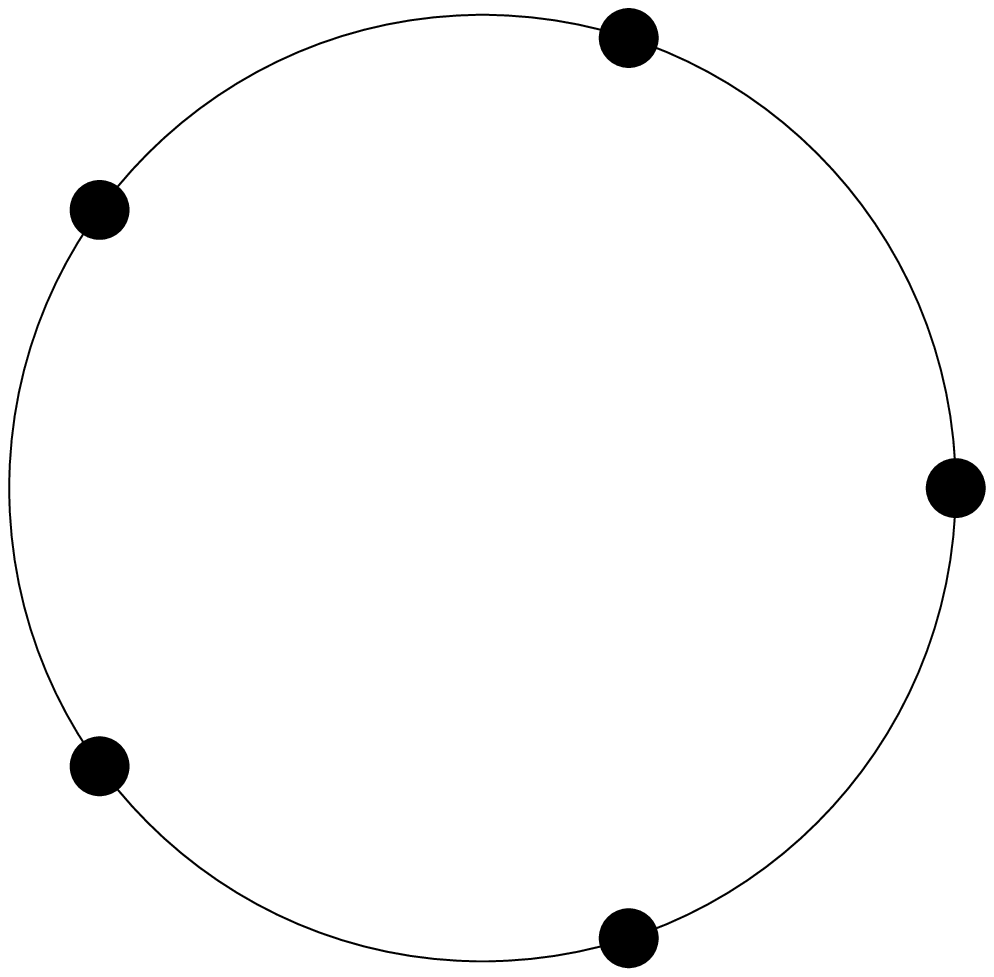}} 
\parbox{1.1in}{\includegraphics[width=1.1in]{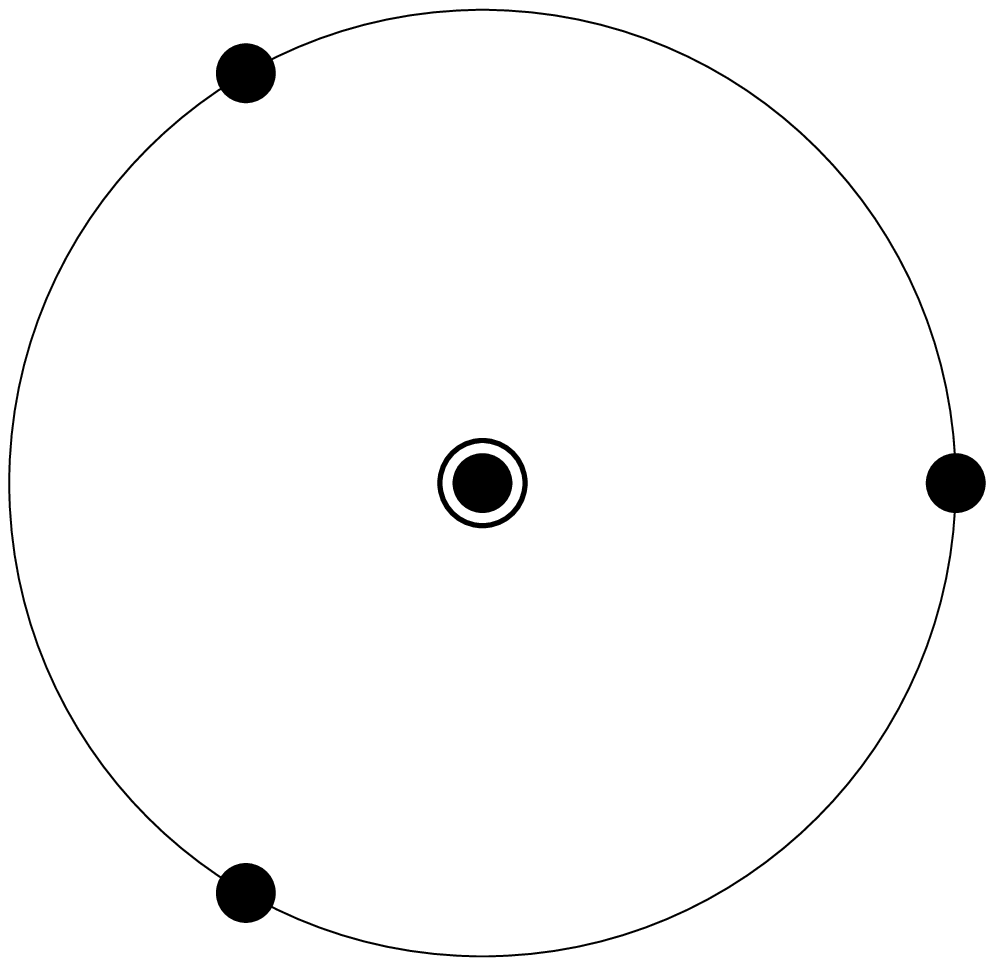}} 
\parbox{1.1in}{\includegraphics[width=1.1in]{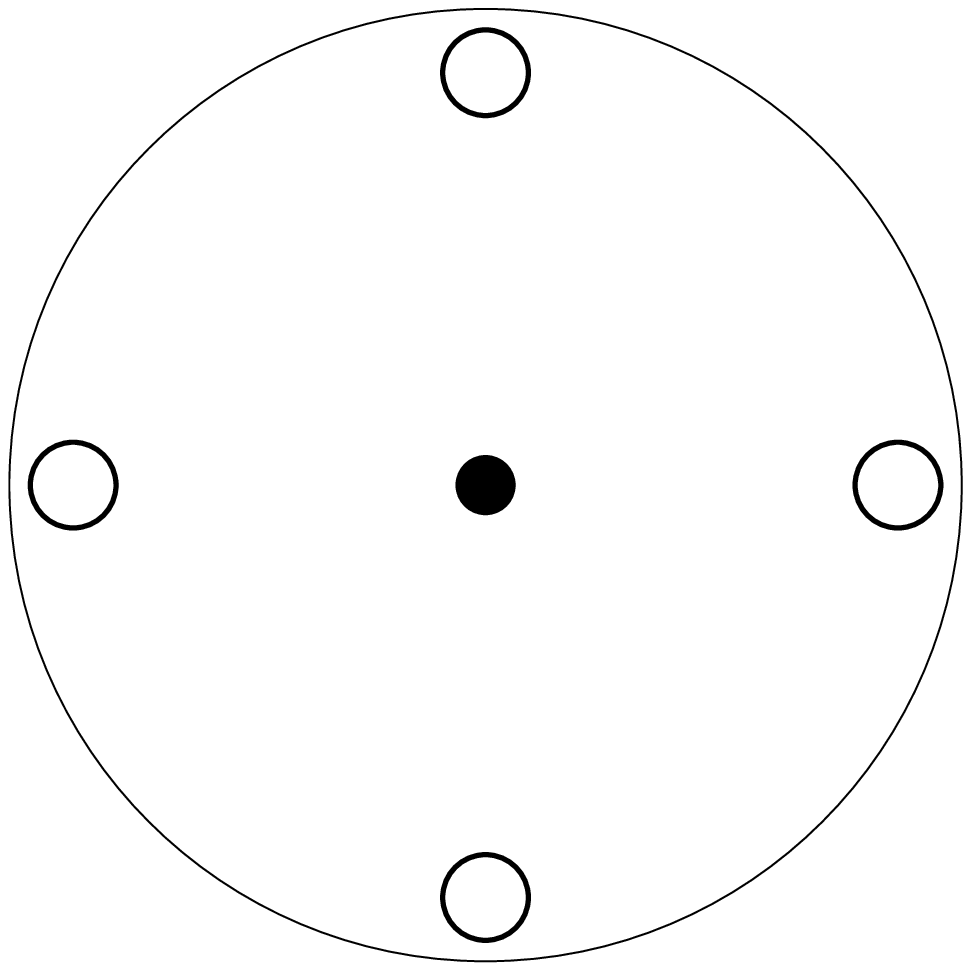}} 
\parbox{1.1in}{\includegraphics[width=1.1in]{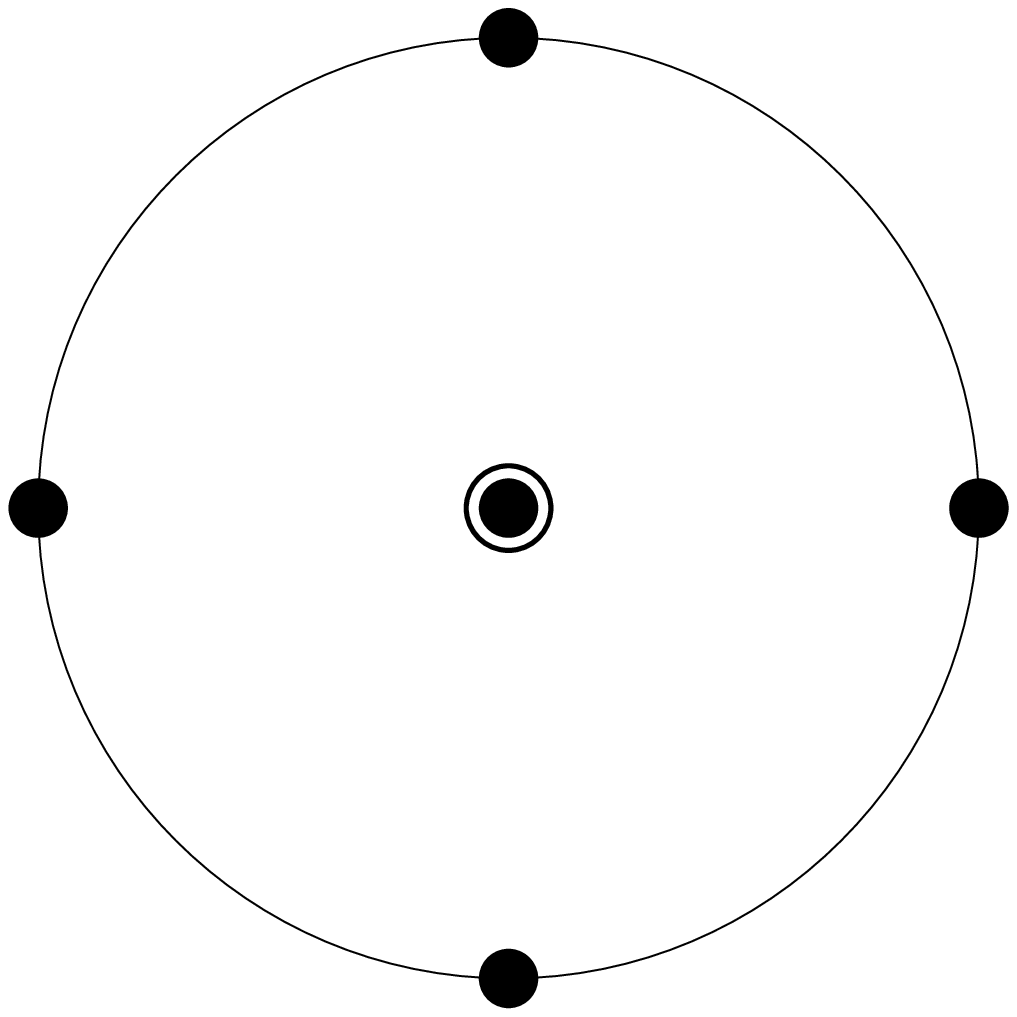}} 
\parbox{1.1in}{\includegraphics[width=1.1in]{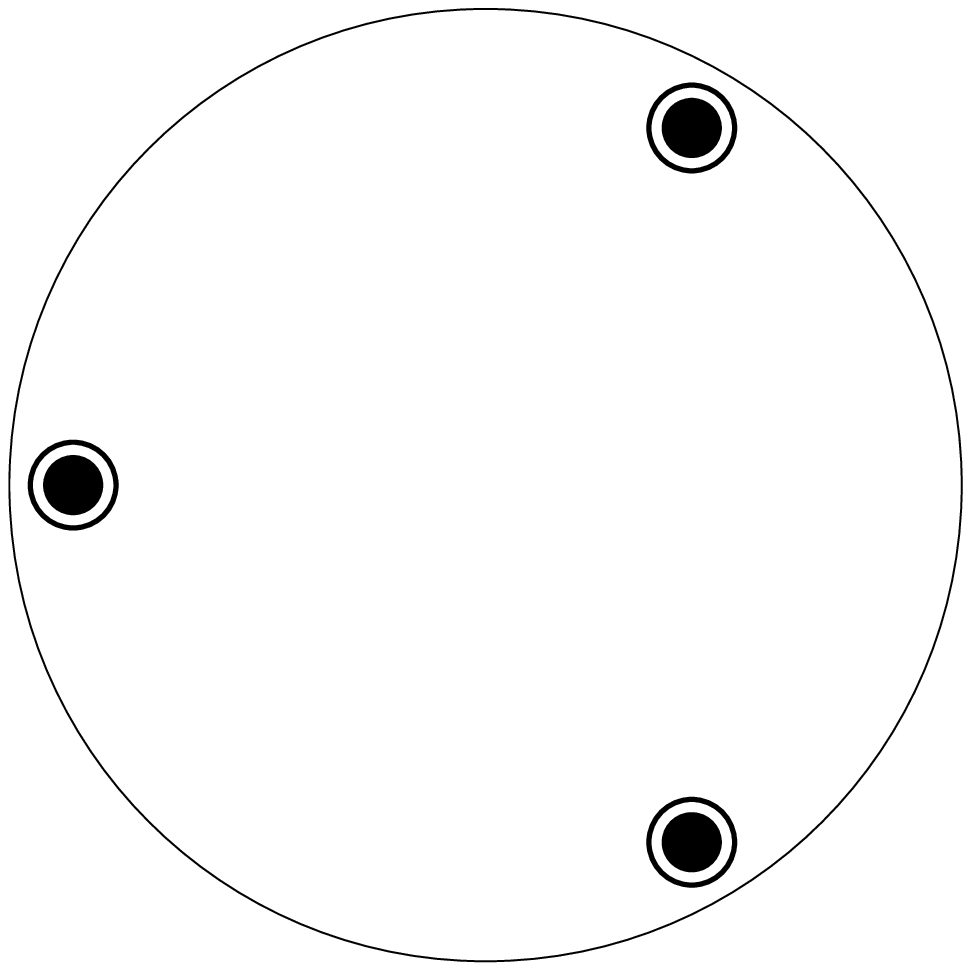}} \\
\parbox{5in}{\phantom{blah}} \\ 
\parbox{1.1in}{\centering \phantom{spacer}  hexagon}
\parbox{1.1in}{\centering pentagonal bipyramid}
\parbox{1.1in}{\centering capped trigonal antiprism}
\parbox{1.1in}{\centering \phantom{spacer}  cube}
\parbox{1.1in}{\centering square antiprism}
\parbox{1.1in}{\includegraphics[width=1.1in]{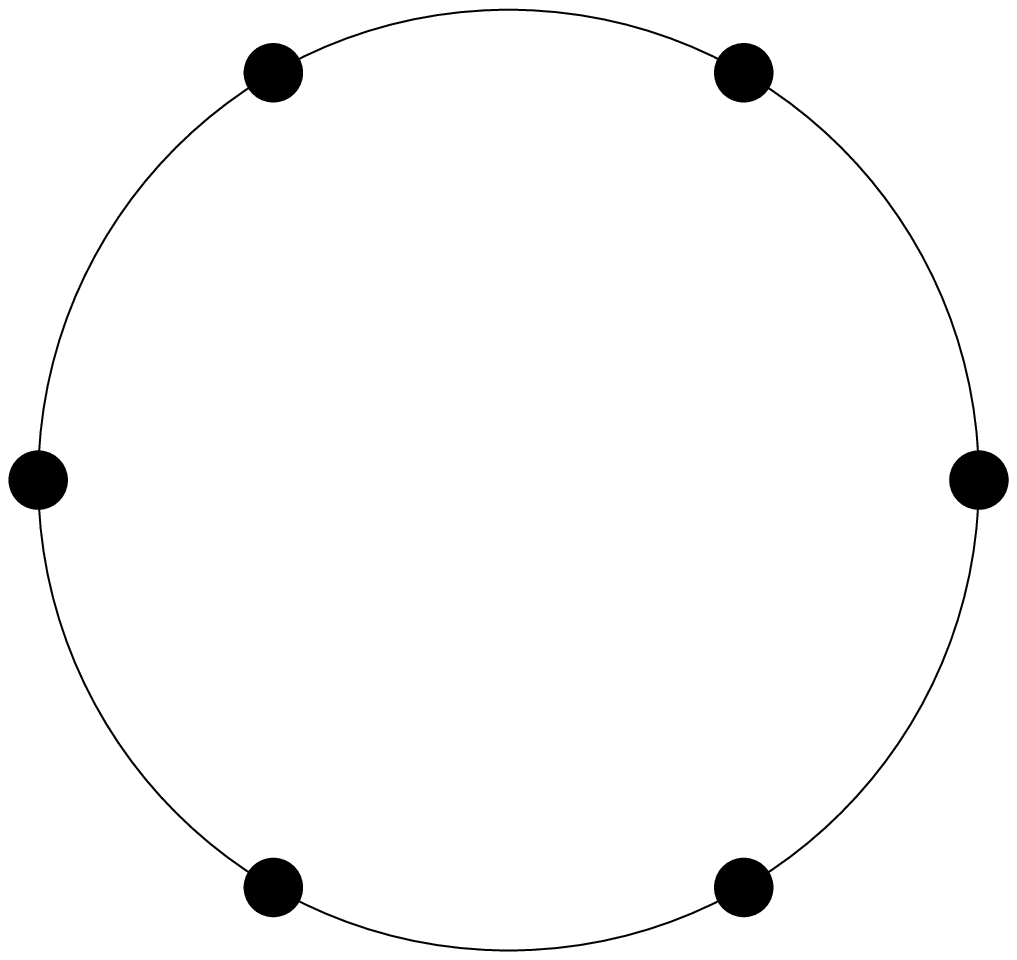}} 
\parbox{1.1in}{\includegraphics[width=1.1in]{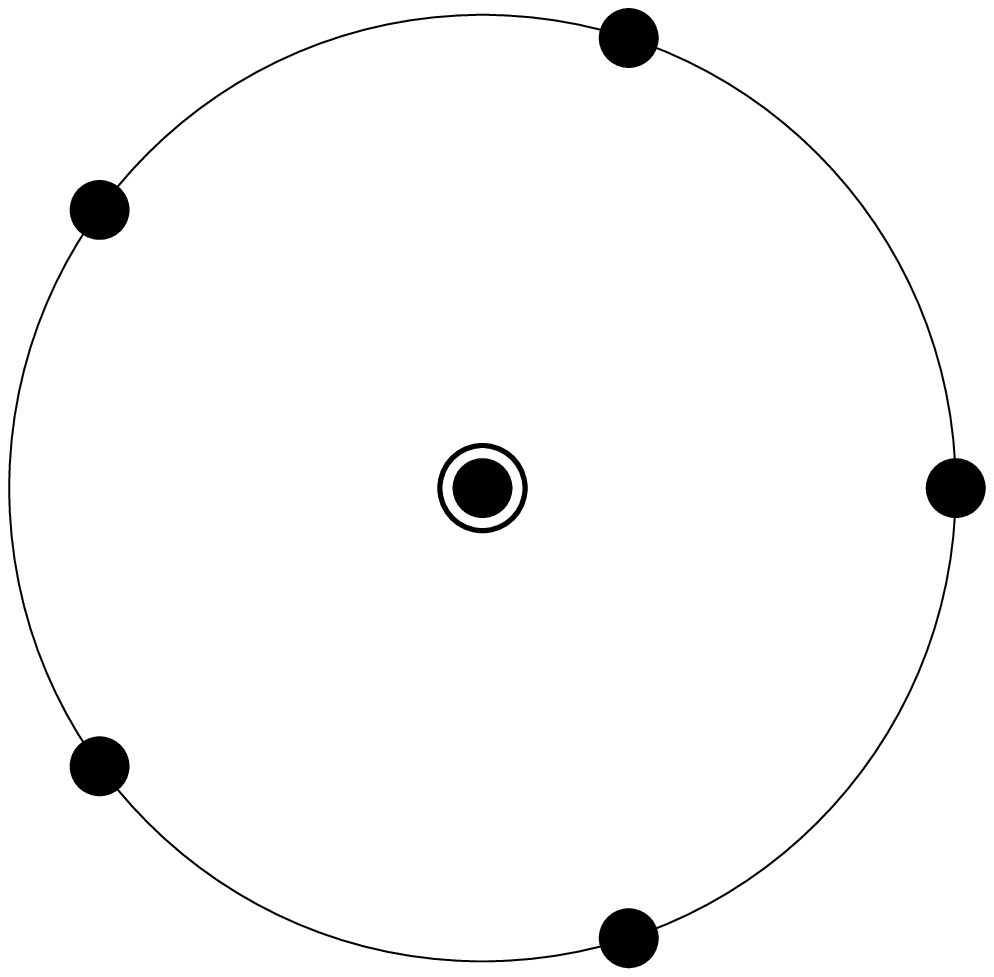}} 
\parbox{1.1in}{\includegraphics[width=1.1in]{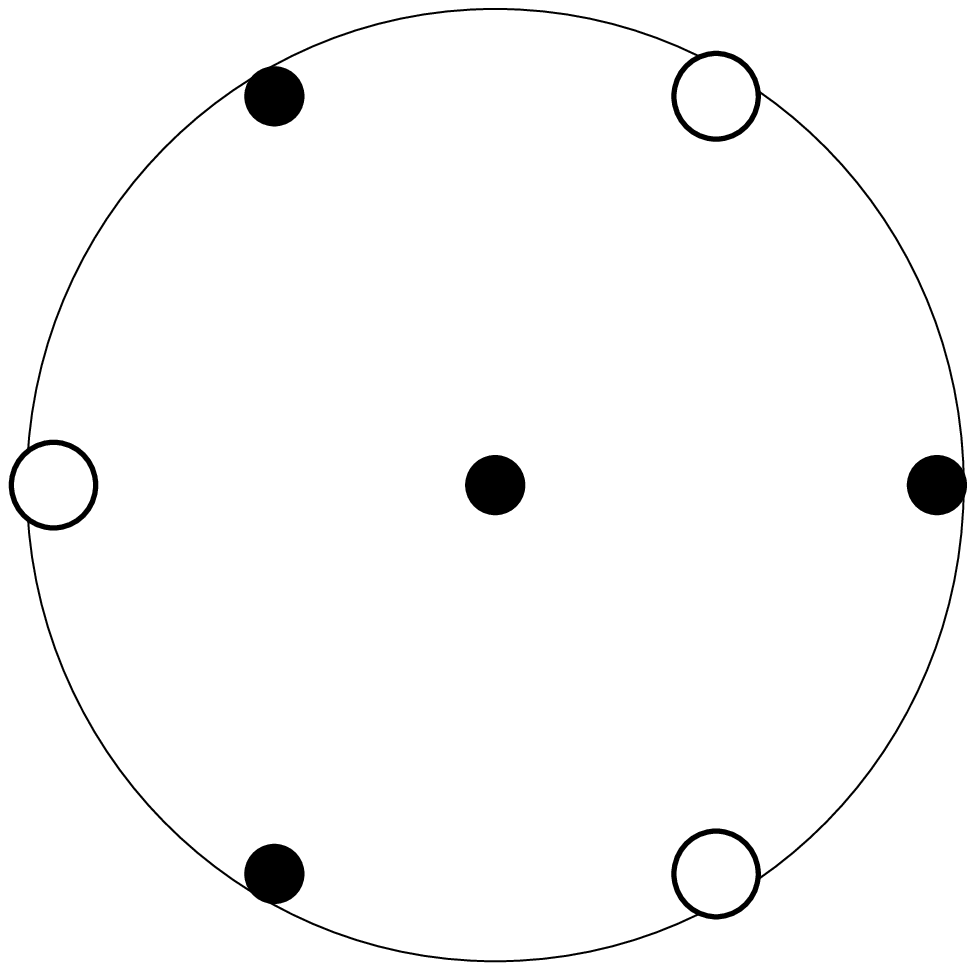}} 
\parbox{1.1in}{\includegraphics[width=1.1in]{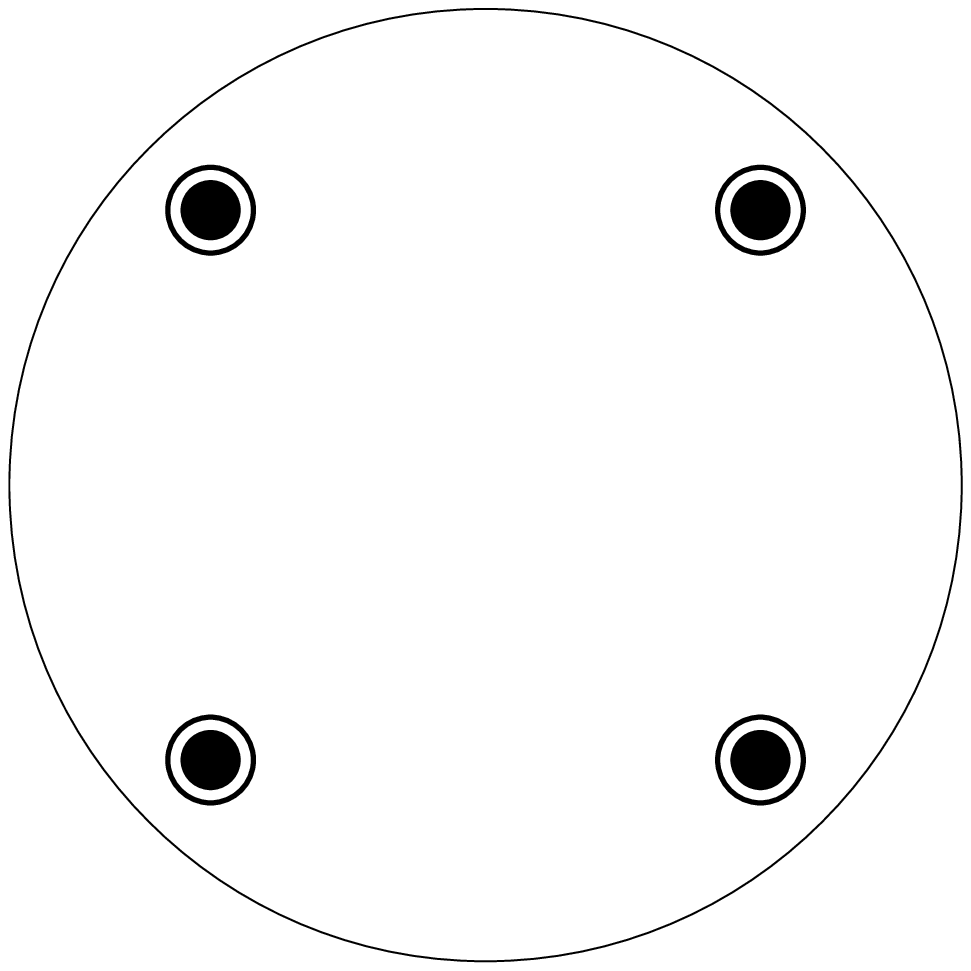}} 
\parbox{1.1in}{\includegraphics[width=1.1in]{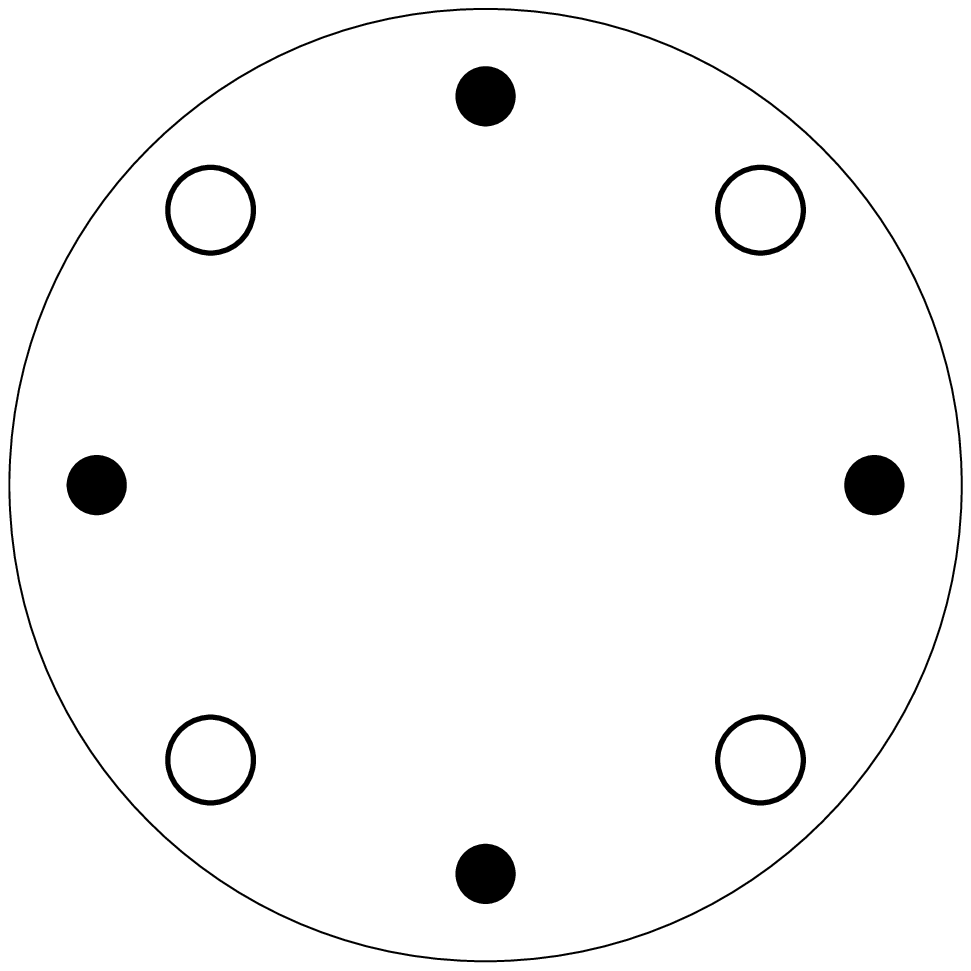}} \\
\parbox{5in}{\phantom{blah}} \\ 
\parbox{1.1in}{\centering \phantom{spacer} hexagonal bipyramid}
\parbox{1.1in}{\centering \phantom{spacer}  augmented trigonal prism}
\parbox{1.1in}{\centering \phantom{spacer}  capped square prism}
\parbox{1.1in}{\centering \phantom{spacer}  capped square antiprism}
\parbox{1.1in}{\centering bicapped square antiprism}
\parbox{1.1in}{\includegraphics[width=1.1in]{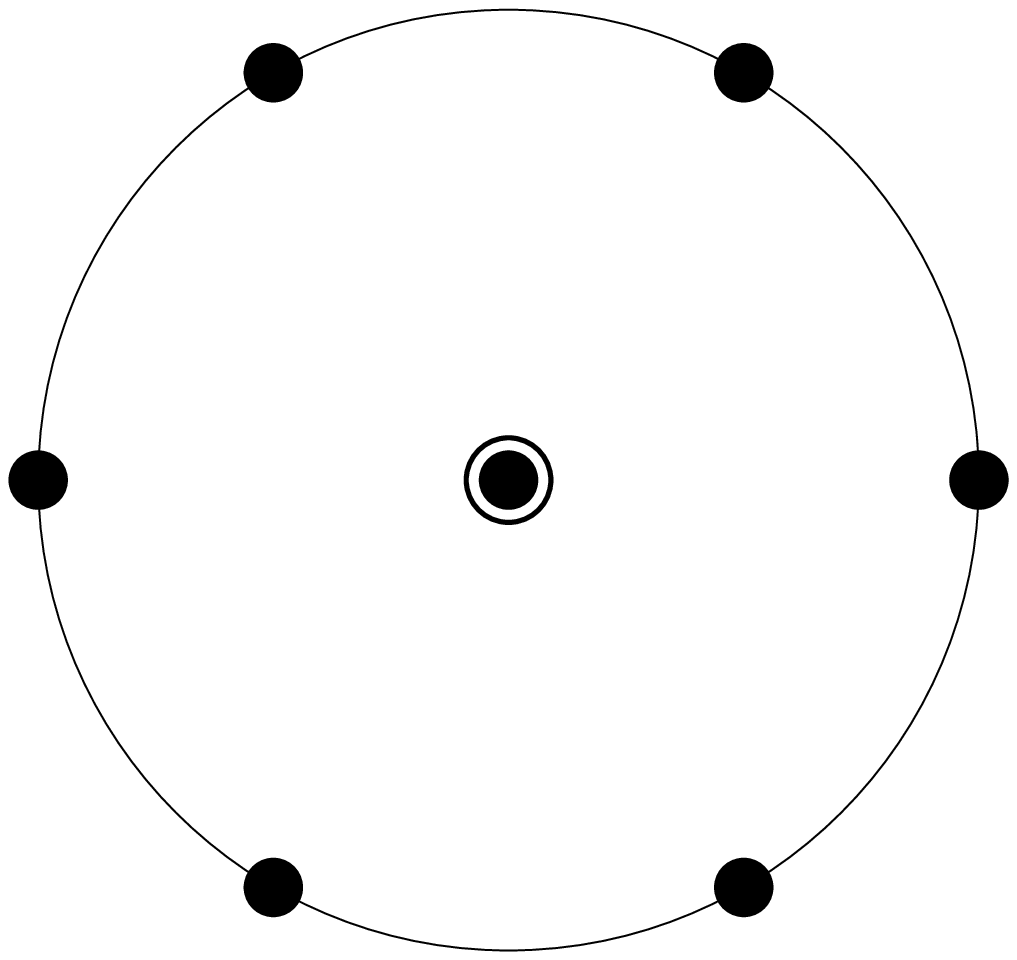}} 
\parbox{1.1in}{\includegraphics[width=1.1in]{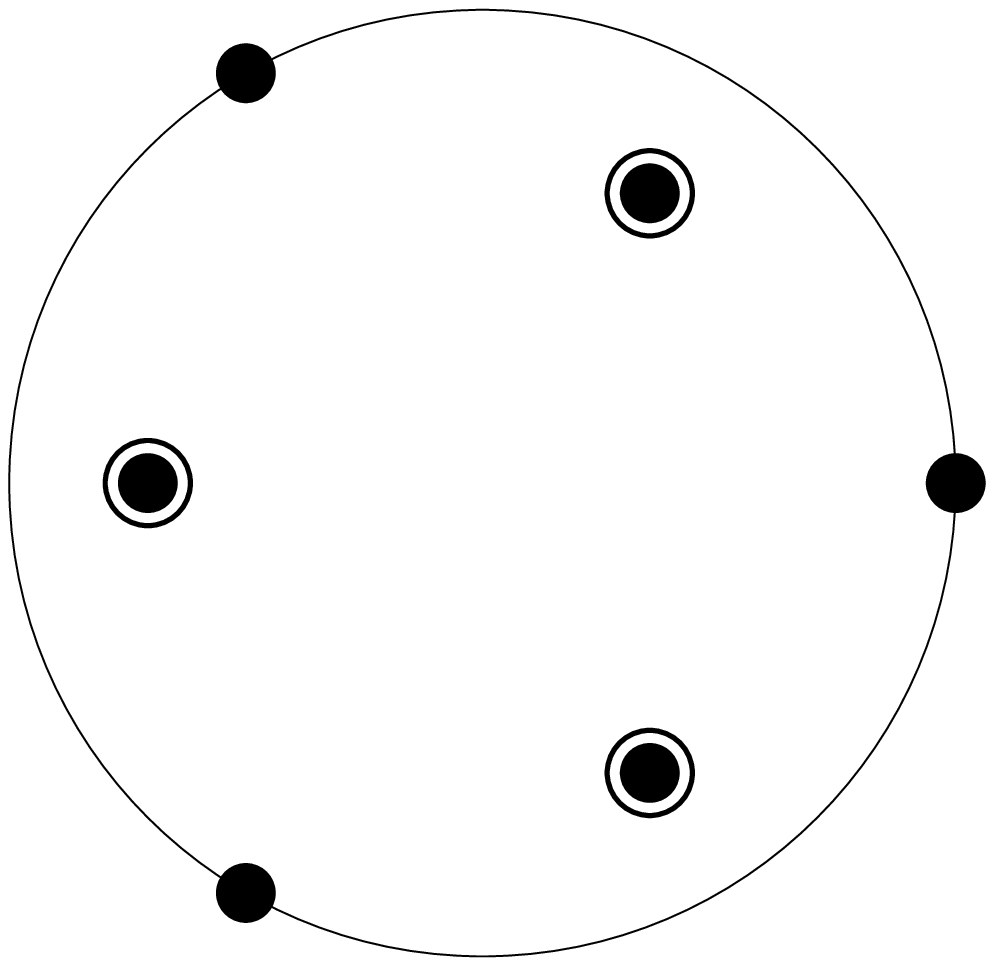}} 
\parbox{1.1in}{\includegraphics[width=1.1in]{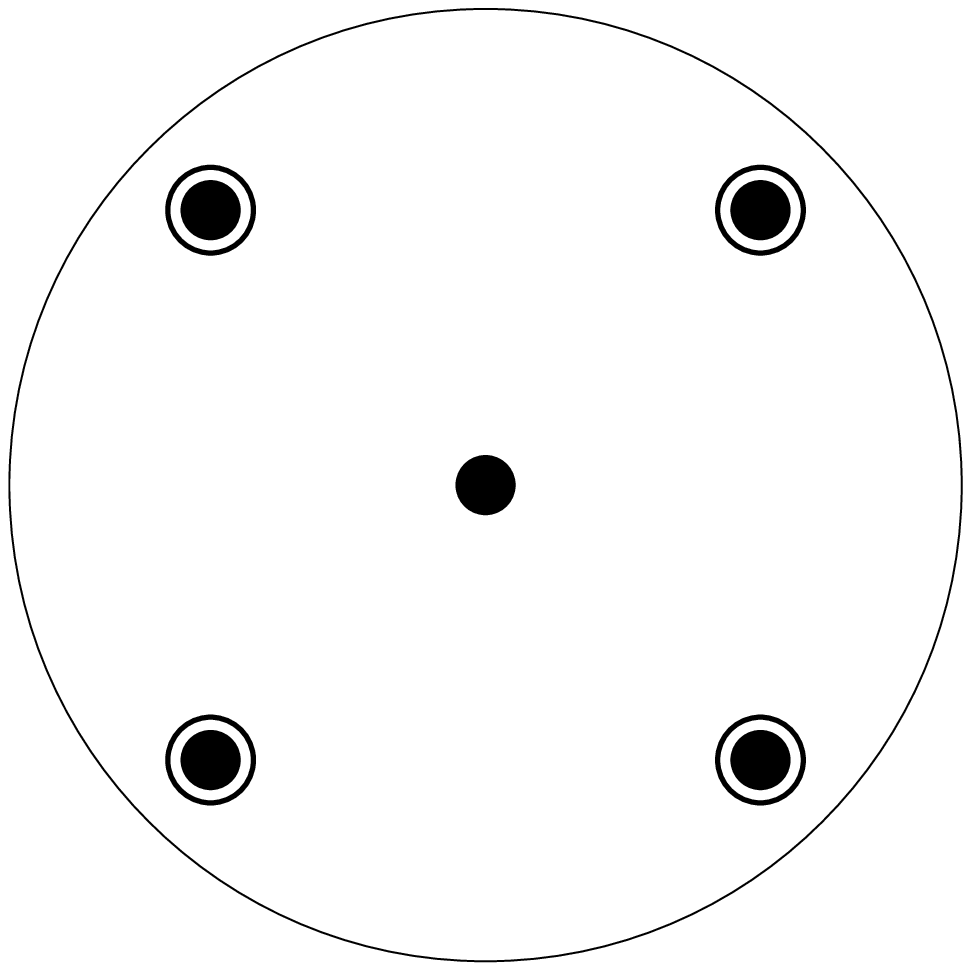}} 
\parbox{1.1in}{\includegraphics[width=1.1in]{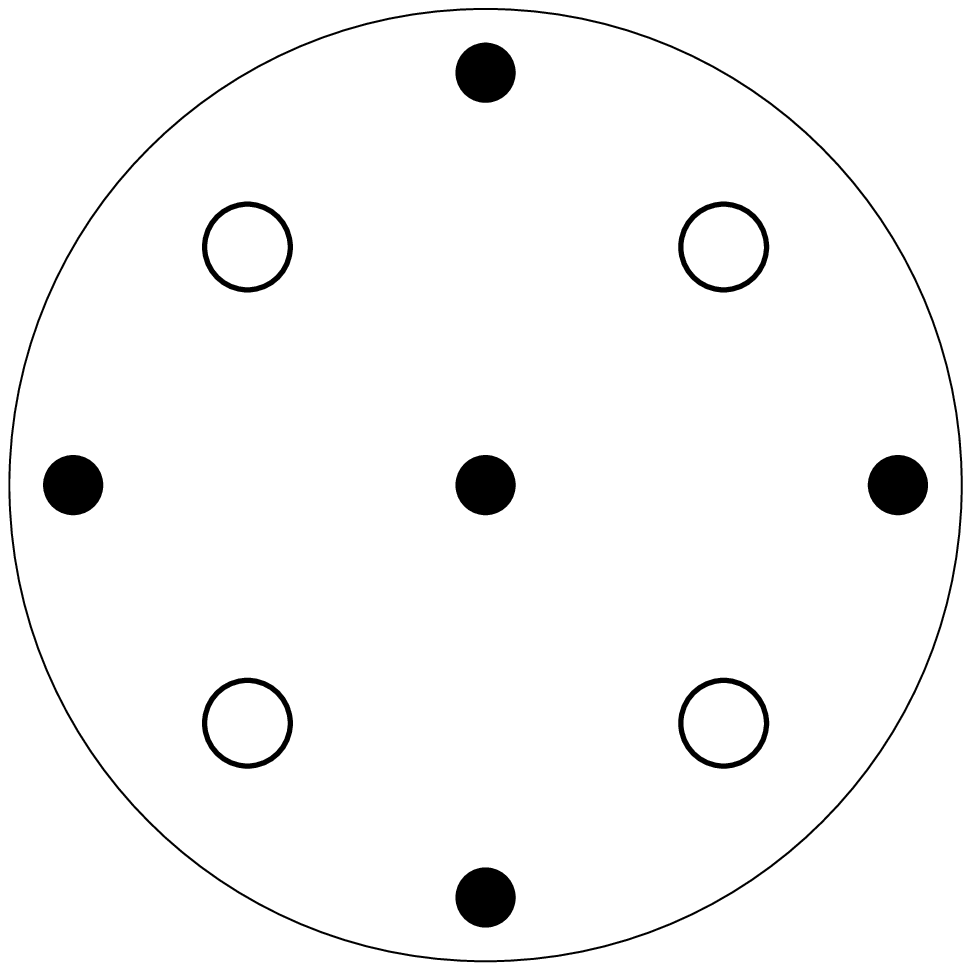}} 
\parbox{1.1in}{\includegraphics[width=1.1in]{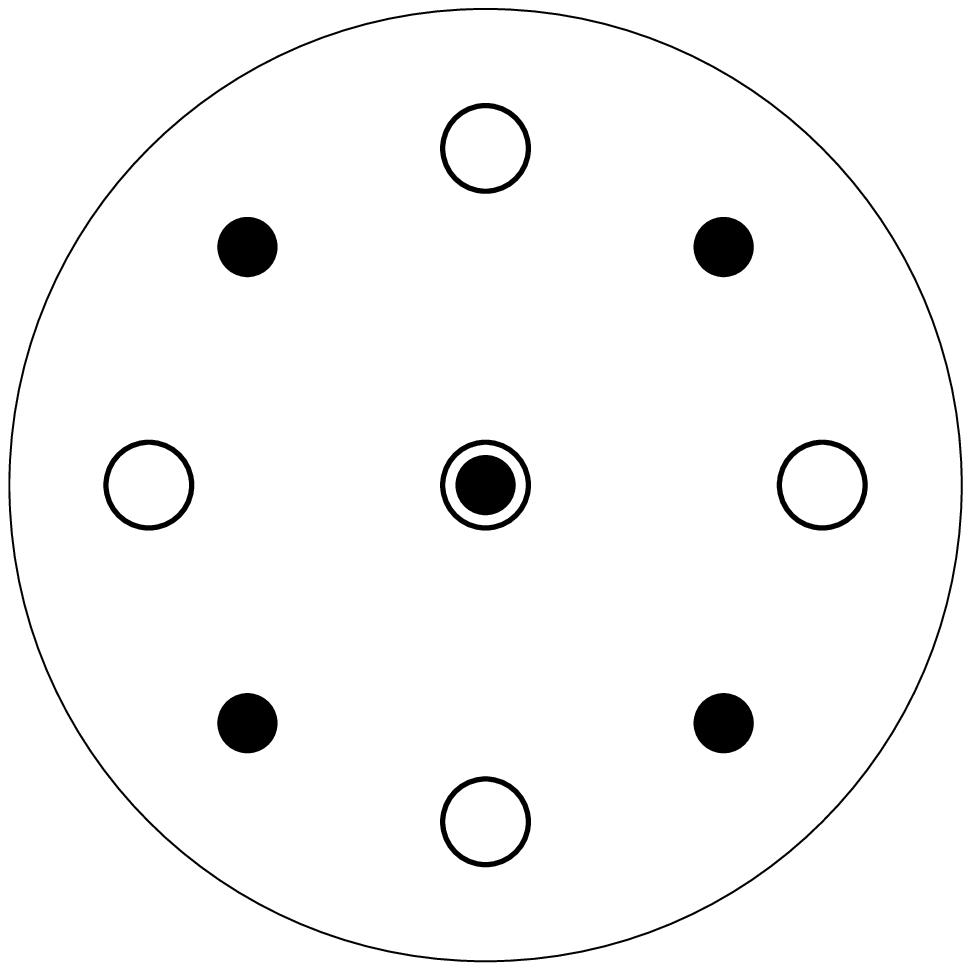}} \\
\parbox{5in}{\phantom{blah}} \\ 
\parbox{1.1in}{\centering icosahedron} 
\parbox{1.1in}{\centering cuboctahedron} 
\parbox{1.1in}{\centering dodecahedron} \\
\parbox{1.1in}{\includegraphics[width=1.1in]{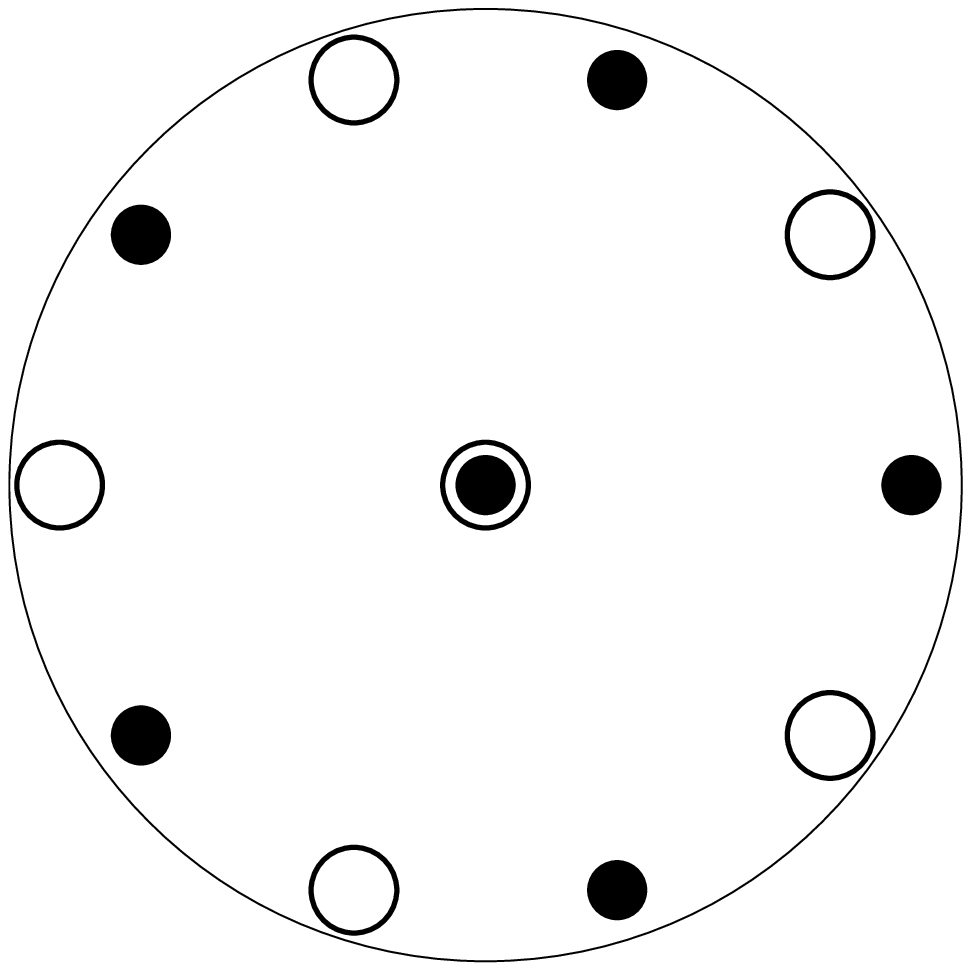}} 
\parbox{1.1in}{\includegraphics[width=1.1in]{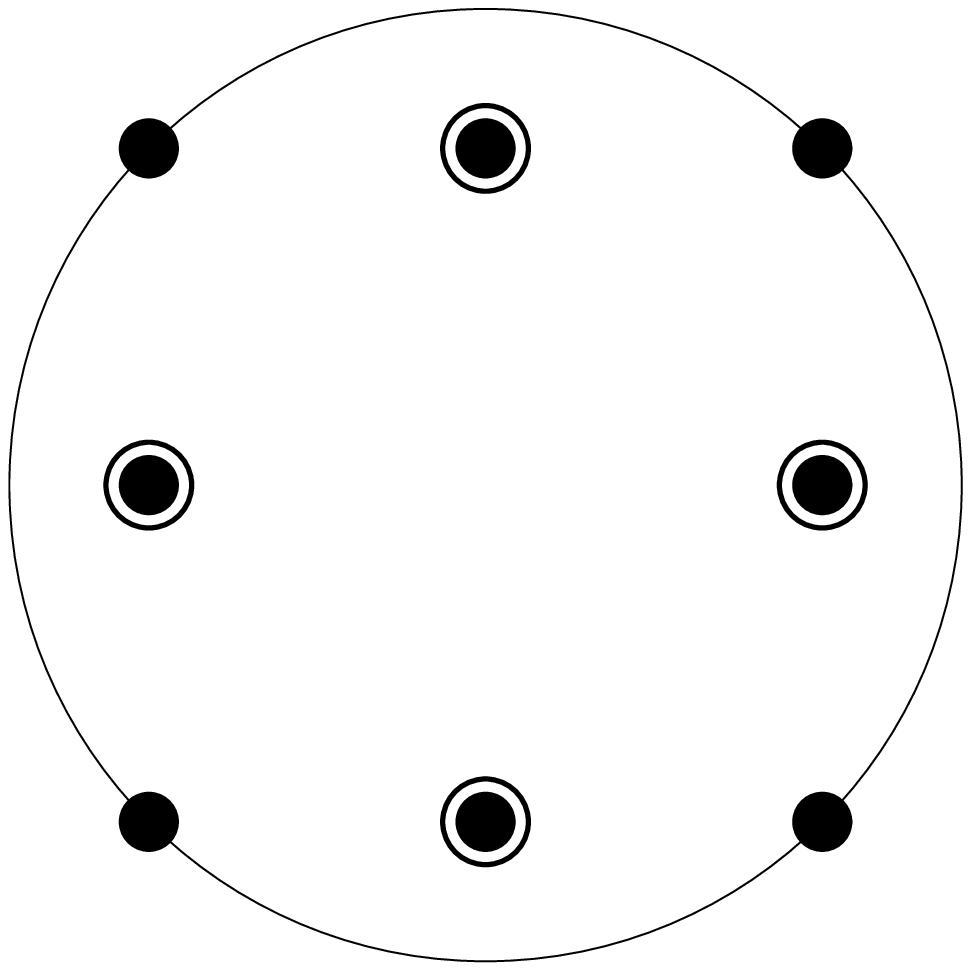}} 
\parbox{1.1in}{\includegraphics[width=1.1in]{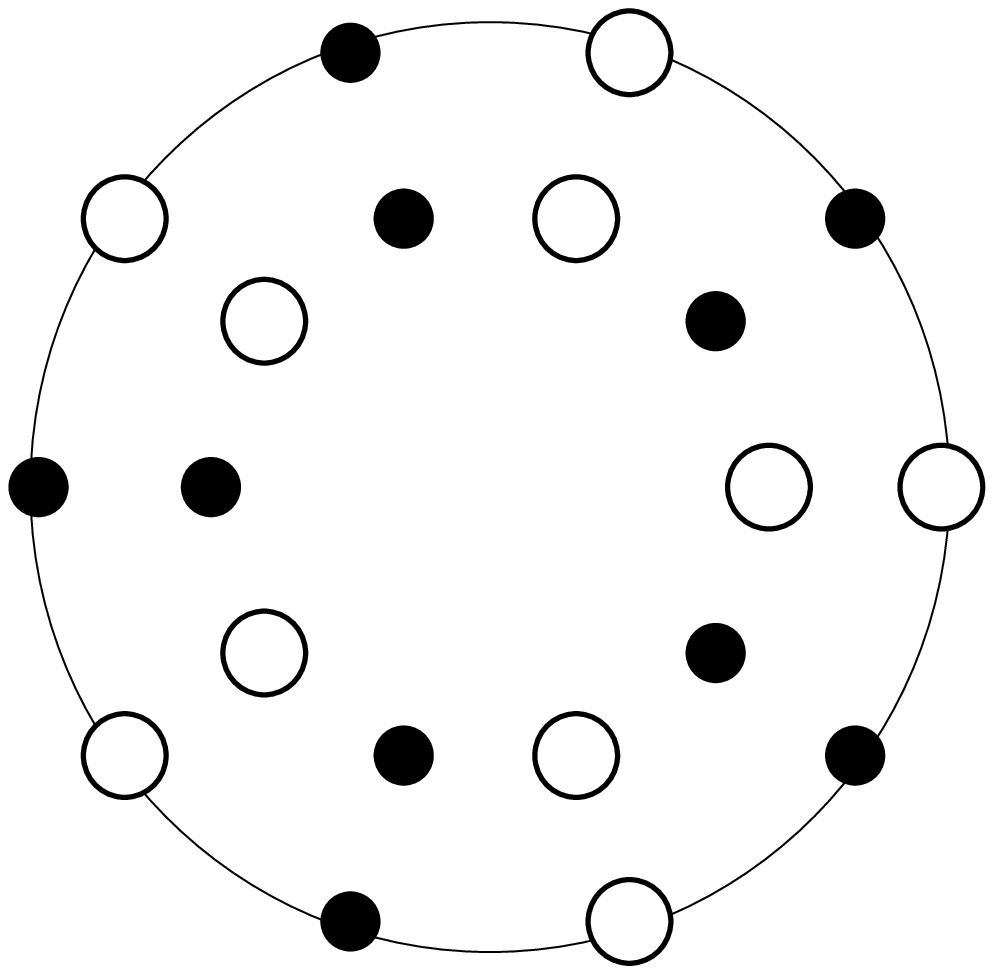}} 
\end{figure}

We have investigated a large number of different multiple-wave
configurations depicted in Fig.~\ref{stereographicfig}
and the results are compiled in Table~1.  The name of
each configuration is the name of a polygon or polyhedron that is
inscribed in a sphere of radius $q_0$; the $P$ vertices of the given
polygon or polyhedron 
then correspond to the $P$ wave vectors in the set
$\mathcal{Q}$.  
%The ``cube'' structure, for example, is the set of
%eight vectors that point from the center of a cube to its eight
%vertices.  
With this choice of nomenclature, keep in mind that what we call the
``cube'' has a different meaning than in much of the previous
literature.  We refer to an eight plane-wave configuration with
the eight wave vectors directed at the eight corners of a cube.
Because this is equivalent to eight vectors directed at
the eight faces of an octahedron --- the cube
and the octahedron are dual polyhedra --- in the nomenclature
of previous literature this eight-wave crystal would
have been called an octahedron, rather than a cube.
Similarly, the crystal that we call the ``octahedron'' 
(six plane waves whose wave vectors point at the six
corners of an octahedron) is the structure that has been
called a cube in the previous literature, because its wave
vectors point at the faces of a cube.

\begin{table}[p]
\caption{ 
Candidate crystal structures with $P$ plane waves,
specified by their symmetry group $\mathcal{G}$ and F\"{o}ppl configuration.
Bars denote dimensionless equivalents: $\bar\beta = \beta\,\dm^2$,
$\bar\gamma = \gamma\,\dm^4$, $\bar\Omega = \Omega/(\dm_2^2 N_0)$
with $N_0 = 2 \bar\mu^2/\pi^2$.  $\bar\Omega_{\min}$ is the 
(dimensionless) minimum
free energy at $\dm = \dm_2$,
obtained from (\ref{DeltaOmegaatdm2}). 
The phase transition (first order for $\bar\beta<0$ and $\bar\gamma>0$, 
second order
for $\bar\beta>0$ and  $\bar\gamma>0$) occurs at $\dm_*$.
}
\vspace{0.1in}
\begin{minipage}{6.5in}
\small
%\begin{center}
\begin{tabular}{rl|c|c|r|r|c|c}
& \multicolumn{1}{l|}{Structure}                   & P  & \multicolumn{1}{c|}{$\mathcal{G}$(F\"{o}ppl)} & \multicolumn{1}{c|}{$\bar\beta$} & \multicolumn{1}{c|}{$\bar\gamma$} & \multicolumn{1}{c|}{$\bar\Omega_{\min}$} & \multicolumn{1}{c}{$\dm_*/\Delta_0$} \\ \hline
%& \multicolumn{1}{l|}{Structure}                   & P  & \multicolumn{1}{c|}{$\mathcal{G}$(F\"{o}ppl)} & \multicolumn{1}{c|}{$\beta \dm_2^2$} & \multicolumn{1}{c|}{$\gamma \dm_2^4$} & \multicolumn{1}{c|}{$\Omega_{\min}/(\dm_2^2 N_0)$} & \multicolumn{1}{c}{$\dm_2^*/\Delta_0$} \\ \hline
1 & point 		   & 1  & $C_{\infty v}$(1) & 0.569 & 1.637 & 0 & 0.754 \\ 
2 &antipodal pair 	   & 2  & $D_{\infty v}$(11) & 0.138 & 1.952 & 0 & 0.754 \\ 
3 &  triangle 		   & 3  & $D_{3 h}$(3)      & -1.976 & 1.687 & -0.452 & 0.872 \\ 
4 &  tetrahedron 	   & 4  & $T_d$(13)    	    & -5.727 & 4.350 & -1.655 & 1.074 \\ 
5 & square 		   & 4  & $D_{4h}$(4)       & -10.350 & -1.538 & --  & -- \\ 
6 & pentagon  		   & 5  & $D_{5h}$(5)       & -13.004 & 8.386 & -5.211 & 1.607 \\ 
7 & trigonal bipyramid     & 5  & $D_{3h}$(131)       & -11.613 & 13.913 & -1.348 & 1.085 \\ 
8 & square pyramid\footnote{Minimum $\gamma$ and $\Omega_{\min}$ obtained for $\theta_2 \simeq 51.4^{\circ}$ (where $\theta_i$ is the polar angle of the $i$th F\"{o}ppl plane).}
         & 5  & $C_{4v}$(14)    & -22.014 & -70.442 & -- & -- \\
9 & octahedron  	   & 6  & $O_h$(141)          & -31.466 & 19.711 & -13.365 & 3.625 \\ 
10 & trigonal prism\footnote{Minimum $\gamma$ 
at $\theta_1=\pi-\theta_2 \simeq 43.9^{\circ}$. }
        & 6  & $D_{3h}$(33)       & -35.018 & -35.202 & -- & -- \\ 
11& hexagon  		   & 6  & $D_{6h}$(6)       & 23.669 & 6009.225 & 0 & 0.754 \\ 
12& pentagonal             & 7  & $D_{5h}$(151)      & -29.158 & 54.822 & -1.375 & 1.143 \\ 
  &  \ bipyramid 	   &    & &  &  & &\\ 
13& capped trigonal	   & 7  & $C_{3v}$($13\bar3$) & -65.112 & -195.592 & -- & -- \\
  & antiprism\footnote{Minimum $\gamma$ at $\theta_2 = \pi-\theta_3 \simeq 70.5^{\circ}$ (a cube with one 
vertex removed).}             & & & & & & \\
14 & cube  		   & 8  & $O_h$(44)          & -110.757 & -459.242 & -- & -- \\ 
%14 & square prism 	   & 8  & $D_{4h}$(44)	& & & & \\
15 & square antiprism\footnote{Minimum $\gamma$ 
at $\theta_1=\pi-\theta_2 \simeq 52.1^{\circ}$.}
  	   & 8  & $D_{4d}$($4 \bar 4$)       & -57.363 & -6.866 & -- & -- \\ 
16 & hexagonal             & 8  & $D_{6h}$(161)       & -8.074 & 5595.528 & $-2.8\times 10^{-6}$ & 0.755 \\
   & \ bipyramid 	   &    & &  &  & &\\ 
17 & augmented  	   & 9  & $D_{3h}$($3\bar 3\bar 3$)       & -69.857  & 129.259 & -3.401 & 1.656 \\ 
   & \ trigonal prism\footnote{Minimum $\gamma$ and $\Omega_{\min}$ at $\theta_1 =\pi-\theta_3 \simeq 43.9^{\circ}$. }  	   &                     &        &  & & & \\
18 & capped                & 9  & $C_{4v}$(144)    & -95.529 &  7771.152  &  -0.0024  & 0.773  \\
   & \ square prism\footnote{Best configuration is degenerate ($\theta_2 = \theta_3$, a square 
pyramid).   Result shown is for $\theta_2 \simeq 54.7^{\circ}$, $\theta_3 \simeq 125.3^{\circ}$ (a capped cube).} & & & & & & \\ 
19 & capped  		   & 9  & $C_{4v}$($14\bar 4$)       & -68.025 & 106.362 & -4.637 & 1.867 \\
   & \ square antiprism\footnote{Minimum $\gamma$ and $\Omega_{\min}$ at $\theta_2 \simeq 72.8^{\circ}$, $\theta_3 \simeq 128.4^{\circ}$.} 	   &    & &  &  & &\\ 
20 & bicapped  	   	   & 10 & $D_{4d}$($14\bar 41$)       & -14.298 & 7318.885 & $-9.1\times 10^{-6}$ & 0.755 \\ 
   & \ square antiprism\footnote{Best configuration is degenerate ($\theta_2 = 0$, an antipodal
pair).  Result shown is for $\theta_2 \simeq 72.8^{\circ}$.} 	   &    & &  &  & & \\ 
%20 & capped augmented   & 10 & $C_{3v}$($13\bar 3\bar 3$) & & & & \\
%   & \ trigonal prism   & & & & & & \\
21 & icosahedron  	   & 12 & $I_h$($15\bar 5 1$)          & 204.873 & 145076.754 & 0& 0.754 \\ 
22 & cuboctahedron  	   & 12 & $O_h$($4\bar 4\bar 4$)          & -5.296 & 97086.514 & $-2.6\times 10^{-9}$ & 0.754 \\
23 & dodecahedron  	   & 20 & $I_h$(5555)          & -527.357 & 114166.566 & -0.0019 & 0.772 \\
\end{tabular}
%\end{center}
\end{minipage}
\end{table} 

Because the LOFF pairing rings have an opening angle $\psi_0 \simeq
67.1^\circ$, no more than nine rings can be arranged on the Fermi
surface without any intersection~\cite{extremal,tammes}.  For this reason we
have focussed on crystal structures with nine or fewer waves,
but we have included several structures with more waves
in order to verify that such structures are not favored.
We have tried to analyze a fairly exhaustive list of candidate structures.
All five Platonic solids are included in Table~1, as is the
simplest Archimedean solid, the cuboctahedron. (All other Archimedean
solids have even more vertices.) We have analyzed many
dihedral polyhedra and
polygons: regular polygons, bipyramids, 
prisms,\footnote{A trigonal prism is two triangles,
one above the other. A cube is an example of a square
prism.} antiprisms,\footnote{An antiprism is a prism
with a twist. For example, a square antiprism is two
squares, one above the other, rotated relative to 
each other by $45^\circ$. The octahedron is an example
of a trigonal antiprism.}
and various capped 
or augmented polyhedra\footnote{Capping a polyhedron
adds a single vertex on the principal symmetry axis of the 
polyhedron (or polygon). Thus, a capped square is a square pyramid
and an octahedron could be called a bicapped square.
Augmenting a polyhedron means adding vertices on the equatorial
plane, centered outside each vertical facet.  Thus, augmenting
a trigonal prism adds three new vertices.}.
For each crystal
structure we list the crystal point group $\mathcal{G}$ and the
F\"{o}ppl configuration of the polyhedron or
polygon.  The F\"{o}ppl configuration
is a list of the number of vertices on circles formed by intersections
of the sphere with consecutive planes perpendicular to the principal
symmetry axis of the polyhedron or polygon.  
We use a modified notation where $a$
or $\bar a$ indicates that the points on a given circle are
respectively eclipsed or staggered relative to the circle 
above.
Note that polyhedra with several different principal symmetry axes,
namely those with $T$, $O$, or $I$ 
symmetry, have several different F\"{o}ppl descriptions: for example,
a cube is $(44)$ along a fourfold symmetry axis or $(1 3 \bar 3 1)$ along a 
threefold symmetry axis.  (That is, the cube
can equally be described as a square prism or
a bicapped trigonal antiprism. 
This should make clear
that the singly capped trigonal antiprism of Fig.~\ref{stereographicfig}
and Table~1 is a cube with one vertex removed.)

We do not claim to have analyzed all possible crystal structures,
since that is an infinite task.  However, there are several classic
mathematical problems regarding extremal arrangements of points
on a sphere and, although we do not know that our problem
is related to one of these, we have
made sure to include  
solutions to these problems. 
For example, 
many of the structures that we have evaluated correspond to solutions
of Thomson's problem~\cite{thomson, extremal} (lowest energy arrangement of 
$P$ point charges on the surface of a sphere) or Tammes's 
problem~\cite{tammes, extremal} (best packing of $P$ equal circles 
on the surface of a sphere without any overlap). In fact, we include
all solutions to the Thomson and Tammes problems for $P \leq 9$.  Our 
list also includes 
all ``balanced'' configurations~\cite{balanced} that are possible for 
nonintersecting rings: a balanced configuration is a set $\mathcal{Q}$
with a rotational symmetry about every $\vq \in \mathcal{Q}$; this 
corresponds to an arrangement of particles on a sphere for which 
the particles are in equilibrium for {\em any} two-particle force law.

For each crystal structure, we have calculated the quartic and sextic
coefficients $\beta$ and $\gamma$ according to 
Eqs.~(\ref{squarehexagonsums}), using methods described
in the appendix to calculate all the $J$ and $K$ integrals.
To further discriminate among the various
candidate structures, we also list the minimum free energy
$\Omega_{\min}$ evaluated at the plane-wave instability point $\dm =\dm_2$
where $\alpha=0$.  
To set the scale, note that the BCS state at $\dm=0$ has 
$\Omega_{\rm BCS}=-\mu^2\Delta_0^2/\pi^2$, corresponding
to $\bar\Omega_{\rm BCS}=-0.879$ in the units of Table~1. 
For those configurations with $\beta>0$ and $\gamma>0$,
$\Omega_{\min} = 0$ at $\dm=\dm_2$  and $\Omega_{\min} < 0$ for
$\dm<\dm_2$, where $\alpha<0$. Thus, we find a second-order
phase transition at $\dm=\dm_2$.
For those configurations with $\beta<0$ and $\gamma>0$,
at $\dm=\dm_2$ the minimum free energy occurs at a
nonzero $\Delta$ with $\Omega_{\min} < 0$.  (The
value of $\Delta$ at which this minimum occurs
can be obtained from (\ref{DeltaOmegaatdm2}).)
Because $\Omega_{\min}<0$ at $\dm=\dm_2$,
if we go to $\dm>\dm_2$, where $\alpha>0$, we lift this minimum
until at some $\dm_*$
it has $\Omega=0$ and becomes degenerate with
the $\Delta=0$ minimum. At $\dm=\dm_*$,
a first-order phase transition occurs. 
For a very weak
first-order phase transition, $\dm_* \simeq \dm_2 \simeq 0.754 \Delta_0$.
For a strong first-order phase transition,
$\dm_* \gg \dm_2$ and the crystalline color superconducting
phase prevails as the favored
ground state over a wider range of $\dm$.

\subsection{Crystal structures with intersecting rings lose}

There are seven configurations in Table~1 with very
large positive values for $\gamma$.  These are precisely the seven
configurations that have intersecting pairing rings: the hexagon,
hexagonal prism, capped square prism, bicapped square antiprism,
icosahedron, cuboctahedron, and dodecahedron. The first two
of these include hexagons, and since $\psi_0>60^\circ$ the
rings intersect. The last four
of these crystal structures have more than nine rings, meaning
that intersections between rings  are also inevitable.  
The capped square prism is an example of a nine-wave structure
with intersecting rings. It has a $\gamma$ which is almost
two orders of magnitude larger than that of the
augmented trigonal prism
and the capped square antiprism which, in contrast, are nine
wave structures with no intersecting rings.  
Because of their very large $\gamma$'s 
all the structures with intersecting rings 
have either second-order phase transitions or
very weak first-order phase transitions occurring
at a $\dm_*\simeq \dm_2$.  At $\dm=\dm_2$, all these
crystal structures have $\Omega_{\min}$ very close to zero.
Thus, as our analysis of two plane waves led us to expect,
we conclude that these crowded configurations with
intersecting rings are disfavored.

\subsection{``Regular'' crystal structures rule, and the cube rules them all}

At the opposite extreme, we see that there are several
structures that have negative values of $\gamma$: 
the square, square
pyramid, square antiprism, trigonal prism, capped
trigonal antiprism, and cube.  
Our analysis demonstrates that the transition to all these
crystal structures (as to those with $\beta<0$ and $\gamma>0$)
is first order.  But, we cannot evaluate $\Omega_{\min}$ or
$\dm_*$ because, to the order we are working, $\Omega$ is
unbounded from below.   For each of these crystal structures,
we could 
formulate a well-posed (but difficult) variational problem in which
we make a variational ansatz corresponding to the structure,
vary, and find $\Omega_{\min}$ without making a 
Ginzburg-Landau approximation.   It is likely, therefore, that within
the Ginzburg-Landau approximation $\Omega$ will be stabilized
at a higher order than the sextic order to which we have
worked.  

Of the sixteen crystal
structures with no intersecting rings, there are seven
that are particularly
favored by the combinatorics of Eqs.~(\ref{squarehexagonsums}).
It turns out that these seven  crystal structures
are precisely the six that we have found with $\gamma<0$,
plus the octahedron, which is the most favored crystal structure
among those with $\gamma>0$.
As discussed earlier, more
terms contribute to the rhombic and hexagonal sums in 
Eqs.~(\ref{squarehexagonsums})
when the $\vq$'s are arranged in such away that
their vertices form rectangles,
trapezoids, and cuboids inscribed in the sphere $|\vq| = q_0$.  Thus
the square itself fares well, as do the square pyramid, square
antiprism, and trigonal prism which contain one, two, and three
rectangular faces, respectively. The octahedron has
three square cross sections.
However, the cube is the outstanding winner
because it has six rectangular faces, six rectangular
cross sections, and also allows the five-corner arrangements
described previously.  The capped trigonal antiprism
in Table~1 is a cube with one vertex removed.
This seven-wave crystal has almost as many waves
as the cube, and almost as many combinatorial
advantages as the cube, and it turns out to have the
second most negative $\gamma$.

With eight rings, the cube is close to the maximum
packing for nonintersecting rings with opening angle
$\psi_0\simeq 67.1^\circ$. Although nine rings of this
size can be packed on the sphere,
adding a ninth ring to the cube and deforming the
eight rings into a cuboid, as we have done with the
capped square prism, necessarily results in
intersecting rings
and the ensuing cost
overwhelms the benefits of the
cuboidal structure.  
To form a nine-ring structure with no intersections
requires rearranging the eight rings, spoiling
the favorable regularities of the cuboid.
Therefore a nine-ring arrangement is actually less favorable
than the cuboid,
even though it allows one more plane wave.
We see from Table~1 that $\gamma$ for the cube is
much more negative than that
for any of the other combinatorially favored structures.
The cube is our winner, and we understand
why.  

To explore the extent to which the 
cube is favored, we can compare it to the octahedron, which
is the crystal structure with $\gamma>0$ for which we found
the strongest first-order phase transition, with
the largest $\dm_*$ and the deepest $\Omega_{\min}$.
The order $\Delta^6$ 
free energy we have calculated  for the cube is far below
that for the octahedron at all values
of $\Delta$.  To take an extreme example, at
$\dm=\dm_2$ the octahedron has $\bar\Omega_{\min}=-13.365$ 
at  $\Delta=1.263\dm_2=0.953\Delta_0$ whereas for the cube
we find that $\bar\Omega=-2151.5$ at this $\Delta$. 
As another example, suppose that we arbitrarily
add $+\quarter 800\Delta^8/\dm^8$
to the $\bar\Omega$ of the cube.  In this case, at $\dm=\dm_2$ 
we find that the cube has $\bar\Omega_{\min}=-32.5$ 
at $\Delta=0.656\Delta_0$ and is
thus still favored over the octahedron, even though
we have not added any $\Delta^8$ term to the free energy
of the octahedron.
These numerical exercises demonstrate the extraordinary
robustness of the cube, but should not be taken as more
than qualitative.
We do not know at what $\Delta$ and 
at what value $\Omega_{\min}$ 
the true free energy for the cube
finds its minimum.  
However, because the qualitative features
of the cube are so favorable
we expect that 
it will have a deeper $\Omega_{\min}$ and a 
larger $\dm_*$ than the octahedron. Within the Ginzburg-Landau
approximation,
the octahedron already has $\Delta=0.953\Delta_0$ 
and a deep $\bar\Omega_{\min}=-13.365$, about fifteen times
deeper than $\bar\Omega_{\rm BCS}=-0.879$ for the 
BCS state at $\dm=0$.

Even if we were to push the Ginzburg-Landau analysis of
the cube to higher order and find a stable
$\Omega_{\min}$, we would not be able to trust such
a result quantitatively.
Because it predicts a strong first-order phase transition,
the Ginzburg-Landau approximation predicts its own
quantitative demise.
What we have learned from it, however,
is that there are qualitative reasons that make the cube
the most favored crystal structure of them all. And, to
the extent that we can trust the quantitative calculations
qualitatively, they indicate that the first-order phase
transition results in a state with $\Delta$ comparable
to or bigger than that in the BCS phase, with $\Omega_{\min}$
comparable to or deeper than that of the BCS phase,
and occurs at a $\dm_*\gg \dm_2$.

\subsection{Varying continuous degrees of freedom}

None of the
regularities of the cube which make it so favorable
are lost if it is deformed 
continuously into
a cuboid, slightly shorter or taller than it is wide, as long
as it is not deformed so much as to cause rings to cross.
Next, we investigate this and some of the other possible 
continuous degrees of freedom present in a number
of the crystal structures we have described above.

So far we have neglected the fact, mentioned at the start of this
section, that some of the candidate structures have multiple orbits
under the action of the point group $\mathcal{G}$.  These structures
include the square pyramid, the four bipyramids, and the five capped
or augmented structures listed in Table~1; all have two orbits
except for the three singly capped crystal structures, which
have three orbits.
For these multiple-orbit structures each orbit should
have a different gap parameter but in Table~1 we have assumed that all
the gaps are equal.  We have, however, analyzed each of these
structures upon assuming different gaps, 
searching for a minimum of the free energy in the
two- or three-dimensional parameter space of gaps.  
In most cases, the deepest minimum
is actually obtained by simply eliminating one of the orbits from the
configuration ({\it i.e.}~let $\Delta = 0$ for that orbit); the resultant
structure with one less orbit appears as another structure in Table~1.  
For example, the bicapped square antiprism has two orbits: the
first is the set of eight $\vq$'s forming a square antiprism, the
second is the antipodal pair of $\vq$'s forming the two ``caps'' of the
structure. Denote the gaps corresponding
to these two orbits as $\Delta_1$ and $\Delta_2$.  
This structure is overcrowded with intersecting rings,
so it is not surprising to find that a lower-energy configuration is
obtained by simply letting $\Delta_2 = 0$, which gives the
``uncapped'' square antiprism.  Configurations with fewer orbits are
generally more favorable, with only three exceptions known to us: the
trigonal bipyramid is favored over the triangle or the antipodal pair;
the square pyramid is favored over the square or the point;
and the capped trigonal antiprism is favored over any of the
structures that can be obtained from it by removing one or
two orbits.  For
these configurations, Table~1 lists the results for $\Delta_1 =
\Delta_2\,(=\Delta_3)$; the numbers 
can be slightly improved with $\Delta_1 \neq \Delta_2\,(\neq \Delta_3)$ 
but the difference is unimportant.

For some configurations in Table~1 the positions of the points are
completely fixed by symmetry while for others the positions of the
points can be varied continuously, while still maintaining the
point-group symmetry of the structure.  For example, with the square
pyramid we can vary the latitude of the plane that contains the
inscribed pyramid base.  Similarly, with the various polygonal prism
and antiprism structures (and associated cappings and augmentations),
we can vary the latitudes of the inscribed polygons (equivalently, we
can vary the heights of these structures along the principal symmetry
axis).  For each structure that has such degrees of freedom, we have
scanned the allowed continuous 
parameter space to find the favored configuration.
Table~1 then shows the results for this
favored configuration, and the latitude angles describing
the favored configuration are given as footnotes.
However, if the structure always has overlapping rings
regardless of its deformation, then either no favorite configuration exists
or the favorite configuration is a degenerate one
that removes the overlaps by changing the structure.  There are two
instances where this occurs: the capped square prism can be deformed
into a square pyramid by shrinking the height of the square prism to
zero, and the bicapped square antiprism can be deformed into an
antipodal pair by moving the top and bottom square faces of the
antiprism to the north and south poles, respectively.  For these
structures, Table~1 just lists results for an arbitrarily chosen
nondegenerate configuration.

\begin{figure}
\centering
\psfrag{x1}[tc][tc]{\small $10^{\circ}$}
\psfrag{x2}[tc][tc]{\small $20^{\circ}$}
\psfrag{x3}[tc][tc]{\small $30^{\circ}$}
\psfrag{x4}[tc][tc]{\small $40^{\circ}$}
\psfrag{x5}[tc][tc]{\small $50^{\circ}$}
\psfrag{x6}[tc][tc]{\small $60^{\circ}$}
\psfrag{x7}[tc][tc]{\small $70^{\circ}$}
\psfrag{x8}[tc][tc]{\small $80^{\circ}$}
\psfrag{x9}[tc][tc]{\small $90^{\circ}$}
\psfrag{x10}[tc][tc]{\small $50^{\circ}$}
\psfrag{x11}[tc][tc]{\small $55^{\circ}$}
\psfrag{x12}[tc][tc]{\small $60^{\circ}$}
\psfrag{x13}[tc][tc]{\small $65^{\circ}$}
\psfrag{y1}[rc][rc]{\small 3000}
\psfrag{y2}[rc][rc]{\small 6000}
\psfrag{y3}[rc][rc]{\small 9000}
\psfrag{y4}[rc][rc]{\small 12000}
\psfrag{y5}[rc][rc]{\small 15000}
\psfrag{y6}[rc][rc]{\small 30}
\psfrag{y7}[rc][rc]{\small 60}
\psfrag{y8}[rc][rc]{\small 90}
\psfrag{y9}[rc][rc]{\small 120}
\psfrag{y10}[rc][rc]{\small 150}
\psfrag{xlabel}{$\theta$}
\psfrag{ylabel}{$\gamma(\theta) \dm^4$}
\includegraphics[width=6in]{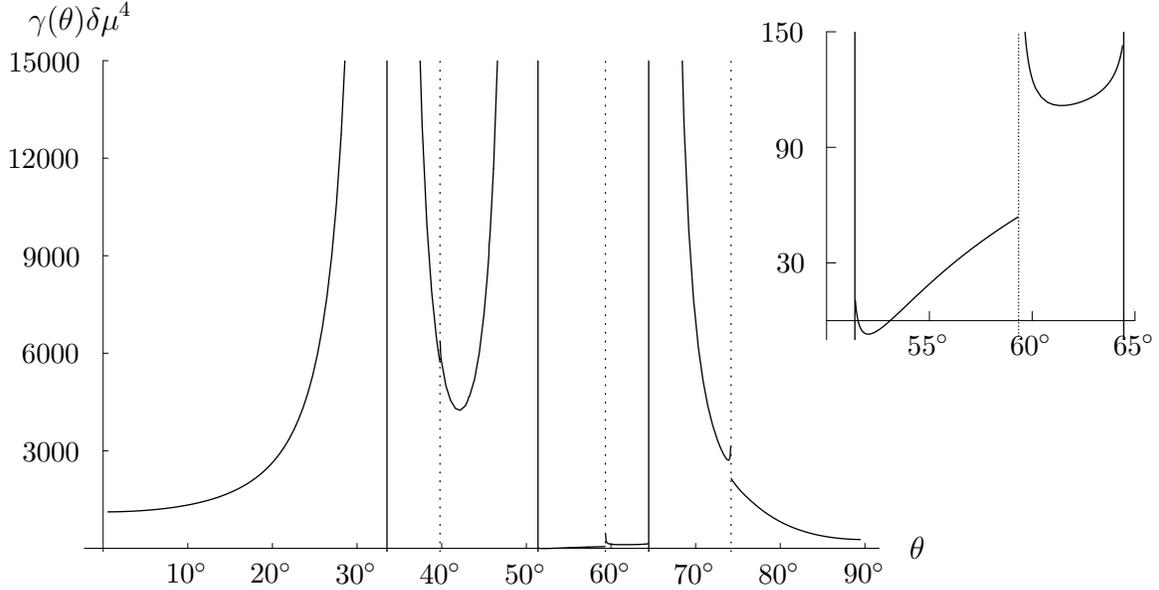}
\caption{
\label{scanfig}
The sextic free energy coefficient $\gamma$ for the square antiprism as a
function of the polar (latitude) 
angle $\theta$ of the top square facet.  (The
polar angle of the bottom square facet is $\pi-\theta$.)
The inset
plot shows the detail in the range of $\theta$ 
where no rings intersect.
Solid and dashed vertical lines indicate the positions of primary and
secondary singularies as discussed in the text (other secondary
singularities occur, but are not discernible on the plot).  }
\end{figure}

A typical parameter scan is shown in Fig,~\ref{scanfig}, where we
have plotted $\gamma$ for the square antiprism as a function of the
polar angle $\theta$ of the top square facet (the polar angle of the
bottom square facet is $\pi-\theta$).  As we expect, $\gamma$ is very
large in regions where any rings intersect, and we search for a
minimum of $\gamma$ in the region where no rings intersect.
The plot has a rather complicated structure of singularities and
discontinuities; these features are analogous to those of 
Fig.~\ref{betagammafig}, and as there they arise as a result
of the double limit we are taking.  
Primary singularities occur at critical angles
where pairing rings are mutually tangent on the Fermi surface.
Secondary singularities occur where rings 
corresponding to harmonic $\vq$'s, 
obtained by taking sums and differences of the fundamental $\vq$'s
that define the crystal structure,
are mutually tangent. Such $\vq$'s arise in the calculation
of $J$ and $K$ because these calculations 
involve momenta corresponding
to various diagonals of the rhombus and hexagon 
in Fig.~\ref{rhombushexagonfig}.

\begin{figure}
\centering
\psfrag{5}[tc][tc]{\small $5^{\circ}$}
\psfrag{10}[tc][tc]{\small $10^{\circ}$}
\psfrag{15}[tc][tc]{\small $15^{\circ}$}
\psfrag{20}[tc][tc]{\small $20^{\circ}$}
\psfrag{25}[tc][tc]{\small $25^{\circ}$}
\psfrag{30}[tc][tc]{\small $30^{\circ}$}
\psfrag{35}[tc][tc]{\small $35^{\circ}$}
\psfrag{40}[tc][tc]{\small $40^{\circ}$}
\psfrag{45}[tc][tc]{\small $45^{\circ}$}
\psfrag{-500}[rc][rc]{\small -500}
\psfrag{-400}[rc][rc]{\small -400}
\psfrag{-300}[rc][rc]{\small -300}
\psfrag{-200}[rc][rc]{\small -200}
\psfrag{-100}[rc][rc]{\small -100}
\psfrag{-100}[rc][rc]{\small -100}
\psfrag{100}[rc][rc]{\small 100}
\psfrag{200}[rc][rc]{\small 200}
\psfrag{xlabel}{$\phi$}
\psfrag{ylabel}{$\gamma(\phi) \dm^4$}
\psfrag{arrowlabel1}[lb][lt]{cube}
\psfrag{arrowlabel2}[tc][rb]{\mbox{square antiprism}}
\includegraphics[width=5in]{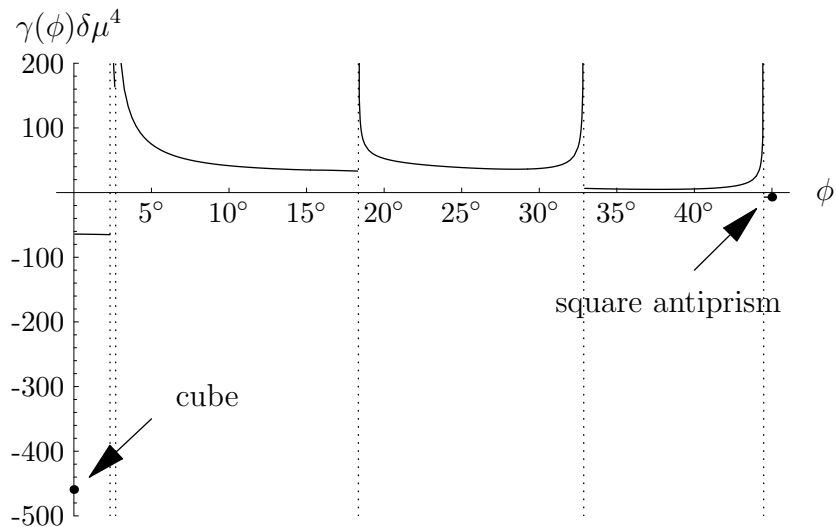}
\caption{
\label{twistfig}
The sextic free energy coefficient $\gamma$ for a scan 
that linearly interpolates from the cube (twist angle
$\phi=0$) to the square antiprism of Table~1 (twist
angle $\phi=45^\circ$). Dashed vertical lines indicate
secondary singularities.  
}
\end{figure}

In addition to varying the latitudes of the F\"oppl planes
in various structures, we varied ``twist angles''.  For
example, we explored the continuous degree of
freedom that turns a cube into a square antiprism, 
by twisting the top square relative to the bottom square
by an angle $\phi$ ranging from $0^\circ$ to $45^\circ$.
In Fig.~\ref{twistfig}, we show a parameter scan
in which we simultaneously vary the twist angle $\phi$ and the
latitudes of the square planes in such a way that
the scan interpolates linearly from the cube to the
most favorable square antiprism of Table~1.
In this parameter scan, we find a collection
of secondary singularities and one striking
fact: $\gamma$ is much more negative when 
the twist angle is zero ({\it i.e.} for the cube itself)
than for any nonzero value. For the cube, $\gamma=-459.2/\dm^4$,
whereas the best one can do with a nonzero twist 
is $\gamma=-64.2/\dm^4$, which is the result for
an infinitesimal twist angle. 
Thus, any nonzero twist spoils the regularities
of the cube that contribute to its combinatorial
advantage, and this has    
a dramatic and unfavorable effect on the free energy.

Finally, we have scanned the parameter space of a generic 
cuboid to see how this compares to the special case of a cube.  
That is, we vary the height of the cuboid relative to its
width, without introducing any twist.  This continuous
variation does not reduce the combinatorial advantage
of the  crystal structure.
As expected, therefore, we find that as long as the cuboid
has no intersecting rings, it has a free energy
that is very similar to that of the cube itself.
Any cuboid with intersecting rings is very
unfavorable.  In the restricted parameter space of nonintersecting
cuboidal arrangements, the cuboid with the most negative free
energy
is a square prism
with a polar angle of $51.4^{\circ}$ for the top square face. (For this
polar angle the pairing rings corresponding to the four corners of the
square are almost mutually tangent.)  This prism is slightly
taller than a perfect cube, which has a polar angle of $54.7^{\circ}$.
The free energy coefficients of the best cuboid
are $\beta = -111.563/(\dm)^2$, $\gamma =
-463.878/(\dm)^4$.  These coefficients differ by less than $1\%$ from those
for the cube, given in Table~1.  There is no significant
difference between the cube and this very slightly 
more favorable cuboid:
all the qualitative arguments that favor the cube favor any 
cuboid with no
intersecting rings equally well.  We therefore expect that
if we could determine the exact (rather than Ginzburg-Landau)
free energy, we would find that the favored crystal structure
is a cuboid with a polar angle somewhere between $51.4^\circ$ and
$56.5^\circ$, as this is the range for which no rings
intersect.   We expect no important distinction between the
free energy of whichever cuboid in this narrow range happens to
be favored and that of the cube itself.

%------------------------------------------------------------------------

\section{Conclusions and Open Questions}
\label{sec:conclusions}

We have argued that the cube crystal structure is the favored ground state at
zero temperature near the plane-wave instability point $\dm = \dm_2$.
By the cube we mean
a crystal structure constructed as the sum of eight
plane waves with wave vectors pointing towards the 
corners of a cube.
The qualitative points (which we have demonstrated in explicit
detail via the analysis of many different crystal structures)
that lead us to conclude that the cube is the winner are:
\begin{itemize}
\item
The quadratic term in the Ginzburg-Landau free energy wants
a $|\vq|$ such that 
the pairing associated with any single choice of $\hat\vq$
occurs on a ring with
opening angle $\psi_0\simeq 67.1^\circ$ on each Fermi surface. 
\item
The quadratic term in the Ginzburg-Landau free energy
favors condensation with many different
wave vectors, and thus many different pairing rings on the Fermi
surfaces.
However, the quartic and sextic terms in the free energy
strenuously prohibit the intersection of pairing rings.
No more than nine rings with opening angle $67.1^\circ$
can be placed on the sphere without overlap.
\item
The quartic and sextic terms favor regular crystal structures,
for example those that include
many different sets of wave vectors whose tips form rectangles.
None of the  nine-wave structures with no intersections between
pairing rings are regular in the required sense.
The cube is a very regular eight-wave crystal structure.
\end{itemize}
Quantitatively, we find that a cube (actually, a cuboid that
is only slightly taller than it is wide) has by far the most
negative Ginzburg-Landau free energy, to sextic order,
of all the many crystal structures we have investigated.

\begin{figure}
\centering
\psfrag{0}{\small 0}
\psfrag{0.25}{\small 0.25}
\psfrag{0.5}{\small 0.5}
\psfrag{0.75}{\small 0.75}
\psfrag{1}{\small 1}
\includegraphics{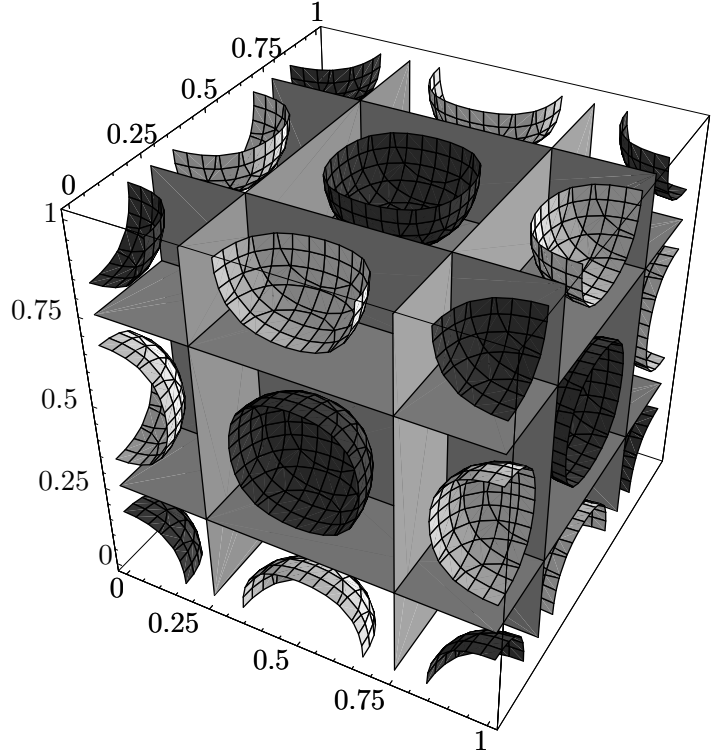}
\caption{
\label{unitcellfig}
A unit cell of the LOFF face-centered-cubic 
crystal.  The gray planes are surfaces where
$\Delta(\vx) = 0$.  The darker surfaces are contours where
$\Delta(\vx) = +4 \Delta$, and the lighter surfaces are contours where
$\Delta(\vx) = -4 \Delta$.  }
\end{figure}

Let us now see what the cubic crystal structure looks like
in position space. 
The eight $\vq$ vectors are the eight shortest vectors in
the reciprocal lattice of a face-centered-cubic crystal.
Therefore, we find that $\Delta(\vx)$ exhibits 
face-centered-cubic symmetry.  
Explicitly,
%the form of the gap function is
\begin{eqnarray}
\Delta(\vx) & = & 2 \Delta \Biggl[ \cos  \frac{2 \pi}{a} (x + y + z) + \cos \frac{2 \pi}{a} (x - y + z) \nonumber \\ 
& & + \cos \frac{2 \pi}{a} (x + y - z) + \cos \frac{2 \pi}{a} (-x + y + z) 
\Biggr]\ ,
%\Delta(\vx) & = & 2 \Delta [ \cos  \frac{2 q_0}{\sqrt{3}} (x + y + z) + \cos \frac{2 q_0}{\sqrt{3}} (x - y + z) \\
%& & + \cos \frac{2 q_0}{\sqrt{3}} (x + y - z) + \cos \frac{2 q_0}{\sqrt{3}} (x - y - z) \Biggr]\ ,
\label{fcccrystal}
\end{eqnarray}
where the lattice constant ({\it i.e.}~the edge length of the unit cube) is
\begin{equation}
a = \frac{\sqrt{3}\pi}{|\vq|} \simeq \frac{4.536}{\dm}
\simeq \frac{6.012}{\Delta_0}\ ,
\end{equation}
where the last equality is valid at $\dm=\dm_2$ and 
where $\Delta_0$ is the gap of the BCS phase that would occur at $\dm=0$.  
A unit cell of the crystal is shown in Fig.~\ref{unitcellfig}.
We have taken $\Delta(\vx)$ to be real by convention: we have
the freedom to multiply $\Delta(\vx)$
by an overall $\vx$-independent phase.
This freedom corresponds to the fact that the condensate spontaneously
breaks the $U(1)$ symmetry associated with conservation of quark number.  

Our Ginzburg-Landau analysis predicts a first-order phase transition
to the cubic crystalline color superconductor at
some $\dm=\dm_*$.  The fact that we predict a first-order
phase transition means that the Ginzburg-Landau analysis
cannot be trusted quantitatively.  Furthermore, at order $\Delta^6$,
which is as far as we have gone, the Ginzburg-Landau free energy
for the cube is unbounded from below.  We therefore 
have no quantitative prediction of
$\dm_*$ or the magnitude of $\Delta$.
The best we can do is to note that the cube is significantly
favored over the octahedron, for which the order $\Delta^6$
Ginzburg-Landau analysis predicts $\dm_*\simeq 3.6\Delta_0$
and predicts that at $\dm=\dm_2=0.754\Delta_0$, the gap
is $\Delta\simeq 0.95\Delta_0$ and the condensation energy is
larger than that in the BCS state by a factor of about fifteen.
As we have warned repeatedly, these numbers should not
be trusted quantitatively: because the Ginzburg-Landau approximation
predicts a strong first-order phase transition, it predicts
its own breakdown. We have learned several qualitative
lessons from it, however:  
\begin{itemize}
\item
We have understood 
the qualitative reasons that make the cube the most
favored crystal structure of them all.
\item
The Ginzburg-Landau analysis indicates that
$\Delta$, the gap
parameter in the crystalline
phase, is comparable to $\Delta_0$, that in
the BCS phase.  
\item
We learn that the condensation energy by which
the crystalline phase is favored over unpaired quark matter
at $\dm\neq 0$
may even be larger than that for the BCS phase at $\dm=0$.
(Note that the spatial average
of $\Delta(\vx)^2$ in (\ref{fcccrystal}) is $8\Delta^2$. It is
not surprising that $\Delta\sim\Delta_0$ corresponds
to an $|\Omega_{\min}|$ in the crystalline phase which is
larger than $|\Omega_{\rm BCS}|$.)
\item
We learn that the crystalline color
superconductivity window $\dm_1<\dm<\dm_*$ is large.  Because
$\dm_2$ is not much larger than $\dm_1$, the window
$\dm_1<\dm<\dm_2$ wherein the single plane-wave condensate
is possible is narrow.  We have learned, however, that
$\dm_*\gg\dm_2$.  Furthermore, because the
condensation energy of the crystalline phase is so robust,
much greater than that for the single plane wave and
likely comparable to that for the BCS phase,
the value of $\dm_1$, the location of the
transition between the crystalline
phase and the BCS phase, will be significantly depressed. 
\end{itemize}

Much remains to be done:

\begin{itemize}

\item
Our Ginzburg-Landau analysis provides a compelling argument
that the crystalline color superconductor is face-centered-cubic.
Given that, and given the prediction of a strong first-order
phase transition, the Ginzburg-Landau approximation should
now be discarded.  What should be taken from our work
is the prediction that the structure (\ref{fcccrystal})
is favored. 
Although $\Delta$ could be estimated by going
to higher order in the Ginzburg-Landau approximation,
a much better strategy is to do 
the calculation of $\Delta$ 
upon assuming the crystal structure (\ref{fcccrystal}) 
but without requiring $\Delta$ to be small.
One possible method of analysis, along
the lines of that applied in Refs.~\cite{FF,Takada,BowersLOFF}
to the single plane wave, would be
to find a variational wave function that incorporates
the structure (\ref{fcccrystal}), vary within this ansatz, 
and find $\Delta$ and $\Omega_{\min}$. Another possibility
is to assume (\ref{fcccrystal}) and then truncate
the infinite set of coupled
Nambu-Gorkov gap equations without assuming $\Delta$ is
small. This strategy would be the analogue of that
applied to the chiral crystal in  Ref.~\cite{RappCrystal}.
A third possibility
is to analyze the state with crystal structure  (\ref{fcccrystal})
using methods like those applied to the two plane-wave 
(antipodal pair) crystal structure in Ref.~\cite{Matsuo}.
Regardless of what method is used, because $\Delta\neq 0$
a complete calculation would require the inclusion of 
condensates with $\vq$'s
corresponding to higher harmonics of the fundamental $\vq$'s that
define the crystal structure.

\item
The cubic crystal structure (\ref{fcccrystal}) breaks $x$-,
$y$- and $z$-translational invariance, and therefore has
three phonon modes.  With the crystal structure (\ref{fcccrystal})
in hand, it would be very interesting to use the methods
developed in Refs.~\cite{LOFFphonon} to derive the dispersion
relations for these phonons and compute the parameters of
the low energy effective Lagrangian describing these phonons.

\item
How could the ``crystalline superfluid'' state be detected
in an ultracold gas of fermionic atoms?  The simplest idea
would be to look for the periodic modulation of the 
atom density, which goes like $\Delta(x)^2\mu^2$.
This means that the ratio of the magnitude of the spatial modulation 
of the density to the density itself is of order $\Delta(x)^2/\mu^2$.
Since the density modulation is only sensitive to $\Delta(x)^2$,
it will not see the sign of $\Delta$ in Fig.~\ref{unitcellfig}.
That is, the crystal structure for $\Delta^2$ is simple cubic,
with a unit cell of size $a/2$.

\item
In the QCD context, 
we need to do a three-flavor analysis, beginning with
the unpaired quark matter of (\ref{UnpairedFermiMomenta})
and letting $ud$, $us$ and $us$ crystalline
condensates form.  Given the robustness of the crystalline
color superconducting condensates we have found, 
we expect the three-flavor crystalline color superconductor
gap, critical temperature 
and condensation energy to be comparable to or even larger
than those
of the CFL phase.  
This means that just as the crystalline
color superconductor will push $\dm_1$ down in our model,
in three flavor QCD it will push
the unlocking transition between the CFL and crystalline
phases up to a higher density than has been estimated to date,
since present estimates have
assumed the condensation energy in the crystalline
phase is negligible.
It also means that
just as in our model the crystalline window $\dm_1<\dm<\dm_*$
is in no sense narrow, crystalline
color superconductivity will be a generic feature of
nature's 
QCD phase diagram, occurring wherever quark matter that
is not color-flavor locked is to be found.

\item
What effects could the presence of a layer of crystalline
color superconducting quark matter have on observable properties of
a compact star?  Could it be the place where
some pulsar glitches originate, as suggested in Ref.~\cite{BowersLOFF}?
The next step toward  answering this question 
and toward studying the properties of glitches that may
originate within crystalline color superconducting quark matter
is to take the crystal structure (\ref{fcccrystal}) and rotate it.
What is the structure of a rotational vortex in this 
phase?  Are vortices pinned?  
It is reasonable to guess that
the vortex core will prefer to live where the condensate
is already weakest, thus along the line of intersection of
two of the nodal planes in Fig.~\ref{unitcellfig}.
If this intuition is correct, the vortices would 
be pinned.
Constructing the vortices
explicitly (and then calculating
the pinning force) will be a challenge.  Vortices are usually
constructed  
beginning with a Ginzburg-Landau free energy functional
written in terms of $\Delta(\vx)$ and $\nabla \Delta(\vx)$.
Instead, we have constructed a Ginzburg-Landau functional 
written in terms of the $\Delta_\vq$'s.  In principle, this contains the
same information. But, it is not well-suited to the analysis
of a localized object like a vortex.
This means that it will take
some thought to come up with
a sensible model within which one can construct
a vortex in the crystalline condensate  (\ref{fcccrystal}).

The greatest difficulties in finding the explicit form of a vortex
will be in determining 
the details of how the vortex core interacts with the crystal structure.
Unfortunately, therefore, calculating the pinning force
will likely be a hard problem.  As always,
it should be easier
to understand the physics of a vortex far from its core. 
The natural expectation is that 
far from its core, a vortex will be described
simply by multiplying
the $\Delta(\vx)$ of (\ref{fcccrystal}) 
by $\exp[i\theta(\vx)]$, where $\theta(\vx)$ is a slowly varying
function of $\vx$ that winds once
from 0 to $2\pi$ as you follow a loop encircling the vortex
at a large distance.  In a uniform superfluid, this slowly varying
phase describes a particle-number current flowing around the vortex.
Here, the nodal planes make this interpretation difficult.
Thus, even the long distance vortex physics will be interesting to
work out.

\item
Our analysis does not apply to QCD at asymptotically high
density, where the QCD coupling becomes weak.  In this
regime, quark-quark scattering is dominated by gluon exchange
and because the gluon propagator is almost unscreened,
the scattering is dominated by forward scattering.
This works in favor of crystalline color 
superconductivity~\cite{pertloff}, but it also has
the consequence of reducing $q_0/\dm$ and hence reducing $\psi_0$.
The authors of Ref.~\cite{pertloff} find  $q_0/\dm$ reduced
almost to 1, meaning $\psi_0$ reduced almost to zero.
However, the authors of Ref.~\cite{Giannakis}
find $q_0/\dm\simeq 1.16$ at asymptotically high density, meaning
that $\psi_0\simeq 61^\circ$.
If the opening angle of the pairing rings on the 
Fermi surface does become very small at asymptotic
densities, then the crystal
structure there is certain to be qualitatively different from that
which we have found.  At present, the crystal structure
at asymptotic densities is unresolved.
This is worth pursuing, since it should ultimately
be possible to begin with asymptotically free QCD
(rather than a model thereof) and calculate the crystal
structure at asymptotic density from
first principles.
(At these densities, the strange quark mass is irrelevant
and a suitable $\dm$ would have to be introduced by hand.)
Although such a first-principles analysis
of the crystalline color superconducting state has a certain
appeal, it should be noted that the asymptotic analysis 
of the CFL state seems to be quantitatively reliable
only at densities that are more than fifteen orders of 
magnitude larger than those reached in compact 
stars~\cite{RajagopalShuster}.
At accessible densities, models like the one we have 
employed are at least as likely to be a good starting point.

\end{itemize}

%------------------------------------------------------------------------

\vspace{3ex}
{\samepage 
\begin{center} Acknowledgements \end{center}
\nopagebreak
We are grateful to Mark Alford, Joydip Kundu, Vincent Liu, Eugene
Shuster, Dam Son and Frank Wilczek for helpful discussions and to Liev
Aleo for editorial assistance.
We are grateful for the support and hospitality
of the Institute for Theoretical Physics at Santa Barbara,
in the case of JAB via the ITP Graduate Fellows program.
This research was supported in part by the U.S. Department of Energy
(D.O.E.) under cooperative research agreement \#DF-FC02-94ER40818,
and by the National Science Foundation under Grant No. PHY99-07949.
The work of JAB was supported in part by a DOD
National Defense Science and Engineering Graduate Fellowship.}

%-------------------------------------------------------------------

\renewcommand{\theequation}{A.\arabic{equation}}
% redefine the command that creates the equation no.
\setcounter{equation}{0}  % reset counter 
\section*{Appendix}  % use *-form to suppress numbering

In this Appendix, we outline the explicit evaluation 
of the loop integrals
in Eqs.~(\ref{integrals}) that occur in $\Pi$,
$J$ and $K$.  For all loop integrals, the
momentum integration is restricted to modes near the Fermi surface by
a cutoff $\omega \ll \bar\mu$, meaning that the density
of states can be taken as constant within the integration
region:
\begin{equation}
\int d^4 p 
= 
\int_{-\infty}^{+\infty} dp^0 \int_{\bar\mu-\omega}^{\bar\mu+\omega} |\vp|^2 d|\vp| \int_{4\pi} d\hat{\vp}
\approx
\bar\mu^2 \int_{-\infty}^{+\infty} dp^0 \int_{-\omega}^{+\omega} ds \int_{4\pi} d\hat{\vp}
\end{equation}
where $s \equiv |\vp|-\bar\mu$.  Each integral is further simplified by
removing the antiparticle poles from the bare propagators $G^{(0)}$
and $\bar G^{(0)}$ that appear in the integrand. (We can disregard
the antiparticles because their effect on the Fermi surface
physics of interest is suppressed by of order $\Delta/\bar\mu$.)
To see how to remove the antiparticle poles,
consider the propagator $(\pslash + 2 \qslash + \muslash_u)^{-1}$ that
appears in the $\Pi$ integral.  Recall that $\mu_u = \bar\mu - \dm$
and we work in the limit where $|\vq|, \dm \ll \omega \ll |\vp|,
\bar\mu$. We are only interested in the behavior of the propagator in
the vicinity of the particle poles where $p^0 \sim \pm(|\vp|-\bar\mu)
\ll \bar\mu$.  Therefore we can factor the denominator and drop
subleading terms proportional to $p^0$, $\dm$, or $|\vq|$ when they
occur outside of the particle pole:
\begin{eqnarray}
\frac{1}{\pslash + 2\qslash + \muslash_u} & = & 
\frac{(p^0 + \mu_u) \gamma^0 - (\vp + 2 \vq)\cdot\boldsymbol{\gamma}}{(p^0 + \mu_u - |\vp + 2\vq|)(p^0 + \mu_u + |\vp + 2 \vq|)} \nonumber \\
 & \approx & \frac{\bar\mu \gamma^0 - \vp\cdot\boldsymbol{\gamma}}{(p^0 + \bar\mu - \dm - |\vp| - 2\vq\cdot\hat{\vp})(2\bar\mu)} \nonumber \\
 & \approx & \frac{1}{2} \frac{\gamma^0 - \hat{\vp} \cdot \boldsymbol{\gamma}}{(p^0 - s - \dm - 2\vq\cdot\hat{\vp})}
\end{eqnarray}
We simplify all of the propagators in this way.  In the numerator
of each integrand we are then left with terms of the form $\gamma_\mu
\gamma^\alpha \gamma^\beta \cdots \gamma^\mu$.  After evaluating these
products of gamma matrices, the $\Pi$ integral can be written as
\begin{equation}
\Pi(\vq)  =  \int \frac{d p^0}{2\pi i} \ \frac{d\hat{\vp}}{4\pi} \int_{-\omega}^{+\omega} ds \ \left[ (p^0 + s - \dm)(p^0 - s - \dm - 2 \vq\cdot\hat{\vp}) \right]^{-1}.
\end{equation}
This integral is straightforward to evaluate: Wick rotate $p^0
\rightarrow i p_4$, do a contour integration of the $p_4$ integral,
and then do the remaining simple integrals to obtain 
Eq.~(\ref{Piandalpha}).

By power counting, we see that while the $\Pi$ integral has a
logarithmic dependence on the cutoff $\omega$, the $J$ and $K$
integrals have $1/\omega^2$ and $1/\omega^4$ cutoff dependences,
respectively.  We can therefore remove the cutoff dependence in the
$J$ and $K$ integrals by taking the limit $\omega/\dm$,
$\omega/|\vq|\rightarrow \infty$.
Then the $J$ and $K$ integrals depend only on
$\dm$ and the $\vq$'s and take the form
\begin{eqnarray}
\label{jkeqns}
\lefteqn{J(\vq_1 \vq_2 \vq_3 \vq_4) = } \nonumber \\
 & & \int \frac{d p^0}{2\pi i}  \ \frac{d\hat{\vp}}{4\pi} \int_{-\infty}^{+\infty} ds \ \prod_{i=1}^{2} \left[ (p^0 + s - \dm + 2\vk_i\cdot\hat{\vp}) (p^0 - s - \dm - 2\vl_i\cdot\hat{\vp}) \right]^{-1} \nonumber \\
\lefteqn{K(\vq_1 \vq_2 \vq_3 \vq_4 \vq_5 \vq_6)  = } \nonumber \\
 & &  \int \frac{d p^0}{2\pi i}  \ \frac{d\hat{\vp}}{4\pi} \int_{-\infty}^{+\infty} ds \ \prod_{i=1}^{3} \left[ (p^0 + s - \dm + 2\vk_i\cdot\hat{\vp}) (p^0 - s - \dm - 2\vl_i\cdot\hat{\vp}) \right]^{-1} \nonumber \\
%\lefteqn{\Pi(\vq)  =   - \frac{2 \lambda \bar\mu^2}{\pi^2} \int \frac{d p^0}{2\pi i} \ \frac{d\hat{\vp}}{4\pi} \ ds \ \left[ (p^0 + s - \dm)(p^0 - s - \dm - 2 \vq\cdot\hat{\vp}) \right]^{-1}}  \nonumber \\ 
%\lefteqn{J(\vq_1 \vq_2 \vq_3 \vq_4) = } \nonumber \\
% & & \int \frac{d p^0}{2\pi i}  \ \frac{d\hat{\vp}}{4\pi} \ ds \ \left[ (p^0 + s - \dm) (p^0 - s - \dm - 2\vq_1\cdot\hat{\vp}) \right. \nonumber \\
% & & \times \left. (p^0 + s - \dm + 2(\vq_1-\vq_2)\cdot\hat{\vp})(p^0 - s - \dm - 2(\vq_1-\vq_2+\vq_3)\cdot\hat{\vp}) \right]^{-1} \nonumber \\
%\lefteqn{K(\vq_1 \vq_2 \vq_3 \vq_4 \vq_5 \vq_6)  = } \nonumber \\
% & &  \int \frac{d p^0}{2\pi i}  \ \frac{d\hat{\vp}}{4\pi} \ ds \ \left[ (p^0 + s - \dm) (p^0 - s - \dm - 2\vq_1\cdot\hat{\vp}) \right. \nonumber \\
% & & \times  (p^0 + s - \dm + 2(\vq_1-\vq_2)\cdot\hat{\vp})(p^0 - s - \dm - 2(\vq_1 - \vq_2 + \vq_3)\cdot\hat{\vp}) \nonumber \\
% & & \times (p^0 + s - \dm + 2(\vq_1-\vq_2+\vq_3-\vq_4)\cdot\hat{\vp}) \nonumber \\
% & & \times \left. (p^0 - s - \dm - 2(\vq_1 - \vq_2 + \vq_3 - \vq_4 + \vq_5)\cdot\hat{\vp}) \right]^{-1}.
\end{eqnarray}
where we have introduced new vectors
\begin{center}
\begin{tabular}{lll}
$\vk_1 = 0$,     & $\vk_2 = \vq_1-\vq_2$,        & $\vk_3 = \vq_1 - \vq_2 + \vq_3 - \vq_4$  \\
$\vl_1 = \vq_1$, \ & $\vl_2 = \vq_1-\vq_2+\vq_3$, \ & $\vl_3 = \vq_1 - \vq_2 + \vq_3 - \vq_4 + \vq_5$.
\end{tabular}
\end{center}
Notice that these vectors are the coordinates of vertices in the
rhombus and hexagon shapes of Fig.~\ref{rhombushexagonfig}. In
particular, $(\vk_1\vk_2)$ and $(\vl_1\vl_2)$ are the pairs of
endpoints for the solid and dashed diagonals of the rhombus figure,
while $(\vk_1\vk_2\vk_3)$ and $(\vl_1\vl_2\vl_3)$ are the triplets of
vertices of the solid and dashed triangles in the hexagon figure.

We now introduce Feynman parameters to combine the denominator
factors in Eqs.~(\ref{jkeqns}).  Two sets of Feynman parameters
are used, one set for the factors involving $\vk_i$'s and one set for
the factors involving $\vl_i$'s.  For the $J$ integral the result is
\begin{eqnarray}
J & = & \int_0^1 dx_1 \ dx_2 \ \delta(x_1+x_2-1) \int_0^1 dy_1 \ dy_2 \ \delta(y_1+y_2-1) \nonumber \\
 & & \times \int \frac{dp_4}{2\pi} \ \frac{d\hat{\vp}}{4\pi} \ ds \ (s-\dm+ip_4+2\vk\cdot\hat{\vp})^{-2} (s+\dm-ip_4 + 2\vl\cdot\hat{\vp})^{-2}
\label{appJeqn}
\end{eqnarray}
where $\vk = \sum_i x_i \vk_i$, $\vl = \sum_i y_i \vl_i$.  Next, we do
the $s$ integral by contour integration, followed by the $\hat{\vp}$ 
and $p_4$ integrals.  For the $p_4$ integral, noting that the 
$s$ integration introduces a sign factor $\mbox{sgn}(p4)$
and that the integrand in (\ref{appJeqn}) depends only on $ip_4$,
we use
\begin{equation}
\int_{-\infty}^{+\infty} dp_4 \ \mbox{sgn}(p_4) \ (\ \cdots\ ) = 2\  \Re \int_{\epsilon}^{\infty} dp_4 \ (\ \cdots\ )
\end{equation}
where $\epsilon$ is an infinitesimal positive number.  The final result 
is
\begin{equation}
\label{Jinteqn}
J = \frac{1}{4} \ \Re \int_0^1 dx_1\ dx_2 \ \delta(\mbox{$\sum$} x - 1) \int_0^1 dy_1\ dy_2 \ \delta(\mbox{$\sum$} y -1) \ \frac{1}{|\vk-\vl|^2 - \dm_+^2}
\end{equation}
where $\dm_+ = \dm + i\epsilon$.  We include the infinitesimal
$\epsilon$ is so that the integral is well-defined even when
$|\vk-\vl|=\dm$ is encountered in the integration region.  This 
is a ``principal value'' specification for a multidimensional 
integral that is not Riemann-convergent.
A similar analysis for the 
$K$ integral gives the result
\begin{equation}
\label{Kinteqn}
K = \frac{1}{8} \ \Re \int_0^1 dx_1\ dx_2\ dx_3 \ \delta(\mbox{$\sum$} x-1) \int_0^1 dy_1\ dy_2\ dy_3 \ \delta(\mbox{$\sum$} y - 1) \ \frac{|\vk-\vl|^2 + 3\dm^2}{(|\vk-\vl|^2 - \dm_+^2)^3}.
\end{equation}

For the case of a single plane wave,
where all the $\vq_i$'s are equal ($\vq_1 = \vq_2 = \cdots =
\vq$), notice that $\vk_1 = \vk_2 = \vk_3 = 0$ and $\vl_1 = \vl_2 =
\vl_3 = \vq$.  Then the integrands in (\ref{Jinteqn}) and (\ref{Kinteqn})
are constants and we immediately obtain the results of (\ref{J0eqn}) and
(\ref{K0eqn}), respectively.

Finally, we must integrate the Feynmann parameters. For the J
integral, two of the integrals can be done using the delta functions,
a third can be done analytically, and the final integration is done
numerically, using an integration contour that avoids the singularity
at $|\vk-\vl|=\dm$.  For the $K$ integral, we do the $x_3$ and $y_3$
integrals using the delta functions, and then make a linear
transformation of the remaining integration variables $x_1$, $x_2$, $y_1$,
$y_2$, introducing new variables
\begin{equation}
r_i = \sum_j a_{ij} x_j + \sum_j b_{ij} y_j + c_i, \ \ \  i = 1,\ldots,4
\end{equation}
with $a_{ij}$, $b_{ij}$ and $c_i$ chosen such that 
\begin{equation}
|\vk-\vl|^2 = r_1^2 + r_2^2 + r_3^2.
\end{equation}
While such a transformation puts the integrand in a convenient simple
form, it complicates the description of the integration 
region considerably.
Therefore 
we use a Fourier-Motzkin 
elimination procedure~\cite{FourierMotzkin} to express
the four-dimensional integration region as a sum of
subregions, for each of which we have an
iterated integral with
``affine'' limits of integration:
\begin{eqnarray}
\lefteqn{\int_0^1 dx_1 \int_0^{1-x_1} dx_2 \int_0^1 dy_1 \int_0^{1-y_1} dy_2 \ ( \ \cdots \ ) \hspace{1in}} \nonumber \\
 & = & \sum_A \left[ \prod_{i=1}^4 \left( \int_{u_{i0}^{(A)} + \sum_{j<i} u_{ij}^{(A)} r_j}^{v_{j0}^{(A)} + \sum_{j<i} v_{ij}^{(A)} r_j} dr_i \right) \left\| \frac{\partial(x,y)}{\partial r} \right\| ( \ \cdots \ ) \right]
\end{eqnarray}
%\begin{eqnarray}
%\lefteqn{\int_0^1 dx_1 \int_0^{1-x_1} dx_2 \int_0^1 dy_1 \int_0^{1-y_1} dy_2 \ ( \ \cdots \ ) } \nonumber \\
% & = & \sum_A \left[ \int_{x_1}^{x_2} dx \int_{m_1 x + b_1}^{m_2 x + b_2} dy \int_{p_1 x + q_1 y + c_1}^{p_2 x + q_2 y + c_2} dz \int_{s_1 x + t_1 y + u_1 z + v_1}^{s_2 x + t_2 y + u_2 z + v_2} dw \ ( \ \cdots \ ) \right]
%\end{eqnarray}
For each subregion $A$, we can immediately do the $r_4$ integration
since the integrand is independent of $r_4$.  We are left with a
three-dimensional integral over the volume of a polyhedron with six
quadrilateral faces.  We then
apply the divergence theorem to turn the three-dimensional
integral into 
a sum of surface integrals over the faces.  For each surface integral,
we convert to plane polar coordinates $(\rho, \phi)$ so that
$|\vk-\vl|^2 = \rho^2 + d^2$, where $d$ is the distance from the
origin $(r_1,r_2,r_3) = (0,0,0)$ to the plane of integration.  Now the
$\phi$ integration can be done because the integrand is independent of
$\phi$.  Finally, the $\rho$ integration is done numerically, using a
deformed integration contour that avoids the singularity at 
$|\vk-\vl| = \dm$.

%We use a Fourier-Motzkin elimination prodecure to 

%------------------------------------------------------------------------

\end{document}